%% file: main.tex
\documentclass{article}

\input{preamble}

\begin{document}

% Import the subfiles in the order in which you want to run them.
% You can use \frontmatter, \mainmatter, \appendix and \backmatter to give small roman numbers, normal numbers, capital roman numbers and no numbers, respectively.

%\subfile{titlepage} % Example of a title page

\frontmatter % We want lowercase roman numerals here. Each *matter automatically forces the stuff after it to start on a new page

\maketitle

\subfile{abstract} % Subfile containing the abstract
\clearpage % Contents on next page
\tableofcontents % The table of contents, always necessary

\mainmatter % We want normal pagenumbers here.
\subfile{00_intro} % Introduction

\subfile{01_spacetime} % Geometry/spacetimes
\subfile{02_integrators} % Integrators
\subfile{03_results} % images
\subfile{04_code} % code

\subfile{05_conclusion} % conclusion

\newpage
\appendix % From here on, sections are numbered with capital letters

\subfile{A_geometry} % Geometry and Hamiltonians
%\subfile{appendixAB} % Contains two 'appendices'
%\subfile{appendixD}

\backmatter % Capital Romans from here on

% Print the bibliography:
\printbibliography[% 
% There are a lot of options for partial bibliographies, but you probably won't use them.
% title = Optional custom title (not recommended)
]

% You can comment/uncomment the following line to makes your references appear in the Table of Contents, or not.
\addcontentsline{toc}{section}{References}

\end{document}

\section{Python Scripts}
\subsection{Code for propagation of rays}
\begin{lstlisting}[language=Python]
#prerequisites: 
#Visual studio 2019.
#CUDA 11.5 (note, earlier versions might work, 11.6 definitely does not).
#Both NumPy and CuPy libraries, CUPY requires the two above.
#Pillow for exporting an image

#in general variables starting with 'the/loc' are local in the function/loop, whereas 'glob/global' are global variables

import sys
import math
import numpy as np
import cupy as cp
import time
from PIL import Image

#The three constants that define our black hole
globalM = 0.5
globalS = 0.24
globalQ = 2.0

#Image resolution and field of view for raytracing

resX = 1000
resY = 1000
#resY must be even
screenResolution = (resX,resY)
globalFOV = 75.0

#function that defines the metric, it automatically uses the global constants defined for the black hole
def metricMatrixNP(theCoord,locM = globalM, locS = globalS, locQ = globalQ):

    theMetricArray = [[0,0,0,0],[0,0,0,0],[0,0,0,0],[0,0,0,0]]

    #Here goes formulas to define the metric

    return(theMetricArray)
    
#Chooses the t component of a vector so that it is lightlike (note that a - sign change might be needed in some cases as it simply solves a quadratic, this needs to be hardcoded
def lightlikeT(theSpaceQdot, theCoord ,theM = globalM, theS = globalS, theQ = globalQ):
    theMatrix = metricMatrixNP(theCoord,theM, theS,theQ)
    tempA = theMatrix[0][0]
    tempB = 0
    for i in range(3):
        tempB = tempB + theSpaceQdot[i]*theMatrix[0][i+1]*2.0
    tempC = 0
    for i in range(3):
        for j in range(3):
            tempC = tempC + theSpaceQdot[i]*theMatrix[i+1][j+1]*theSpaceQdot[j]
    theTQdot = (-tempB + np.sqrt(tempB*tempB - 4.0*tempA*tempC))/(2.0*tempA)
    return(theTQdot)

#The real calculation. Note that once data is sent to the GPU there is no more communication until it is done.
#No constants from "outside" can be used or altered in any way. The only data the GPU knows is what is given in the function.
#The method used for geodesics is Hamiltonian. As follows:
#\partial_\tau q^\mu = g^{\mu\nu}p_\nu
#\partial_\tau p_\mu = \partial_\mu g^{\nu\sigma}p_\nu p_\sigma
#Here upper indices act as vectors, lower indices act as covectors, such as the momentum
#g^{\mu\nu} is given by theMetricArray[\mu][\nu]
#\partial_\mu g^{\nu\sigma} is given by theDerMetricArray[\mu][\nu][\sigma]
#Timesteps are normalized by \dot{q_t} so that the simulation is in coordinate time.
#A cutoff is made once the timestep reaches a certain threshold, this is considered reaching the event horizon
#A cutoff is also made if q_r, the radius, reaches 1.5 times the starting radius. This is considered flying off to infinity
#In the end only the end position of the input geodesic is returned

geodesicKernel = cp.ElementwiseKernel(
   'float32 theT, float32 theR, float32 theTheta, float32 thePhi, float32 thePT, float32 thePR, float32 thePTheta, float32 thePPhi, float32 locM, float32 locS, float32 locVarQ, int32 theN',
   'float32 theEndT, float32 theEndR,float32 theEndTheta,float32 theEndPhi',
   '''

//Variable initialisation, code within this kernel is not in python, it is in C
double theQ[4] = {theT, theR, theTheta, thePhi};
double theP[4] = {thePT, thePR, thePTheta, thePPhi};
double theR0 = 1.5*theR;

double derPrecision = 0.00000001;
double stepPrecision = 0.00015;

double Metric[4][4] = {{0,0,0,0},{0,0,0,0},{0,0,0,0},{0,0,0,0}};
double dRMetric[4][4] = {{0,0,0,0},{0,0,0,0},{0,0,0,0},{0,0,0,0}};
double dThetaMetric[4][4] = {{0,0,0,0},{0,0,0,0},{0,0,0,0},{0,0,0,0}};
double DerMetric[4][4][4] = {{{0,0,0,0},{0,0,0,0},{0,0,0,0},{0,0,0,0}},{{0,0,0,0},{0,0,0,0},{0,0,0,0},{0,0,0,0}},{{0,0,0,0},{0,0,0,0},{0,0,0,0},{0,0,0,0}},{{0,0,0,0},{0,0,0,0},{0,0,0,0},{0,0,0,0}}};

double theDeltaP1[4] = {0,0,0,0};
double theDeltaQ1[4] = {0,0,0,0};
double theDeltaP2[4] = {0,0,0,0};
double theDeltaQ2[4] = {0,0,0,0};
double theDeltaP3[4] = {0,0,0,0};
double theDeltaQ3[4] = {0,0,0,0};
double theDeltaP4[4] = {0,0,0,0};
double theDeltaQ4[4] = {0,0,0,0};

//The RK4 integrator in a loop
for (int i = 0 ; i < theN ; i++) { 
    
//---------------------- K1
    metricAtPoint(Metric,theQ[1],theQ[2],locM,locS,locVarQ);
    metricAtPoint(dRMetric,theQ[1] + derPrecision,theQ[2],locM,locS,locVarQ);
    metricAtPoint(dThetaMetric,theQ[1],theQ[2] + derPrecision,locM,locS,locVarQ);
    
    for(int mu = 0 ; mu < 4 ; mu++) {
    for(int nu = 0 ; nu < 4 ; nu++) {
        DerMetric[1][mu][nu] = (dRMetric[mu][nu] - Metric[mu][nu])/derPrecision;
        DerMetric[2][mu][nu] = (dThetaMetric[mu][nu] - Metric[mu][nu])/derPrecision;
    };
    };
    
    for(int mu = 0 ; mu < 4 ; mu++) {
        theDeltaP1[mu] = 0;
    for(int nu = 0 ; nu < 4 ; nu++) {
    for(int sigma = 0 ; sigma < 4 ; sigma++) {
        theDeltaP1[mu]+= -0.5*DerMetric[mu][nu][sigma]*theP[nu]*theP[sigma];
    };
    };
    };
    
    for(int mu = 0 ; mu < 4 ; mu++) {
        theDeltaQ1[mu] = 0;
    for(int nu = 0 ; nu < 4 ; nu++) {
        theDeltaQ1[mu]+= Metric[mu][nu]*theP[nu];
    };
    };
    
    double timeScale = abs(stepPrecision/theDeltaQ1[0]); //not sure what to do here, just went 1st order

//---------------------- K2
    metricAtPoint(Metric,theQ[1]+0.5*timeScale*theDeltaQ1[1],theQ[2]+0.5*timeScale*theDeltaQ1[2],locM,locS,locVarQ);
    metricAtPoint(dRMetric,theQ[1]+0.5*timeScale*theDeltaQ1[1] + derPrecision,theQ[2]+0.5*timeScale*theDeltaQ1[2],locM,locS,locVarQ);
    metricAtPoint(dThetaMetric,theQ[1]+0.5*timeScale*theDeltaQ1[1],theQ[2]+0.5*timeScale*theDeltaQ1[2] + derPrecision,locM,locS,locVarQ);
    
    for(int mu = 0 ; mu < 4 ; mu++) {
    for(int nu = 0 ; nu < 4 ; nu++) {
        DerMetric[1][mu][nu] = (dRMetric[mu][nu] - Metric[mu][nu])/derPrecision;
        DerMetric[2][mu][nu] = (dThetaMetric[mu][nu] - Metric[mu][nu])/derPrecision;
    };
    };
    
    for(int mu = 0 ; mu < 4 ; mu++) {
        theDeltaP2[mu] = 0;
    for(int nu = 0 ; nu < 4 ; nu++) {
    for(int sigma = 0 ; sigma < 4 ; sigma++) {
        theDeltaP2[mu]+= -0.5*DerMetric[mu][nu][sigma]*(theP[nu]+0.5*timeScale*theDeltaP1[nu])*(theP[sigma]+0.5*timeScale*theDeltaP1[sigma]);
    };
    };
    };
    
    for(int mu = 0 ; mu < 4 ; mu++) {
        theDeltaQ2[mu] = 0;
    for(int nu = 0 ; nu < 4 ; nu++) {
        theDeltaQ2[mu]+= Metric[mu][nu]*(theP[nu]+0.5*timeScale*theDeltaP1[nu]);
    };
    };

//---------------------- K3
    metricAtPoint(Metric,theQ[1]+0.5*timeScale*theDeltaQ2[1],theQ[2]+0.5*timeScale*theDeltaQ2[2],locM,locS,locVarQ);
    metricAtPoint(dRMetric,theQ[1]+0.5*timeScale*theDeltaQ2[1] + derPrecision,theQ[2]+0.5*timeScale*theDeltaQ2[2],locM,locS,locVarQ);
    metricAtPoint(dThetaMetric,theQ[1]+0.5*timeScale*theDeltaQ2[1],theQ[2]+0.5*timeScale*theDeltaQ2[2] + derPrecision,locM,locS,locVarQ);
    
    for(int mu = 0 ; mu < 4 ; mu++) {
    for(int nu = 0 ; nu < 4 ; nu++) {
        DerMetric[1][mu][nu] = (dRMetric[mu][nu] - Metric[mu][nu])/derPrecision;
        DerMetric[2][mu][nu] = (dThetaMetric[mu][nu] - Metric[mu][nu])/derPrecision;
    };
    };
    
    for(int mu = 0 ; mu < 4 ; mu++) {
        theDeltaP3[mu] = 0;
    for(int nu = 0 ; nu < 4 ; nu++) {
    for(int sigma = 0 ; sigma < 4 ; sigma++) {
        theDeltaP3[mu]+= -0.5*DerMetric[mu][nu][sigma]*(theP[nu]+0.5*timeScale*theDeltaP2[nu])*(theP[sigma]+0.5*timeScale*theDeltaP2[sigma]);
    };
    };
    };
    
    for(int mu = 0 ; mu < 4 ; mu++) {
        theDeltaQ3[mu] = 0;
    for(int nu = 0 ; nu < 4 ; nu++) {
        theDeltaQ3[mu]+= Metric[mu][nu]*(theP[nu]+0.5*timeScale*theDeltaP2[nu]);
    };
    };

//---------------------- K4
    metricAtPoint(Metric,theQ[1]+timeScale*theDeltaQ3[1],theQ[2]+timeScale*theDeltaQ3[2],locM,locS,locVarQ);
    metricAtPoint(dRMetric,theQ[1]+timeScale*theDeltaQ3[1] + derPrecision,theQ[2]+timeScale*theDeltaQ3[2],locM,locS,locVarQ);
    metricAtPoint(dThetaMetric,theQ[1]+timeScale*theDeltaQ3[1],theQ[2]+timeScale*theDeltaQ3[2] + derPrecision,locM,locS,locVarQ);
    
    for(int mu = 0 ; mu < 4 ; mu++) {
    for(int nu = 0 ; nu < 4 ; nu++) {
        DerMetric[1][mu][nu] = (dRMetric[mu][nu] - Metric[mu][nu])/derPrecision;
        DerMetric[2][mu][nu] = (dThetaMetric[mu][nu] - Metric[mu][nu])/derPrecision;
    };
    };
    
    for(int mu = 0 ; mu < 4 ; mu++) {
        theDeltaP4[mu] = 0;
    for(int nu = 0 ; nu < 4 ; nu++) {
    for(int sigma = 0 ; sigma < 4 ; sigma++) {
        theDeltaP4[mu]+= -0.5*DerMetric[mu][nu][sigma]*(theP[nu]+timeScale*theDeltaP3[nu])*(theP[sigma]+timeScale*theDeltaP3[sigma]);
    };
    };
    };
    
    for(int mu = 0 ; mu < 4 ; mu++) {
        theDeltaQ4[mu] = 0;
    for(int nu = 0 ; nu < 4 ; nu++) {
        theDeltaQ4[mu]+= Metric[mu][nu]*(theP[nu]+timeScale*theDeltaP3[nu]);
    };
    };

    //-----------Adding final positions
    for(int mu = 0 ; mu < 4 ; mu++) {
        theP[mu] += timeScale*0.1667*(theDeltaP1[mu] + 2*theDeltaP2[mu] + 2*theDeltaP3[mu] + theDeltaP4[mu]);
        theQ[mu] += timeScale*0.1667*(theDeltaQ1[mu] + 2*theDeltaQ2[mu] + 2*theDeltaQ3[mu] + theDeltaQ4[mu]);
    };

    //Gives exit parameters, in this case a maximum radius and minimum timestep
    if(theQ[1] > theR0) break;
    if(timeScale < stepPrecision*0.00003) break;
};

#Saves the end positions to the return variables
theEndT = theQ[0];
theEndR = theQ[1];
theEndTheta = theQ[2];
theEndPhi = theQ[3];
   ''',
   'geodesic',
   preamble='''
__device__ void metricAtPoint(double ioMetric[4][4], double locR, double locTheta, double locM, double locS, double locQ)
{

    //Here goes formulas to define the INVERSE metric
 
    ioMetric[0][0] = -1/locF + locF*locOmega*locOmega/(locDelta*locSinTheta*locSinTheta);
    ioMetric[1][1] = locF*locDelta/(locExpGamma*locRho2);
    ioMetric[2][2] = locF/(locExpGamma*locRho2);
    ioMetric[3][3] = locF/(locDelta*locSinTheta*locSinTheta);
    ioMetric[0][3] = locF*locOmega/(locDelta*locSinTheta*locSinTheta);
    ioMetric[3][0] = ioMetric[0][3];
};
   ''')

#The below kernel simply turns output positions into colours. Large R becomes blue, small R becomes black, with some fancy lines being drawn in the large R case to give the picture a background.

colourKernel = cp.ElementwiseKernel(
   'float32 theT, float32 theR, float32 theTheta, float32 thePhi',
   'int32 colors',
   '''
   float theQ[4] = {theT,theR,theTheta,thePhi};
if (theQ[1] > 8.0) {
    if (__sinf(theQ[3])*__sinf(theQ[3]) < 0.01) {
        colors = 0xff0000ff;
    } else if (__cosf(theQ[3])*__cosf(theQ[3]) < 0.01) {
        colors = 0xff00ff00;
    } else if (__sinf(2.0*theQ[2])*__sinf(2.0*theQ[2]) < 0.01) {
        colors = 0xffff0000;
    } else {
        colors = 0xff3c0000;
    }
//} else if (theQ[1] > 5.0*0.5) {
//        colors = 0xff00ff00;
} else {
    colors = 0xff000000;
}   ''',
   'colours')

start_time = time.time()
print("initializing positions")

#The position from which we view
globQ0 = [0.0,8.0,cp.pi*0.5,0.0]

#The position from which we view, but as an array of resX times resY times the same position, this is needed for correct input into the GPU
globT0 = cp.ones(screenResolution,dtype=cp.float32)*globQ0[0]
globR0 = cp.ones(screenResolution,dtype=cp.float32)*globQ0[1]
globTheta0 = cp.ones(screenResolution,dtype=cp.float32)*globQ0[2]
globPhi0 = cp.ones(screenResolution,dtype=cp.float32)*globQ0[3]

#Calculating the correct horizontal and vertical field of view, usually just the same, but if the image is not of an aspect ratio of 1 this might differ
verticalRange = np.tan(0.5*globalFOV/180*np.pi)/globQ0[1]
horizontalRange = np.arctan(np.tan(verticalRange)*resX/resY)

#The starting direction of our rays, in essence a rasterized screen in the tangent space of R3
pr = cp.ones(screenResolution,dtype=cp.float32)*-1.0
pphiPREMESH = cp.linspace(-horizontalRange,horizontalRange,screenResolution[0],dtype=cp.float32)
pthetapphiPREMESH = cp.linspace(-verticalRange,verticalRange,screenResolution[1],dtype=cp.float32)
ptheta, pphi = cp.meshgrid(pthetapphiPREMESH,pphiPREMESH)

#Getting the correct \dot{q_t} to be lightlike
pt = lightlikeT([pr,ptheta,pphi],globQ0,globalM,globalS,globalQ)

#Turning starting directions into starting momenta (which are covectors)
curvatureQ0 = metricMatrixNP(globQ0,globalM,globalS,globalQ)

globPT0 = pt*curvatureQ0[0][0] + pr*curvatureQ0[0][1] + ptheta*curvatureQ0[0][2] + pphi*curvatureQ0[0][3]
globPR0 = pt*curvatureQ0[1][0] + pr*curvatureQ0[1][1] + ptheta*curvatureQ0[1][2] + pphi*curvatureQ0[1][3]
globPTheta0 = pt*curvatureQ0[2][0] + pr*curvatureQ0[2][1] + ptheta*curvatureQ0[2][2] + pphi*curvatureQ0[2][3]
globPPhi0 = pt*curvatureQ0[3][0] + pr*curvatureQ0[3][1] + ptheta*curvatureQ0[3][2] + pphi*curvatureQ0[3][3]

#the constants, which also need to be given to the GPU per core
globalML = cp.float32(globalM)
globalSL = cp.float32(globalS)
globalQL = cp.float32(globalQ)

print(time.time()-start_time)
start_time = time.time()
print("beginning GPU calculation")

#The actual calculation. The input is 8 cupy arrays of the exact same shape (resX times resY) and 3 constants and an integer, with the output being 4 cupy arrays of that shape 
Tboard,Rboard,ThetaBoard,PhiBoard = geodesicKernel(globT0,globR0,globTheta0,globPhi0,globPT0,globPR0,globPTheta0,globPPhi0,globalML,globalSL,globalQL,500000)
cp.cuda.Stream.null.synchronize()

#Saving the out positions as 4 numpy arrays, for further analysis without re-running an hours worth of simulation
np.save("outputTBoard.npy",Tboard)
np.save("outputRBoard.npy",Rboard)
np.save("outputThetaBoard.npy",ThetaBoard)
np.save("outputPhiBoard.npy",PhiBoard)

#Turning the output positions into a bmp
colourBoard = np.transpose(colourKernel(Tboard,Rboard,ThetaBoard,PhiBoard))
cp.cuda.Stream.null.synchronize()

print(time.time()-start_time)
start_time = time.time()
print("beginning drawing")

#Turning out array back to numpy to be able to draw it
colourBoardNP = cp.asnumpy(colourBoard)

#Saves the array as an image
rgb_img = im = Image.fromarray(colourBoardNP, 'RGBA')
rgb_img.save('image.png')

print(time.time()-start_time)

\end{lstlisting}

\newpage
\section{C++ code}
{\tiny

}

%% file: preamble.tex
\author{Frank Imbens \\  ITF, Utrecht University, Princetonplein 5, 3584 CC Utrecht}%, under supervision of Tanja Hinderer and Alvaro del Pino Gomez at Utrecht University Institute of Theoretical Physics}
\date{\today}
\title{Graphical Processing of Geodesic Propagation in Python}

\usepackage{
		subfiles,		% For separate main and sub documents
		graphicx,		% For image modifications and the figure environment
		amsmath,		% For the AMS math styles
		amssymb,		% The extended AMS math symbol list
		amsthm,			% For use of theorems (works together with thmtools)
		fancyhdr,		% For fancy headers and footers on pages
		subcaption,		% For easy subfigures in a plot (with nice captions)
		tikz,			% Difficult drawing of awesome vector plots
		listings,		% For listing pieces of code in a nice and neat way
		multicol,		% For easy local multicolumn use
		color,			% For handy colour definitions (used cause of styling)
		thmtools,		% Lets you define your own theorem style (used for all the fancy theorems, definitions etc.)
		etoolbox,		% Allows adjustment of commands (used to reset the claim counter at the end of a proof).
		xspace,			% Makes latex not eat spaces after commands
		dsfont,         %MATH SYMBOLS
        hyperref		% Makes links, references, the Table of Contents, etc. clickable.
        }
\usepackage[english]{babel} % Correct language setting, 'british', 'american'='english' or 'dutch'.
\usepackage[autostyle]{csquotes} % Fixes quotes to correspond to the babel language.
 \usepackage[margin=2.5cm]{geometry}
\declaretheorem[style=definition,numberwithin=subsection]{definition} %If you want your theorems to be counted per section instead of subsection, then just remove the sub from the numberwithin
\declaretheorem[style=plain,sibling=definition]{theorem}
\declaretheorem[style=plain,sibling=definition]{lemma}

\declaretheorem[style=definition]{claim}

\AtEndEnvironment{proof}{\setcounter{claim}{0}} % Sets the claim number to 0 after ending a proof

\usepackage[% Options
style = alphabetic, % 
sorting = nty % 
]{biblatex}

\newcommand\frontmatter{%
    \cleardoublepage
    \pagenumbering{roman}} %small Roman numbers

\newcommand\mainmatter{%
    \cleardoublepage
    \pagenumbering{arabic}} %normal numbers

\newcommand\backmatter{%
    \cleardoublepage %% double page style
    \pagenumbering{Roman}} %capital Roman numbers

%% file: abstract.tex
\begin{abstract}
\noindent %% Prevents an indent, like the jumps at the start of a paragraph
Black holes and other compact objects are powerful tools to observationally test Einsteins theory of General Relativity. We develop raytracing code to create visual images of compact objects that are solutions of Einsteins field equations. These include Kerr black holes and Manko-Novikov spacetimes with extra independent multipole moments. Using parallel processing on a graphical processing unit we can trace the millions of geodesics required to create such images of compact objects. The loss of symmetry in Manko-Novikov spacetime leads to strange features such as chaotic regions in the dynamics. We also analyse other properties of geodesics, such as their exit times from the strong field nearby the object, their redshift, or behaviour near capture regions. We also add objects such as accretion disks to our spacetime, and determine how they are lensed by the object. The main draw of this code is it's efficiency and simplicity, as it allows for creating these images even on older hardware, while still allowing any time independent metric to be studied. The code is publicly available on github.
\end{abstract}

\thispagestyle{plain} %% Removes the header for this page; a header on a page without a well-defined section is strange.

\restoregeometry %% pairs with the \newgeometry at the top of the page.

%% file: 00_intro.tex
% Add a new section: The enumeration of this section will continue in the mainfile, even though it's alone here. So if this is the 5th section including earlier subfiles, this will print: '5  First section' in the main file.
\newpage % Use newpage to start on a new page
\section{Introduction}
\subsection{Relativity}
Einsteins theory of relativity has revolutionized the way we view gravity. Rather than seeing gravity as a force it posits gravity as a curvature in space and time, bending paths of particles not by force, but by changing what a straight line is. This theory was first observationally checked by the perihelion precession of the planet Mercury, a movement that was at first thought to be caused by another planet.

The theory also gave rise to some strange objects that were more unexpected than small variations in the orbits of planets. The most well known of these is the black hole as discovered by Schwarzschild in 1916 \cite{schwarzschild}. This object consisted of a singularity in spacetime, shrouded in an event horizon. This event horizon marks the border in space where the pull of the singularity becomes so strong that time itself points inwards to the singularity, meaning not even light can escape. After this more solutions of this type were discovered. Descriptions of a spinning black hole were made by Kerr \cite{Kerr}. This spinning black hole, besides just pulling space in towards itself also twists space, dragging it along with the rotation. This so-called frame drag acts on test particles in much the way a spinning whirlpool would on a ship, allowing it to orbit much closer along with the rotation than against it. Furthermore its internal structure changes significantly. It has an inner and an outer event horizon, and it has a circular singularity rather than a point singularity. Black holes such as these are defined by only their mass, charge and spin, and the no-hair theorem conjectures that these are the only possible characteristics a black hole can have. \cite{misnerGrav}

However more solutions were discovered that contradict this idea. One such solution is the Manko-Novikov spacetime \cite{MankoNovikov}. This spacetime is an extension to the Kerr spacetime adding additional independent multipole moments, on top of the dipole moment of the Kerr spin. This solution is often considered non-physical due to both some if its internal characteristics and the fact that it violates the no-hair theorem \cite{nohair}. It is however still an exact solution to the Einstein field equations, meaning it could form a counterexample to the no-hair theorem if observed.

With the advent of technologies like gravitational wave detection and better telescopes we have started being able to observe the effects of these compact objects. For example in 1998 proof was found of a black hole in the centre of our galaxy \cite{ghez}. Rather than directly observing the black hole this was done using infrared telescopes to observe the orbits of stars around the black hole. In 2015 the first gravitational waves were observed using the LIGO \cite{merger}, adding to the observational proof of Einsteins relativity. In 2019 the first visual images of what we believe to be a black hole were constructed from data by the EHT (Event Horizon Telescope) collaboration \cite{EHT}. This black hole is the supermassive black hole at the center of Messier 87. More recently the EHT collaboration also created visual images of the black hole at the centre of our own galaxy. It is clear that these images can be powerful tools to test relativity when compared to simulations of shadows. For example the M87 black hole shadow was used to compute the mass of the black hole in \cite{M87Shadow2}. However to test the nature of the black holes it is important to also test alternative examples of black holes, such as \cite{M87Shadow}. Such simulations are the goal of the code written for this project.

\subsection{Theoretical Framework}
To be able to theoretically describe such objects, we will need to delve into differential geometry. This is the theory of manifolds, and the structure on those manifolds. The manifold structure that describes the spacetimes of relativity is given by pseudo-Riemannian geometry. Riemannian geometry was first created to be able to give a local concept of distance to manifolds. This concept of distance is given in the form of a smooth family of inner products on the tangent space. This concept of distance was connected by Levi Civita \cite{levicivita} to that of parallel transport and covariant derivatives. However the structure of pure Riemannian metrics is too rigid the describe the manifolds of general relativity, so we turn to pseudo-Riemannian manifolds.

A pseudo-Riemannian manifold is given by a manifold $Q$ and a metric $g$. This metric is a symmetric and non-degenerate $2$-tensor $g:TQ\times TQ\rightarrow \mathbb{R}$, where $TQ$ is the tangentspace of the spacetime $Q$. In the case that $g$ is positive definite it forms a smooth family of inner products on $TQ$, and the manifold is called Riemannian. This will however not be the case for the manifolds we find in general relativity. Here the metrics are Lorentzian, which means that locally they have one negative direction and three positive ones. The canonical example of this is $\mathbb{R}^4$ with the Minkovski metric $\eta = \text{diag}(-1,1,1,1)$. This is often called flat space, and is the subject of study in the special theory of relativity.

We can see that this metric splits vectors or directions into three distinct sets. Vectors $v$ that have $g(v,v) < 0$ are called timelike, and these directions are those that particles and objects with mass can travel. Directions with $g(v,v) = 0$ are called lightlike, these are the directions that light can travel, and form the lightcone. Lastly directions with $g(v,v) > 0$ are spacelike, these paths cannot be followed by physical objects. This splitting gives rise to the lightcone, which encompasses the points reached by light- and timelike directions from some starting points. This gives all the points in the causal future of our starting point, which is all the points that can be reached and affected.

Once the space becomes more complicated it helps to look not just at the spacetime manifold, but at the entire phase space. First one can do this by going from the spacetime manifold $Q$ to it's tangent space $TQ$. A point in this space encodes not just a position but a velocity. Now one can define a Lagrangian, which is a function $\mathcal{L}:TQ\rightarrow\mathbb{R}$. On a pseudo-Riemannian manifold this Lagrangian is often given by $\mathcal{L}(q,v) = g_q(v,v)$ (where $g_q$ is the metric at the point $q$). The dynamics can be extracted from this Lagrangian using the Euler-Lagrange equations which follow from calculus of variations. 

However it is often simpler to move from a Lagrangian description to a Hamiltonian description. In this case we have not the tangent space but the cotangent space $T^*Q$ as our phase-space, and a Hamiltonian $H:T^*Q\rightarrow\mathbb{R}$. The reason we go to the cotangent space is because it allows a canonical symplectic structure. A symplectic structure on a manifold $M$ is a $2$-form $\omega$ which is closed, non-degenerate and anti-symmetric. This symplectic form allows us to extract dynamics from the Hamiltonian by the definition $dH = \omega(X_H,\cdot)$. A deep treatment of all the geometric properties of symplectic spaces can be found in \cite{mcduff}, which includes many theorems on symmetries of systems, such as Noethers theorem and Marsden-Weinstein quotients. These theorems form powerful tools to simplify our system of geodesics. Noethers theorem can be used to find symmetries and conserved quantities of the system, allowing for qualitative analysis. Marsden-Weinsteins theorem extends on this by allowing a reduction of the dimension of the system, to both simplify calculations and allow for easier intuition into the dynamics.

\subsection{Numerics}
Analytical calculations become very hard for more complicated spacetimes such as rotating black holes. For this we turn to numerically solving the geodesic equation. The standard way of solving differential equations is using numerical integrators. The simplest of these is Eulers method, which is first order. This means that every timestep is defined only by the first derivative. It is clear that as the timestep $h$ gets smaller, this more precisely models the function, where the error is of the order $O(h^2)$. We can however do better, Runge-Kutta integration schemes can get an error of $O(h^n)$ with as high an order as one wants, but in general the best compromise between speed and precision is found at order $4$, with an error of $O(h^5)$. We could also look at symplectic integrators, these are special types of integrators that are built so they conserve static quantities. For instance in the case of a Hamiltonian equation they conserve the levelset of the Hamiltonian exactly, where Eulers method, or Runge-Kutta schemes would drift. However in our case these schemes remain implicit, which while not making them unusable, significantly limits their execution speed.

We want to use these numerical calculations to create visual images of the black hole. To do this we have to generate a large surface with initial conditions which will form our perspective. After this we propagate these as lightlike geodesics, colouring the pixel they represent on the screen a different colour based on the final position after propagation. While modern processor hardware has come a long way in parallel processing, allowing us to propagate upwards of 10 rays at the same time, this pales in comparison to using Graphical processors. These are made specifically for the kind of task of doing the same calculations thousands of times in parallel, with slightly varying inital conditions. We will be using such hardware to be able to create visual results of black holes at a resolution not easily acquired on a home computer before.

%Using a similar technique, but with a different set on inputs we can also look at periodic orbits around a black hole, specifically periodic orbits of light. This can be done by starting the orbits near an expected periodic orbit and seeing how far they diverge from their starting positions. For example in Schwarszschild there is only one simple set of periodic light orbits on the lightring. These correlate strongly with the edge of the shadow. From a mathematical perspective the periodic orbits of Hamiltonian systems are of great interest. 

\subsection{Organisation of this document}
In this document I will first be describing the mathematical formulation of geodesics we used in section \ref{mathSection}. Subsequently I will discuss the numerical methods used in section \ref{numericalSection}. After this I will show the results of the simulation for the Kerr and Manko-Novikov spacetimes, for a variety of inputs in section \ref{visualSection}. Shadows of Manko-Novikov spacetimes were simulated before in \cite{MNShadow}, we aim to significantly increase the resolution of these simulations, while using simple computational techniques that are more easily expanded to complicated spacetimes.

%% file: 01_spacetime.tex
% Add a new section: The enumeration of this section will continue in the mainfile, even though it's alone here. So if this is the 5th section including earlier subfiles, this will print: '5  First section' in the main file.
\newpage % Use newpage to start on a new page
\section{The Mathematical Description}\label{mathSection}
\subsection{The Equation of Motion}
The mathematical description of geodesics is often given in the form of a Lagrangian. For a metric $g_{\mu\nu}$ this is given by
\begin{align}
	\mathcal{L} = \tfrac{1}{2} g_{\mu\nu} \dot{q}^\mu \dot{q}^\nu
\end{align}
Here the dots mean the derivative to the parametrisation time $\tau$, and $q$ is the coordinate. Greek letters $\mu\nu\rho\sigma$ in superscript indicate indices of vectors, in subscript they indicate indices of co-vectors (or 1-forms). Calculating the Euler-Lagrange equations we obtain
\begin{align}
	\frac{d^2q^\mu}{d\tau^2} + \Gamma^{\mu}_{\rho\sigma}\frac{dq^\rho}{d\tau}\frac{dq^\sigma}{d\tau} = 0
\end{align}
A more complete version of this description of geodesics can be found in \cite{carroll}. For our purposes however, the Hamiltonian description of geodesics is much more convenient. This gives us an 8 dimensional first order differential equation, which is computationally easier to deal with than a 4 dimensional second order equation. This is due to the better studied integration schemes.

The Hamiltonian we obtain for our metric $g_{\mu\nu}$ is given by
\begin{align}
	H(q,p) = \frac{g^{\mu\nu}p_\nu p_\mu}{2}
\end{align}
Here $p_\mu$ is the momentum conjugate to $\dot{q}^\mu$. For the full calculation showing that this is the case I will refer to the appendix \ref{appendixMath}. Here we also find that the equations of motion of our geodesic are given by:

\begin{align}\label{hamGeodesic}
	\frac{d}{d\tau} q^\mu &= g(q)^{\mu\nu}p_\nu\\
	\frac{d}{d\tau} p_\mu &= -\partial_\mu g(q)^{\nu\sigma}p_\nu p_\sigma\\
	q^\mu(0) &= (q^\mu)_0\\
	p_\mu(0) &= (p_\mu)_0
\end{align}

Note that we use the shorthand notation $\partial_\mu$ for $\frac{\partial}{\partial q}$. 

\subsection{The Spacetimes}
\subsubsection{The Schwarzschild Metric}
The first spacetime we will look at will be the black hole solution first posed by Schwarzschild \cite{schwarzschild}. This metric is given in spherical coordinates, in units with $c=1$ below:
\begin{align}\label{schwMetric}
	g_{\mu\nu} = -\Bigg(1-\frac{2GM}{r}\Bigg)dt^2 + \Bigg(1-\frac{2GM}{r}\Bigg)^{-1}dr^2 + r^2 d\theta^2 + r^2\sin^2\theta d\phi^2
\end{align}
on the manifold $\mathbb{R}_t\times\{r>0|r\neq 2GM\}_r\times (0,\pi)_\theta\times\mathbb{R}_\phi$. This spacetime describes the exterior spacetime of a non-rotating blackhole. It has a single parameter $M$, the mass, and a horizon at $2GM$. The spacetime is also very symmetric, spherical and time symmetry lead to converved energy and angular momenta, which together with the conserved Hamiltonian make the system completely integrable.

\subsubsection{The Kerr Metric}
Where the Schwarzschild solution gives the spacetime of a non-rotating black hole, the Kerr spacetime gives a more general solution of a black hole that is rotating, rather than entirely spherically symmetric like the Schwarzschild case. The metric is given by
\begin{align}
	g_{\mu\nu} &= -\Bigg(1-\frac{2GMr}{\rho}\Bigg)dt^2 +\frac{\rho^2}{\Delta}dr^2 + \rho^2 d\theta^2 \\
	 &+ \frac{2GMar\sin^2\theta}{\rho^2}(d\phi dt + dt d\phi) \\
	 &+ \frac{\sin^2\theta}{\rho^2}\Big((r^2+a^2)^2 - a^2\Delta\sin^2\theta\Big)d\phi^2
\end{align}
with
\begin{align}
	\Delta(r) &= r^2-2GMr + a^2\\
	\rho^2 &= r^2+a^2\cos^2\theta\\
	a &= S/M
\end{align}
where $S$ is the spin of the black hole. This spacetime gives a black hole for the values $-M^2 < S < M^2$. Outside that range the event horizon disappears, and we no longer classify the spacetime as a black hole, as it violates the censorship conjecture \cite{censor}. The coordinates of the metric are called Boyer-Lindquist coordinates, these are given by
\begin{align}
	t &= t\\
	x &= \sqrt{r^2+a^2}\sin\theta\cos\phi\\
	y &= \sqrt{r^2+a^2}\sin\theta\sin\phi\\
	z &= r\cos\theta\\
\end{align}
We use the coordinates as used in the book \cite{carroll}, which are not the same as in the original version by \cite{Kerr}, the reason for this is the coordinates are significantly easier to work with for our simulation.

Like the Schwarzschild spacetime this spacetime is very symmetric. There is a single conserved angular momentum, a conserved energy and Hamoltonian. In addition to this there is a less obvious conserved quantity known as the Carter constant. So this system is still entirely integrable.

\subsubsection{The Manko Novikov Metric}
As Kerr expanded the Schwarzschild spacetime by adding a dipole current-moment $S$, Manko and Novikov expanded this spacetime further by adding additional higher order multipole mass moments \cite{MankoNovikov}. Like the Kerr solution this is still a stationary and axi-symmetric vaccuum solution of Einsteins equation. However it has some issues that could dequalify it from being a physical solution, such as that it has a curvature singularity on the horizon, meaning it does not follow the censorship conjecture \cite{censor}. It also loses integrability of the geodesic dynamics, because unlike Kerr it does not have an extra constant of motion called the Carter constant \cite{MNSpace}. In our case we will only be expanding to the quadrupole moment $Q$, that is one step further than Kerr.

This spacetime is given by the following equation in Boyer-Lindquist coordinates:
\begin{align}
	g_{\mu\nu} = -f (dt - \omega d\phi)
	+ \frac{e^{2\gamma}\rho^2}{f\Delta}dr^2
	+ \frac{e^{2\gamma}\rho^2}{f}
	+ \frac{\Delta \sin^2\theta}{f} d\phi^2
\end{align}
Where analogously to Kerr:
\begin{align}
	\rho^2 &= (r - M)^2 - k^2 \cos^2\theta\\
	\Delta &= (r-M)^2 - k^2
\end{align}
And $f$, $k$, $\omega$ and $\gamma$ are defined by:
\begin{align}
	%S &= aM\\
	\alpha &= \frac{-M+\sqrt{M^2-(S/M)^2}}{S/M}\\
	k &= M\frac{1-\alpha^2}{1+\alpha^2}\\
	\beta &= q\frac{M^3}{k^3}\\
	f &= e^{2\psi}\frac{A}{B}\\
	\omega &= 2ke^{-2\psi} \frac{C}{A} - 4k\frac{\alpha}{1-\alpha^2}\\
	e^{2\gamma} &= e^{2\gamma'}\frac{A}{(x^2-1)(1-\alpha^2)^2}\\
	A &= (x^2-1)(1+\mathfrak{a}\mathfrak{b})^2 - (1-y^2)(\mathfrak{b}-\mathfrak{a})^2\\
	B &= \Big( (x+1) + (x-1)\mathfrak{a}\mathfrak{b} \Big)^2 \Big( (1+y)\mathfrak{a} + (1-y)\mathfrak{b} \Big)\\
	C &= (x^2-1)(1+\mathfrak{a}\mathfrak{b})\Big( (\mathfrak{b}-\mathfrak{a})-y(\mathfrak{a}+\mathfrak{b}) \Big)
	+(1-y^2)(\mathfrak{b}-\mathfrak{a})\Big( (1+\mathfrak{a}\mathfrak{b}) + x(1-\mathfrak{a}\mathfrak{b}) \Big)\\
	\psi &= \beta \frac{P_2}{R^3}\\
	\gamma' &= \ln\sqrt{\frac{x^2-1}{x^2-y^2}} + \frac{3\beta^2}{2R^6}(P_3^2-P_2^2)
	+ \beta\Bigg( -2 \sum_{l=0}^2 \frac{x-y+(-1^{2-l})(x+y)}{R^{l+1}}P_l \Bigg)\\
	\mathfrak{a} &= -\alpha \exp\Bigg( -2\beta \Bigg( -1 + \sum_{l=0}^2 \frac{(x-y)P_l}{R^{l+1}} \Bigg) \Bigg)\\
	\mathfrak{b} &= \alpha \exp\Bigg( 2\beta \Bigg( 1 + \sum_{l=0}^2 \frac{(-1)^{3-l}(x+y)P_l}{R^{l+1}} \Bigg) \Bigg)\\
	R &= \sqrt{x^2+y^2-1}\\
	P_l &= P_l\bigg(\frac{xy}{R}\bigg)\\
	x &= \frac{r-M}{k}\\
	y &= \cos\theta
\end{align}
Here $S$ is the spin, the unitless spin is given by $S/M^2$ and $q$ is the quadrupole deviation from the Kerr metric, and $P_l$ are the legendre polynomials. Here $q<0$ gives a prolate deviation, and $q>0$ gives an oblate deviation. This means that when seeing the $z$ direction or the axis of rotation, as vertical, the event horizon gets taller for $q<0$ and less tall for $q>0$. This will later be visible when we draw the shadows generated by these spacetimes.

This spacetime however has a few patholigical properties. Firstly it has a curvature singularity on the equator of the event horizon. Secondly an analogue to the Carter constant of Kerr no longer exists, meaning the system is no longer integrable. Because of this chaotic behaviour may arise in strong-field regions of the spacetime, especially near the singularity on the horizon.

%% file: 02_integrators.tex
\newpage % Use newpage to start on a new page
\section{Numerical Methods}\label{numericalSection}
\subsection{Integration Techniques}
We have shown that we can reduce the problem of calculating geodesics to a Hamiltonian system given by equation \ref{hamGeodesic}. This gives us a first-order dynamical system to integrate. In some cases, such as a flat metric or the Schwarzschild black hole from equation \ref{schwMetric}, this system is simple enough to integrate analytically. However that is quickly lost when we wish to integrate more complicated metrics. To be able to calculate geodesics on general metrics we will use numerical integration schemes.

\subsubsection{Explicit Euler Integration}
The Euler method for solving a differential equation $x' = f(x)$ is based on the idea that for a small enough  $h$:
\begin{align}
	x(t+h) \approx x(t) + h x'(t)
\end{align}
This method piecewise linearly approximates $x$, and thus (assuming $x$ is sufficiently differentiable) its error is at most of order $h^2$. We see this because:
\begin{align}
	\lim_{h\rightarrow 0} (x(t+h) - x(t) + h x'(t))/h = x'(t) - x'(t) = 0
\end{align}
This integrator is very simple to calculate per step, but we will see later that it's rather low accuracy will cause problems. Because of this we will be using a more complicated integrator.

\subsubsection{Runge Kutta Integrators}
A Runge-Kutta integrator is in essence a higher order version of an explicit Euler integrator. In the way that Eulers method only uses the linear term of the  local Taylor expansion of the function, Runge Kutta takes higher order terms into account as well. This uses the principles of fitting a polynomial to a set of points. The integration scheme that is used in our simulation is RK4, or fourth order Runge Kutta. This scheme gives a good tradeoff between calculation speed and accuracy.  This scheme goes as follows, given an equation
\begin{align}
	\frac{dx}{dt} &= f(x,t)\\
	x(t_0) &= x_0
\end{align}
We integrate as follows:
\begin{align}
	t_{n+1} &= t_n+h\\
	x_{n+1} &= x_n + \frac{1}{6}\big(k_1 + 2k_2 + 2k_3 + k_4\big)
\end{align}
where
\begin{align}
	k_1 &= f(x_n,t_n)\\
	k_2 &= f(x_n+\tfrac{1}{2}h,t_n + \tfrac{1}{2}hk_1)\\
	k_3 &= f(x_n+\tfrac{1}{2}h,t_n + \tfrac{1}{2}hk_2)\\
	k_4 &= f(x_n+h,t_n + hk_3)\\
\end{align}

\subsection{Simulation techniques}

\subsubsection{Orbits}
\begin{figure}[h]
	\centering
	\begin{subfigure}[b]{0.30\textwidth}
		\centering
		\includegraphics[width=\textwidth]{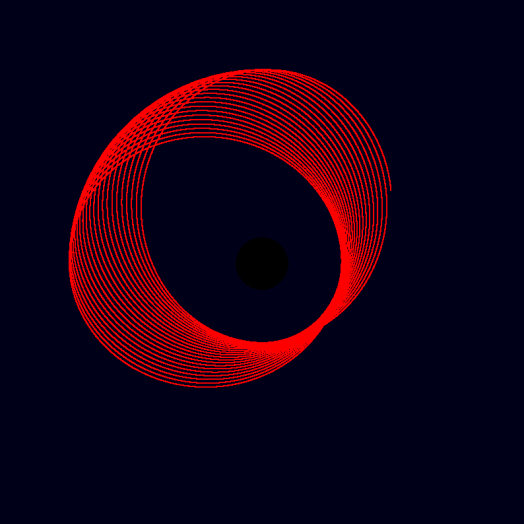}
		\caption{An orbit around a Schwarzschild black hole}
		\label{orbittestruna}
	\end{subfigure}
	\hfill
	\begin{subfigure}[b]{0.30\textwidth}
		\centering
		\includegraphics[width=\textwidth]{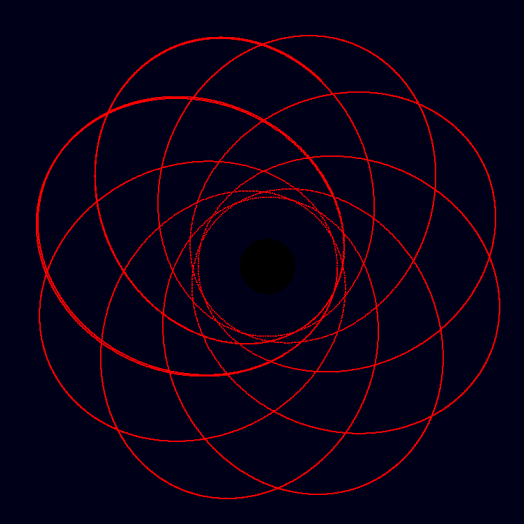}
		\caption{A resonant orbit around a Schwarzschild black hole}
		\label{orbittestrunb}
	\end{subfigure}
	\hfill
	\begin{subfigure}[b]{0.30\textwidth}
		\centering
		\includegraphics[width=\textwidth]{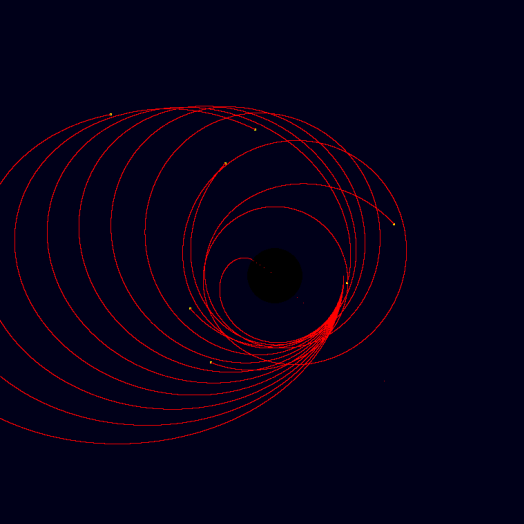}
		\caption{Multiple orbits run in parallel, to test performance}
		\label{orbittestrunc}
	\end{subfigure}
	\caption{Various test orbits run on the simulation program. These are timelike orbits around the equator of a Schwarzschild black hole.}
	\label{orbittestrun}
\end{figure}

To simulate geodesics for a given metric we want to solve the Hamiltonian equation, which is an 8 dimensional first order differential equation \ref{hamGeodesic}.  We can run this simulation for a few timelike orbits to check for behaviour we expect of orbits in general relativity. In figure \ref{orbittestrun} we see some equatorial orbits of test particles around a Schwarzschild black hole. Note the black circle in the centre is not to scale with the event horizon. We see in especially the left \ref{orbittestruna} and middle \ref{orbittestrunb} the perihelion precession, which was one of the first experimental proofs of general relativity. 
In the right picture \ref{orbittestrunc} we see multiple orbits at the same time. These orbits were run in parallel to improve the performance of the simulation.

\subsubsection{Parallelisation}
We want to expand our simulation to be able to generate visual pictures. To do this we need to build a ray-tracer. This is a type of graphics engine that creates a 2d image of 3d space by calculating where the ray of light corresponding to each pixel goes. However our ray-tracing engine will be much more mathematically intensive than a general graphics engine, as we no longer consider a flat space. As such we have to run the above calculation for each pixel, which means that for a full-HD image we would have to run the simulation around 2 million times. This simulation took about 10 seconds, meaning that we would take more than half a year to generate a single image.

While the above simulation allowed us to calculate a few orbits in parallel, it is rather limited in its scaling potential. A general modern desktop computer contains a CPU with about 8 cores, allowing us to only run only a few handful of orbits in parallel. This would still leave us with almost a month of calculation time.

So we turn away from our CPU and towards the Graphical processor or GPU. Such a chip is built exactly for this kind of parallel processing, and allows us to calculate upwards of 1000 geodesics in parallel. The library used for this was CuPy. This is an extension of the NumPy library for python, with capabilities of creating and running kernels on the GPU. 

\subsection{Raytracing}
\subsubsection{Initial positions}
\begin{figure}[h]
	\centering
	\includegraphics[height=0.25\textheight]{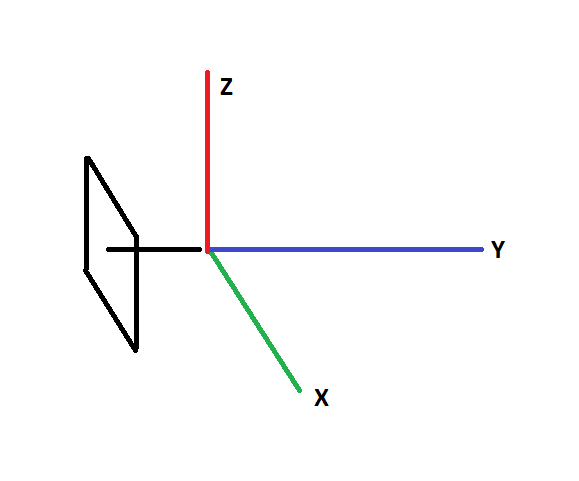}
	\caption{The screen (black rectangle) that defines the $q$ position for each pixel. The origin is centered on the black hole.}
	\label{raytraceScreen1}
\end{figure}

To create a raytraced image we first need to build a camera. This means that for every pixel of our image, we assign a starting position for our geodesic in our spacetime. The way we do this is as follows. We start in $3$-dimensional euclidean space, with coordinates $x,y,z$ where $z$ will be the direction of the axis of spherical or Boyer-Lindquist coordinates. We then put the screen in the space on the $xz$-plane and offset it in the $y$ direction. This is seen in figure \ref{raytraceScreen1}.

\begin{figure}[h]
	\centering
	\includegraphics[height=0.25\textheight]{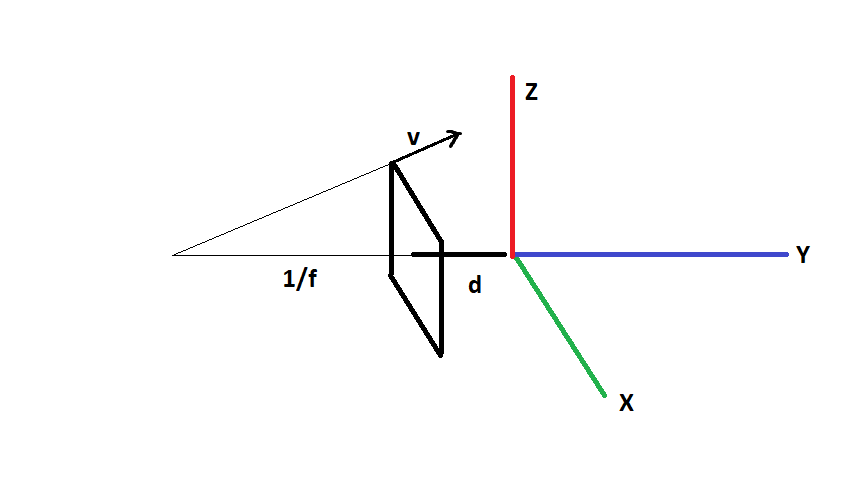}
	\caption{The screen that defines the velocity for each pixel. Here $d$ is the width of the screen and $f$ is the inverse focal distance.}
	\label{raytraceScreen2}
\end{figure}

These $xyz$ positions will be converted to Boyer-Lindquist coordinates, setting $t=0$. After this one can set all velocity directions to $v_y = \dot{y} = -1$ and have an image with focal distance infinity. However should we want to change this focal distance, we can add the $x$ and $z$ position to the velocities $v_x$ and $v_z$, multiplying by a factor $2f/d$ where $f$ and $d$ are as seen in figure \ref{raytraceScreen2}.

After having these two $3$-vectors, we can rotate and move the point of view. For our purposes because of the symmetry of the space the only relevant change would be rotating around the $x$ axis to view from a different $\theta$.

We then convert the position and velocity into Boyer-Lindquist $3$-vectors. We know
\begin{align}
	x &= \sqrt{r^2+a^2}\cos\phi\sin\theta\\
	y &= \sqrt{r^2+a^2}\sin\phi\sin\theta\\
	z &= r\cos\theta
\end{align}
This coordinate transformation is easily inverted by solving a quadratic. We can then take the derivative of this inverse transformation to transform the velocity vector.

Having converted our $3$-vectors to Boyer Lindquist we have to turn them to $4$-vectors, which form out $q^\mu$ and $\dot{q}^\mu$. For $q^\mu$ this is a simple case of setting $q^t=0$. For $\dot{q}^\mu$ this is a little more involved. We first set $\dot{q}^t$ such that the vector is lightlike. We do this by solving
\begin{align}
	g_{\mu\nu}\dot{q}^\mu \dot{q}^\nu = 0
\end{align}
as a quadratic equation in $\dot{q}^t$. We then multiply each ray by a scalar, such that $\dot{t}$ becomes $-1$. This is done to more easily calculate the redshift a geodesic undergoes in the code. Lastly we convert the $4$-velocity gained to a momentum by multiplying with the metric:
\begin{align}
	p_\mu = g_{\mu\nu}v^\nu
\end{align}

We have now calculated the initial position $(q^\mu,p_\mu)$ of our differential equation for each pixel.

\subsubsection{Propagation}
Propagation of the geodesics is a straightforward process, as it simply involves applying the Runge-Kutta integration scheme a large number of times. In most of the pictures shown later on the number of steps varies from $50.000$ to $1.000.000$. Boundary checks were also used. These are to make sure the geodesics stay inside a certain bounding box, eg $r<20M$ for a black hole of mass $M$. This assumes that geodesics that stray this far from the black hole will never fall in, and can save massively on calculation time by exiting in fewer than $1.000$ steps for a large portion of the geodesics at the edge of our generated screen. Exit parameters such as these can also be used to place objects around the black hole, such as an accretion disk, or other orbiting objects.

%The propagation script can be used on any set of input geodesics, not just a raytrace camera. It can for instance also be used to search for periodic or pseudo-periodic geodesics. Examples of this can be found in the results chapter.

\subsubsection{Redshift}
We also want to be able to calculate the redshift of light from our raycast. For instance because we want to add objects into the simulation like a ring or a star close to the black hole, and we want to see how its light is shifted by the travel around the black hole, and the speed of the object. To be able to do this we first have to calculate the velocity of the object. 

For general objects calculating the velocity is difficult, but for an accretion disk we will make a few simple assumptions to be able to do a calculation. This assumption is that the disk is circular, and is in a stable circular orbit at every point of the disk. Mathematically this means we assume:
\begin{align*}
	\dot{p_\mu} &= 0\\
	p_r &= 0\\
	p_\theta &= 0
\end{align*}
After this we set $p_t = -1$ and calculate $p_\phi$ from the Hamiltonian equations of motion. This will come down to solving
\begin{align}
	p_\phi^2 \partial_q g_{\phi\phi} + p_t^2 \partial_q g_{tt} + 2 p_t p_\phi \partial_q g_{t\phi} = 0
\end{align}
We calculate derivatives numerically to make the code work for any metric without having to calculate the derivative by hand. Once we have $p_\phi$ and $p_t$ we renormalise such that $p^2 = -1$ to avoid scaling issues with our redshift calculation. After this we can calculate redshift $z$ by the equation
\begin{align}
	z = \frac{\lambda_O}{\lambda_S}
	=
	\frac{g_{\mu\nu}\kappa_S^\mu S^\nu}{g_{\mu\nu}\kappa_O^\mu O^\nu}
\end{align}
Where $O$ is the observer and $O^\mu$ its $4$-velocity, and $S$ is the source. $\kappa$ is the $4$-velocity of the light when it hits the observer/source. To make our simulation easier we can normalise the inverse lightrays sent out from the observer such that $g_{\mu\nu}\kappa^\mu O^\nu = 1$, this is why we rescaled out geodesics initial positions such that $\kappa_O^t = -1$ in the generation of the raytracing camera.

\section{Resolution and Calculation Speed}
\begin{figure}[h]
	\centering
	\begin{subfigure}[b]{0.30\textwidth}
		\centering
		\includegraphics[width=\textwidth]{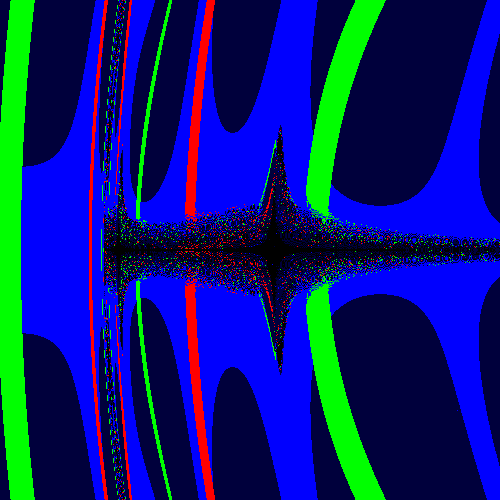}
		\caption{Integrator: RK4}
	\end{subfigure}
	\hfill
	\begin{subfigure}[b]{0.30\textwidth}
		\centering
		\includegraphics[width=\textwidth]{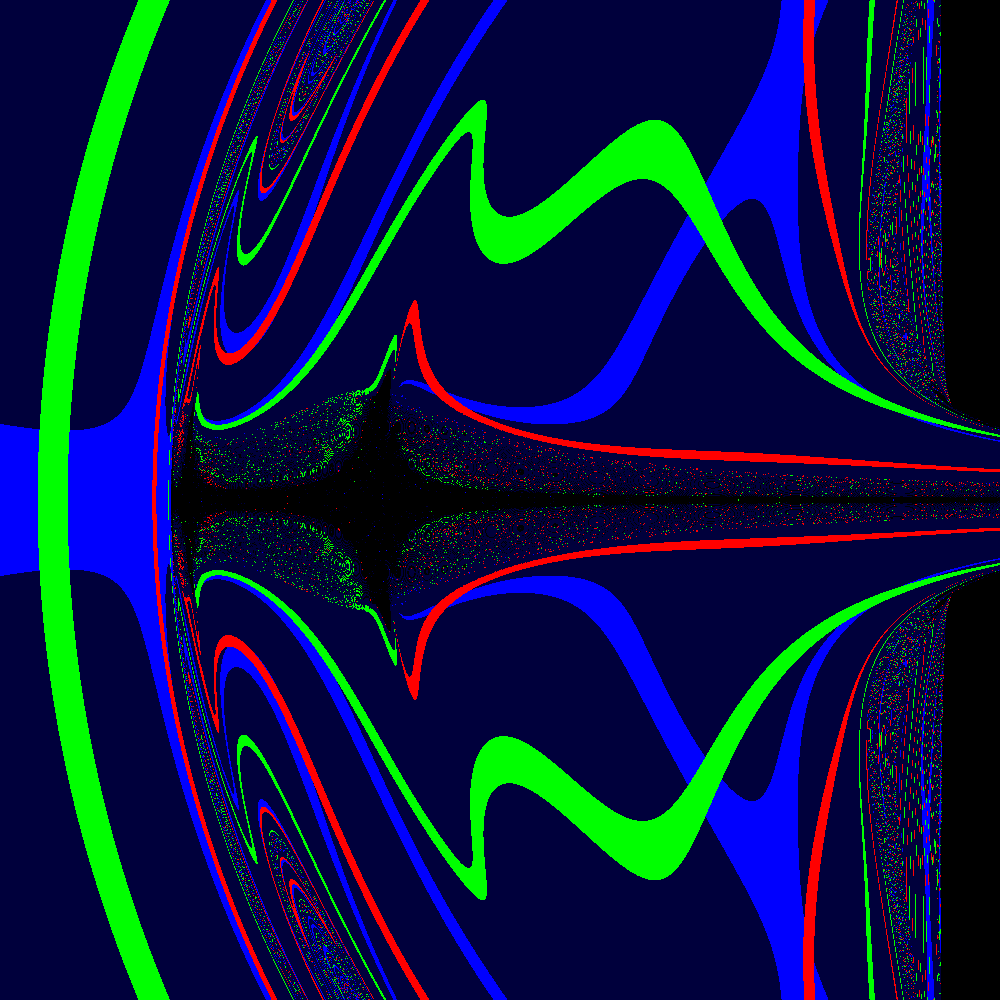}
		\caption{Integrator: Euler}
	\end{subfigure}
	\hfill
	\begin{subfigure}[b]{0.30\textwidth}
		\centering
		\includegraphics[width=\textwidth]{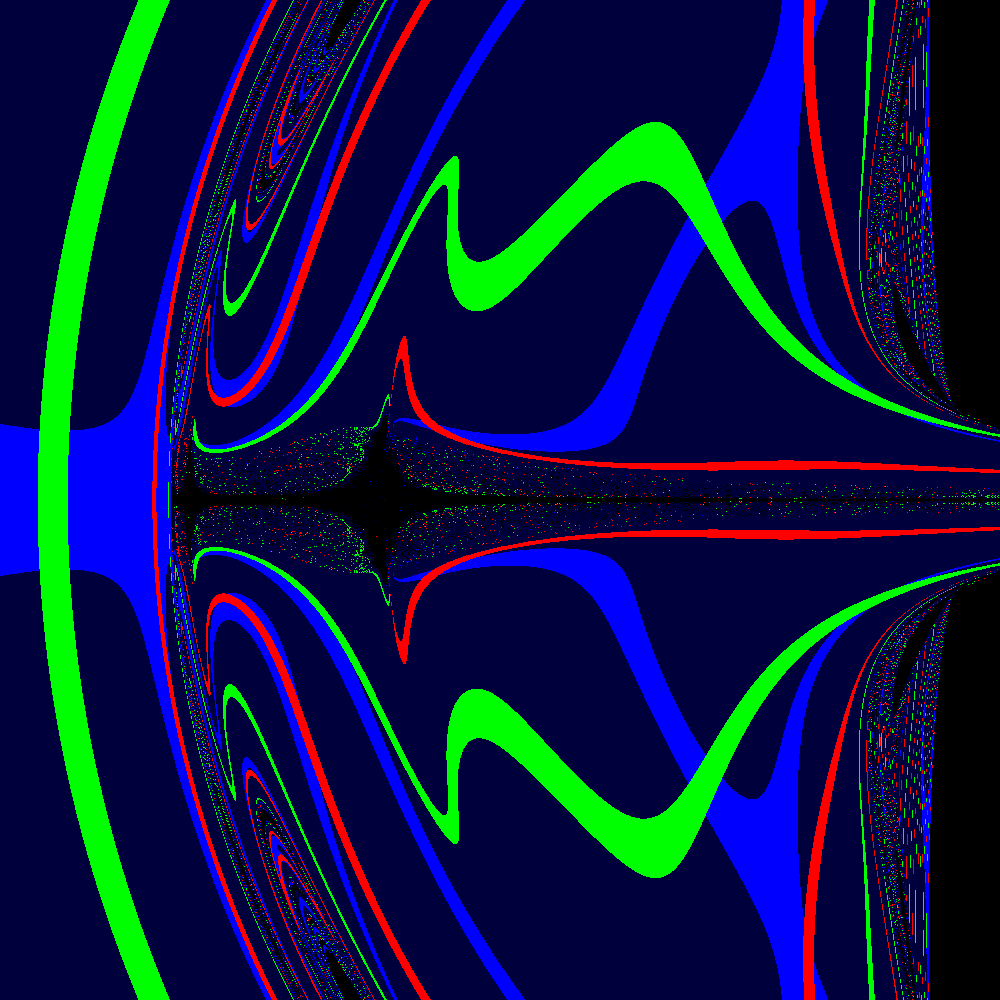}
		\caption{Integrator: Euler, smaller stepsize.}
	\end{subfigure}
	\caption{Various integrators applied to the raytracing software. The left gives a result much closer to the desired result. Due to symmetry constraints on the system the picture should contain a horizontal blue band. Furthermore the black region in the centre is a calculation error due to a coordinate singularity. We see this get smaller both with stepsize and a better integrator.}
	\label{resolutionIMG}
\end{figure}

The main intents of the python code is ease of use, and performance on modern home computers. But for the performance increase to be worth it the resolution and precision need to be sufficient. This was tested by running a raytrace on the most volatile region of a visual shadow of a Manko-Novikov black hole. For the Euler integrator the shape of this region varied wildly with the stepsize used for the integrator, whereas for the RK4 scheme this varied less. This can be seen in figure \ref{resolutionIMG}. We see here that the green line moves significantly with the stepsize difference in the Euler integrator. What this green line is is explained later in figures \ref{perspective} and \ref{flatspace}. We see in the case of the RK4 integrator a blue line around the chaotic region. This blue line is drawn on the equator of the bounding sphere of the simulation. Analytically we would expect there to be a blue line here, due to the symmetry of the space in the equator.

As for performance, because we use graphical processors rather than conventional ones, we get a massive decrease in runtime for parallel tasks, such as raytracing. This can yield upwards of 100x reduction in calculation times, on top of the code being moved from python to C. The images seen later will generally be either of a resolution of $1000\times 1000$, sometimes smaller. In the case of a Kerr shadow, the propagation of all the geodesics would take around $300$ seconds, or $5$ minutes. Performance for more difficult spacetimes such as Manko-Novikov would take between $1200$ and $3000$ seconds for a single megapixel image, depending on the input parameters. This dependency is because the size of the shadow can vary, meaning more rays have to take more steps before terminating. The fact that the Manko-Novikov calculations are slower is simply due to the metric being more complicated.

%% file: 03_results.tex
% Add a new section: The enumeration of this section will continue in the mainfile, even though it's alone here. So if this is the 5th section including earlier subfiles, this will print: '5  First section' in the main file.
\newpage % Use newpage to start on a new page
\section{Visual results}\label{visualSection}
In this section we will discuss the code used to visualise the results.
\subsection{The shadow}
\begin{figure}[h]
\centering
\includegraphics[height=0.25\textheight]{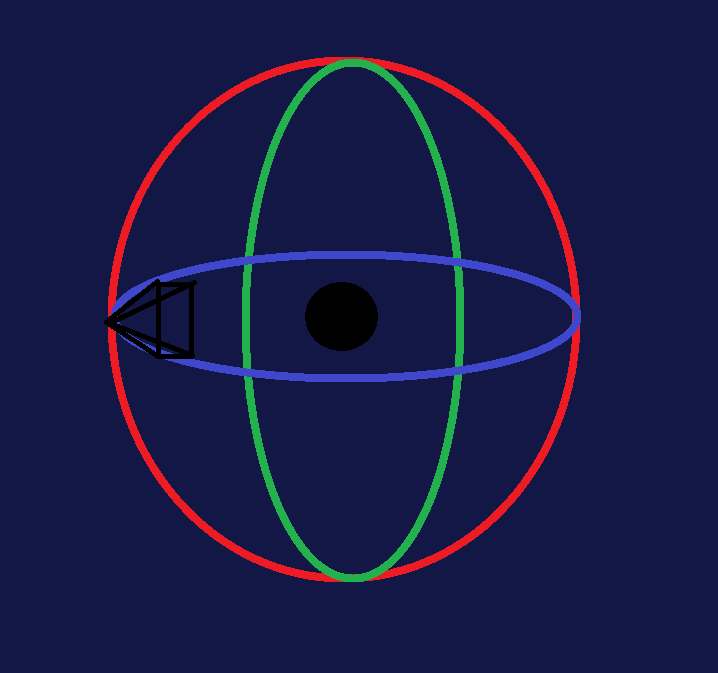}
\caption{The perspective of the raytracing simulation, red green and light blue are lines drawn on a larger sphere around the origin, the black triangle gives the position of the camera in the sphere.}
\label{perspective}
\end{figure}

\begin{figure}[h]
\centering
\includegraphics[height=0.25\textheight]{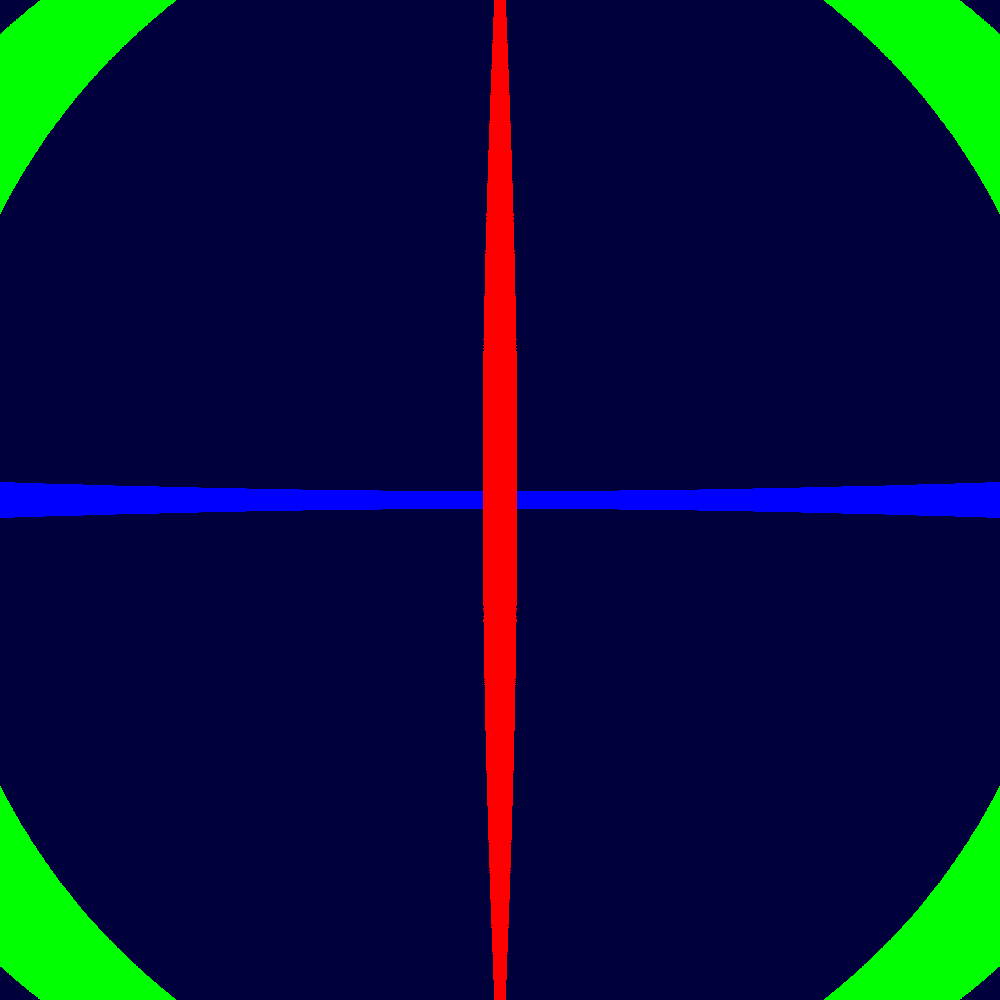}
\caption{Flat space as seen from the perspective of the camera. Dark blue is the background, light blue is the line on the equator, red is the line vertically around the origin and green is the line that forms a circle around the origin from the perspective of the camera.}
\label{flatspace}
\end{figure}

The code discussed in the previous section outputs positions and momenta. This is not human readable data, so we will use another script to create visual images to make sense of the results. The simplest visualisation splits the geodesics into $2$ groups. The first group leaves the simulation area (usually defined by the $r$ coordinate) and is assumed to fly off to the background at infinity. The second group does not leave the simulation area, in which case they are considered to have been captured by the black hole. Giving these two groups a separate colour will allow one to draw the shadow of the black hole. Another addition to this can be to give the background simple features, such as $3$ lines, drawn as seen in figure \ref{perspective}. These will give some sense of how the background becomes warped by the presence of a black hole. In this case the lines are simply defined as small intervals in the $\theta$ and $\phi$ Boyer-Lindquist coordinates. For flat space the visual image we get from this looks like figure \ref{flatspace}.

% For some pictures we also add a background to the picture of the milky way, this background was made by the European Southern observatory:
% \begin{figure}[h]
% \centering
% \includegraphics[height=0.25\textheight]{figures/backgroundLD.png}
% \caption{The milky way panorama as made by the ESO.}
% \label{flatspace}
% \end{figure}

\newpage

We will now draw the visual image of both the shadow and distorted background for a few black holes. We will be limiting ourselves to Schwarzschild and Kerr black holes, and the mass quadrupole restriction of Manko Novikov spacetimes.

\begin{figure}[h]
     \centering
     \begin{subfigure}[b]{0.32\textwidth}
         \centering
         \includegraphics[width=\textwidth]{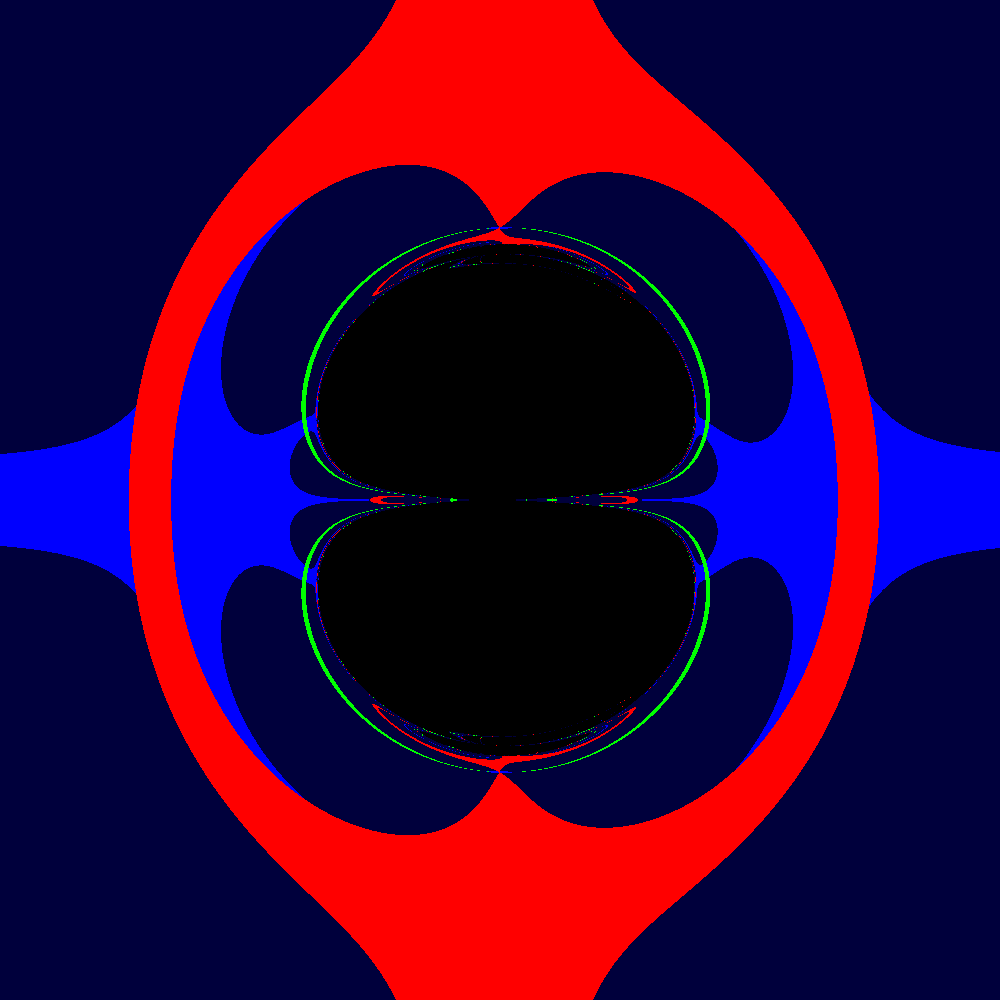}
         \caption{$q=-2,S=0.1M^2$}
     \end{subfigure}
     \hfill
     \begin{subfigure}[b]{0.32\textwidth}
         \centering
         \includegraphics[width=\textwidth]{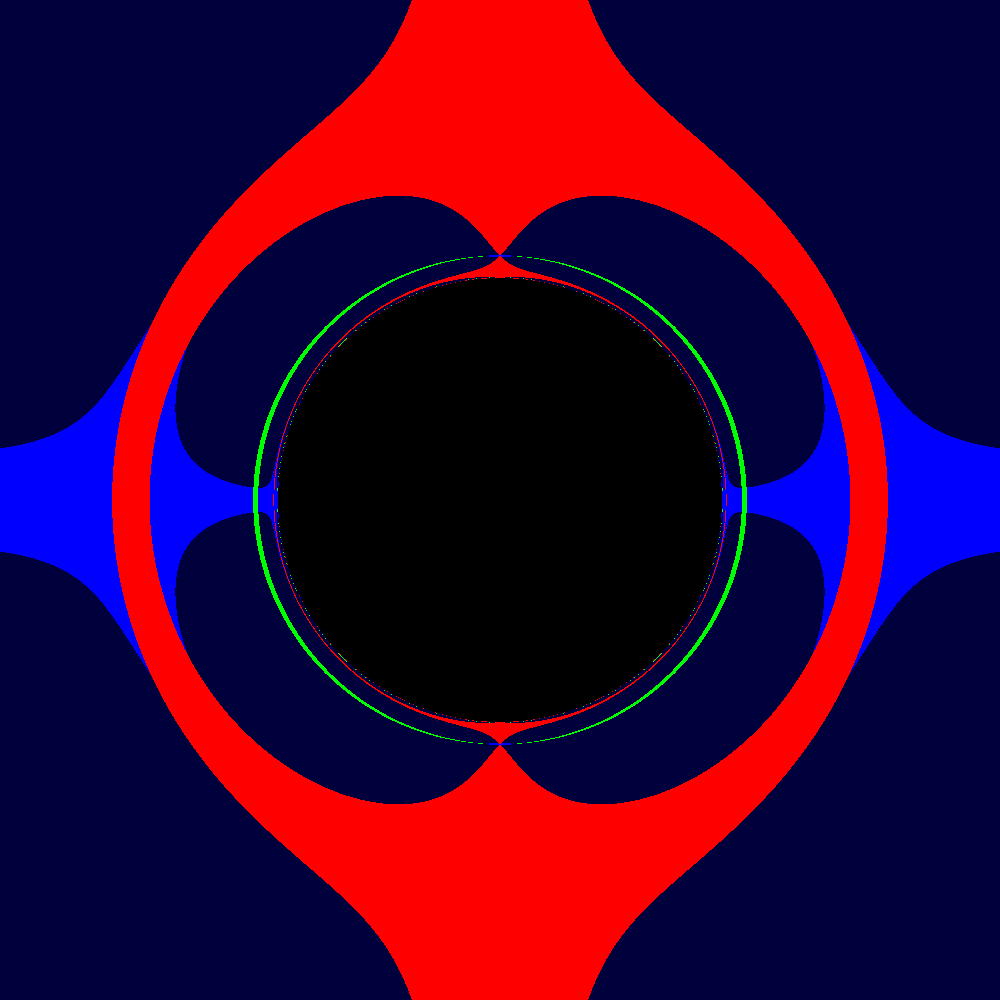}
         \caption{$q=0,S=0.0M^2$}
     \end{subfigure}
     \hfill
     \begin{subfigure}[b]{0.32\textwidth}
         \centering
         \includegraphics[width=\textwidth]{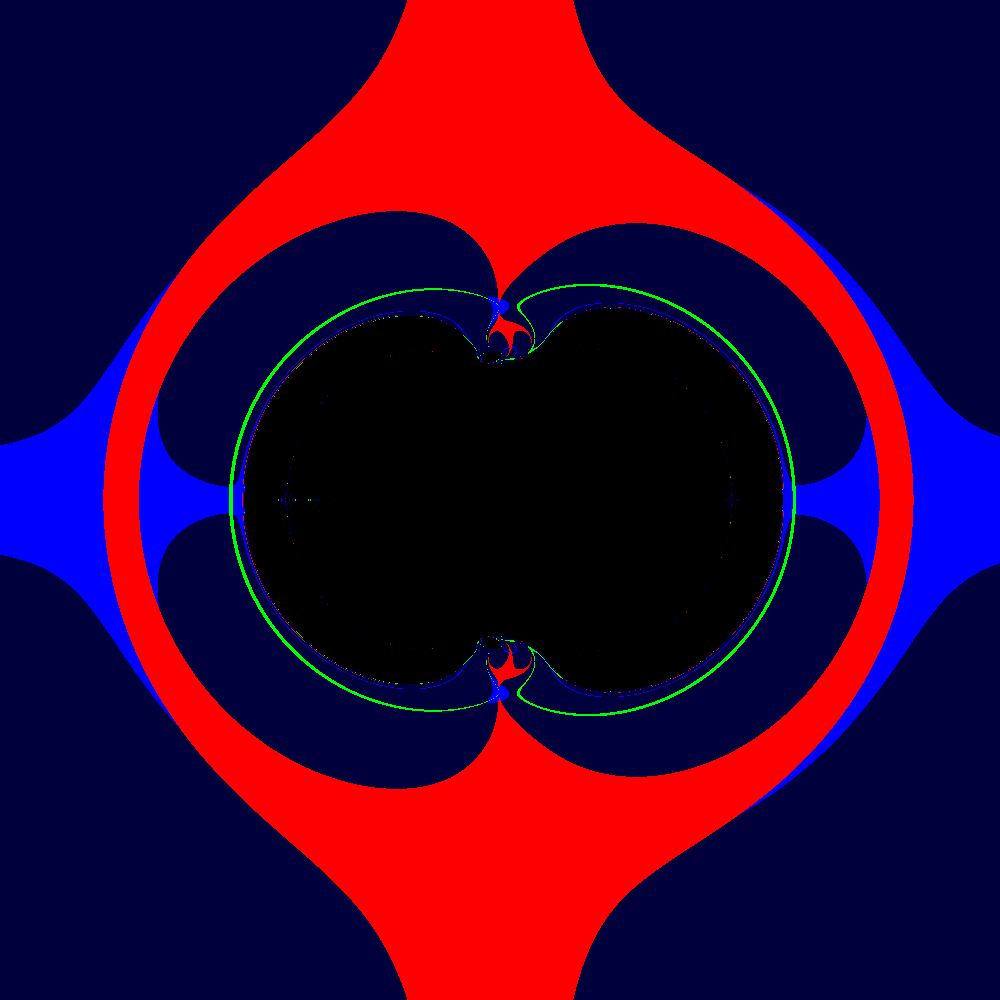}
         \caption{$q=2,S=0.1M^2$}
     \end{subfigure}
      \begin{subfigure}[b]{0.32\textwidth}
         \centering
         \includegraphics[width=\textwidth]{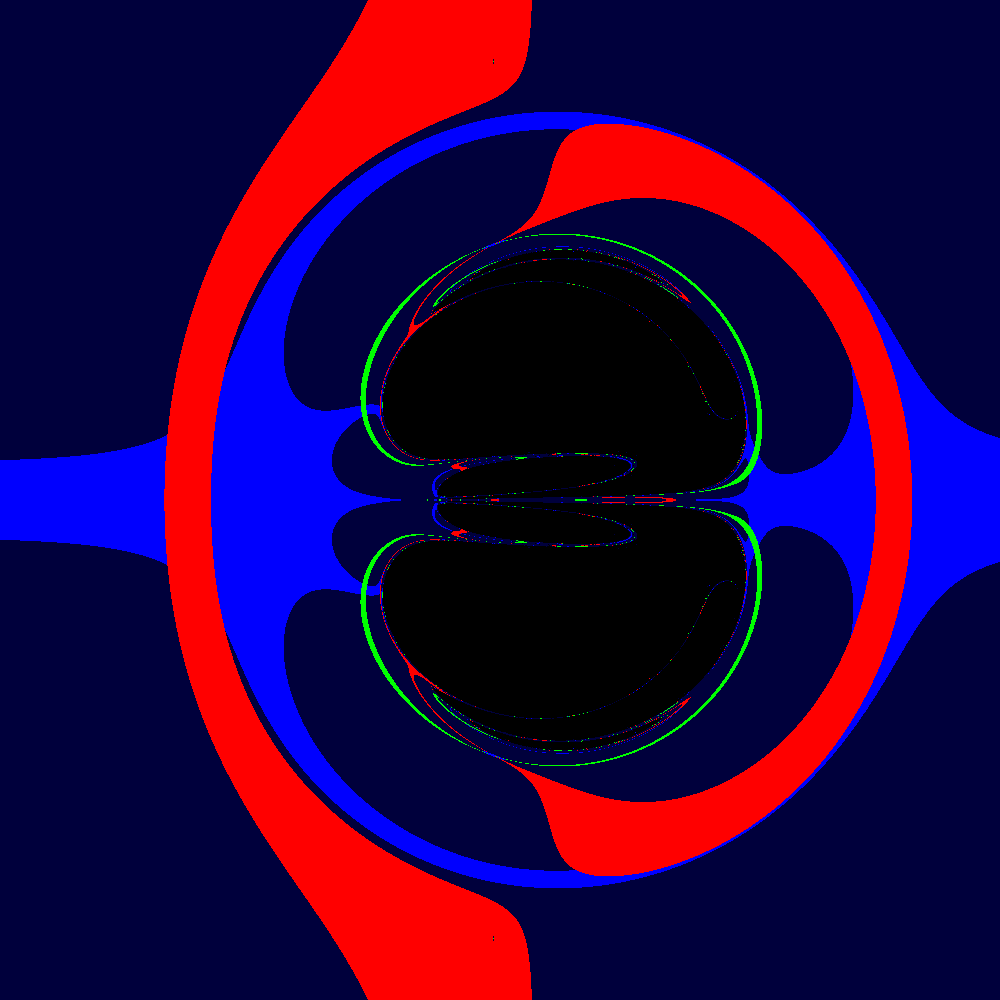}
         \caption{$q=-2,S=0.98M^2$}
     \end{subfigure}
     \hfill
     \begin{subfigure}[b]{0.32\textwidth}
         \centering
         \includegraphics[width=\textwidth]{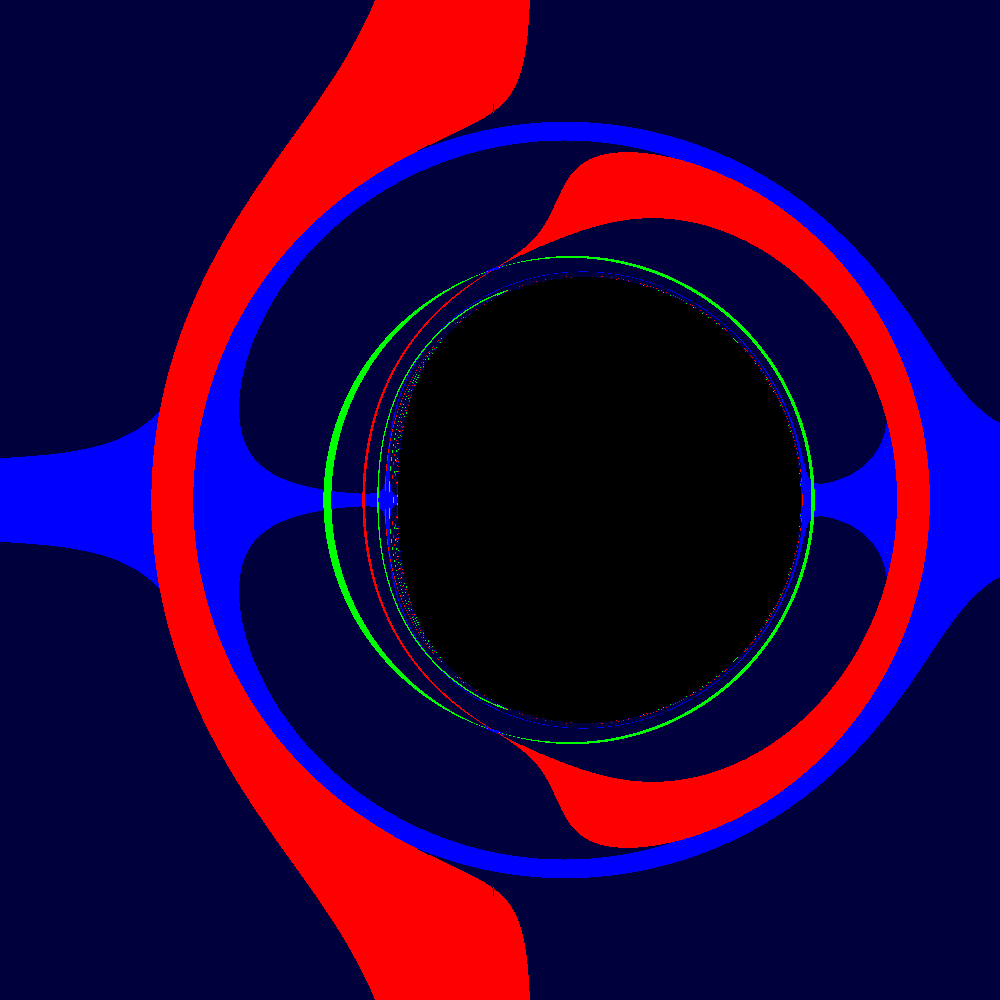}
         \caption{$q=0,S=0.98M^2$}
     \end{subfigure}
     \hfill
     \begin{subfigure}[b]{0.32\textwidth}
         \centering
         \includegraphics[width=\textwidth]{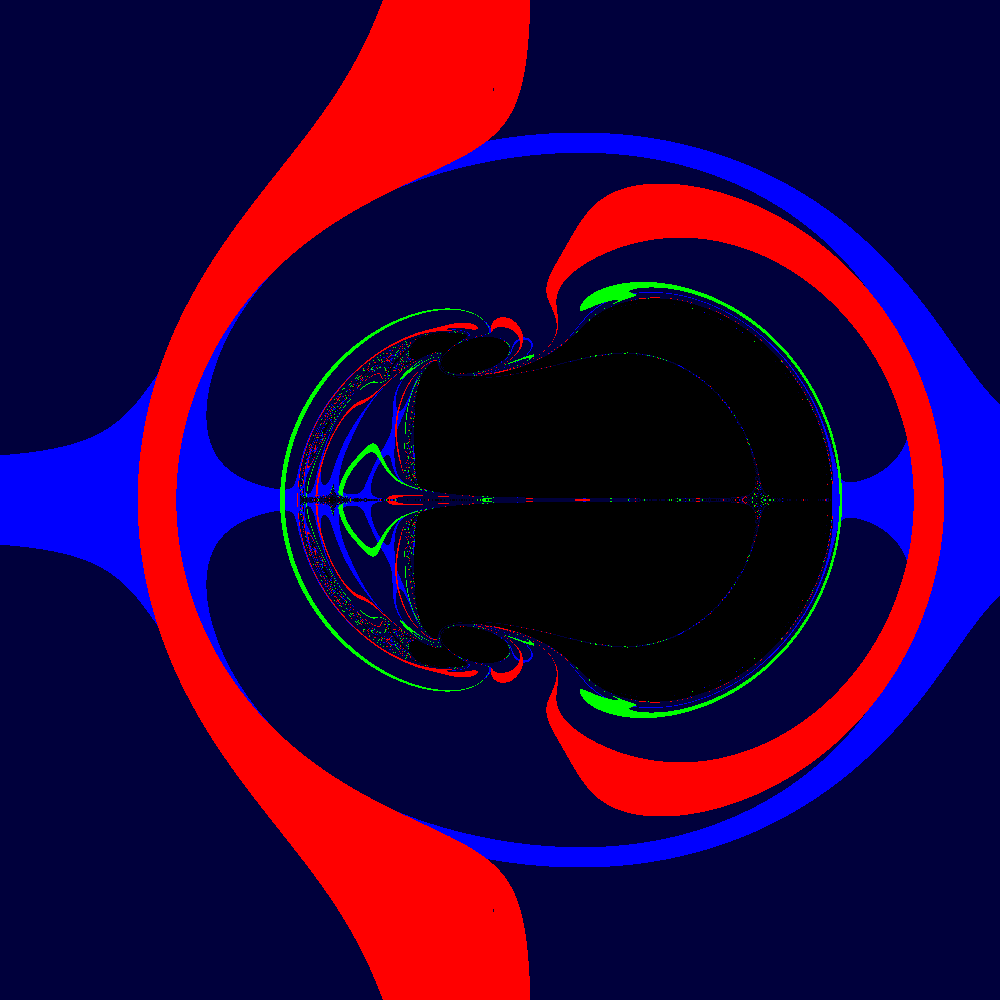}
         \caption{$q=2,S=0.98M^2$}
     \end{subfigure}
     \caption{Our black hole zoo used to test the raytracing calculations. We see that as $q>0$ our shadow becomes more oblate, or wide, whereas as $q<0$ the shadow becomes more prolate or tall.}
     \label{BHzooSpace}
\end{figure}

We now draw the shadow of the black hole for some different input parameters. Note that for the Manko Novikov spacetime we cannot set the spin to exactly $0$ in the calculation as that leads to division by zero, so in those cases it is set to $S = 0.1M^2$. The visual representation we get of the shadows of these black holes, and how they warp their background can be seen in figure \ref{BHzooSpace}. The images given are $1000\times1000$ pixels each. We can separate several regions of interest in these pictures. Drawn in black is the shadow of the objects, for Manko-Novikov these can become disjointed. We also see some higher order reflections of the lines on the background (red, green and blue) close to the shadow. In specifically the $Q=2,S=0.98M^2$ case we see a strange chaotic region on one side of the shadow. This will later be explored a little further.\\

We can compare our calculations of the Manko-Novikov shadow to calculations in \cite{MNShadow}. We see a difference in precision especially in cases with high positive $Q$, where the chaotic region of the image seems to have less errors. We also were able to achieve this resolution and precision on a home computer with parts sold in 2016, as opposed to using a computing cluster.\\

\begin{figure}[h]
	\centering
	\begin{subfigure}[b]{\textwidth}
		\centering
		\includegraphics[width=\textwidth]{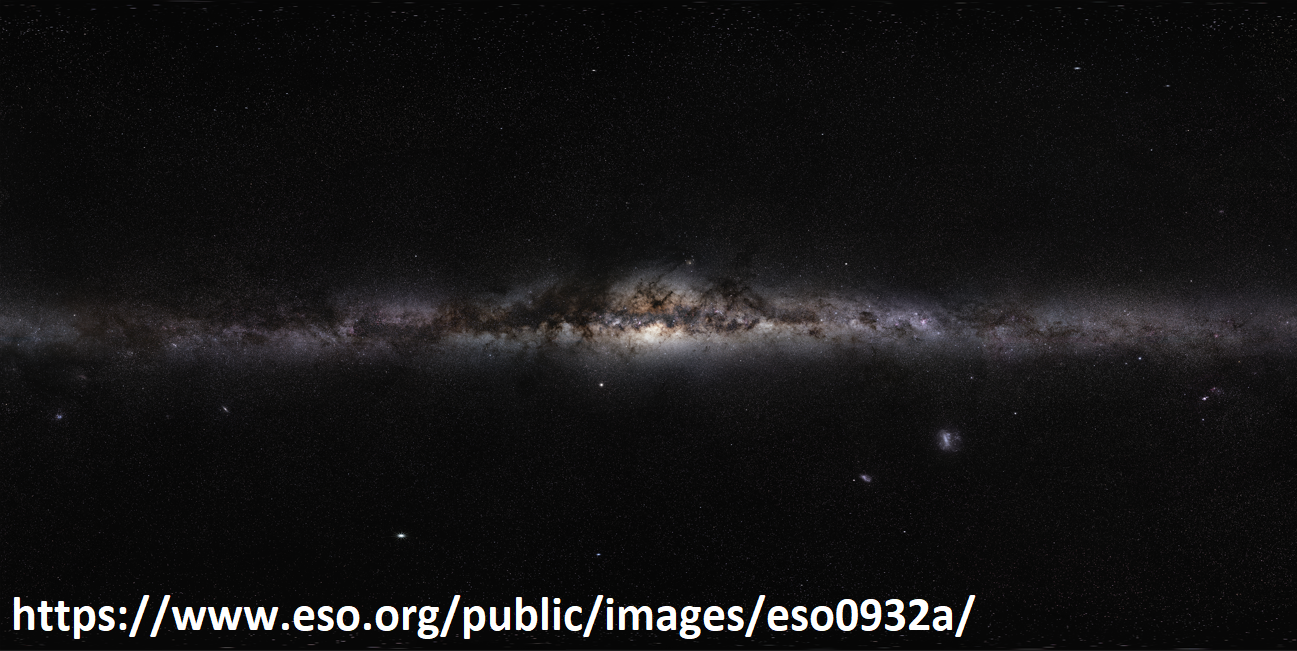}
		\caption{Milky Way panorama by \cite{panorama}, to provide a more realistic background.}
	\end{subfigure}
	\hfill
	\begin{subfigure}[b]{0.32\textwidth}
		\centering
		\includegraphics[width=\textwidth]{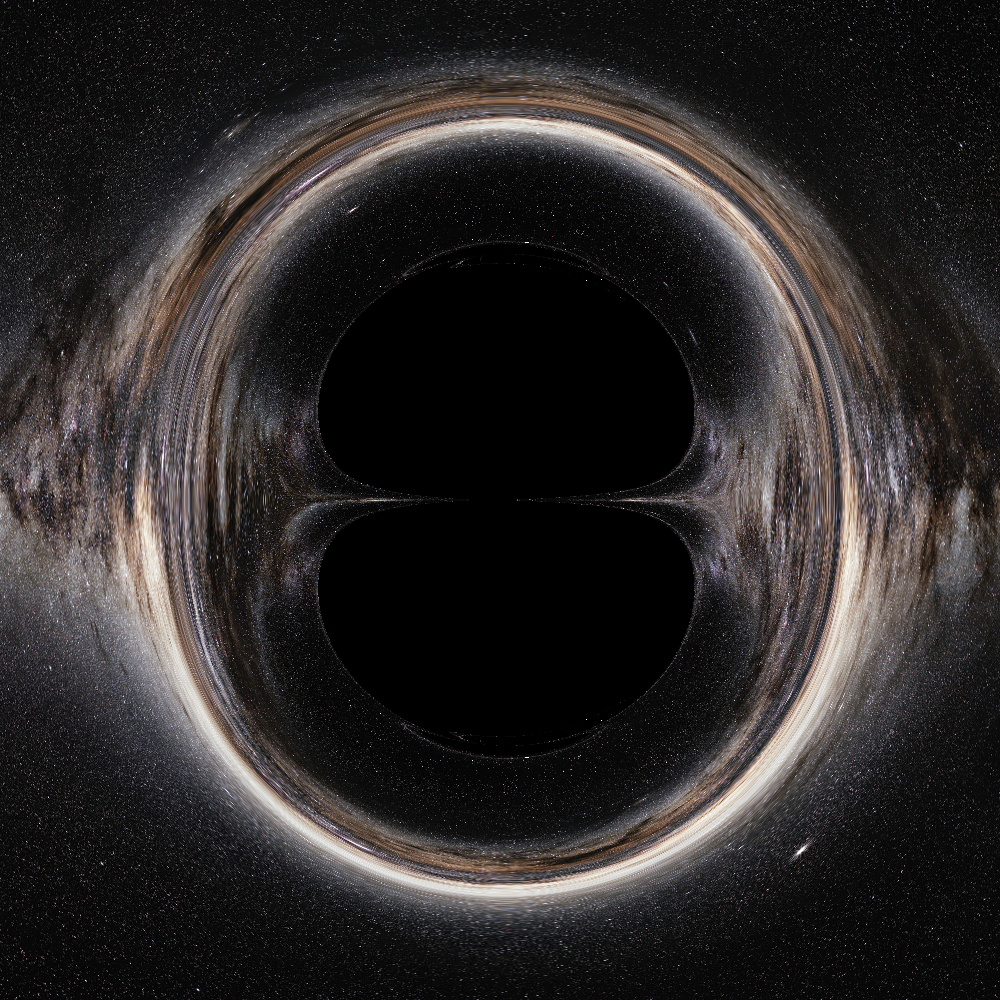}
		\caption{$q=-2,a=0.1$}
	\end{subfigure}
	\hfill
	\begin{subfigure}[b]{0.32\textwidth}
		\centering
		\includegraphics[width=\textwidth]{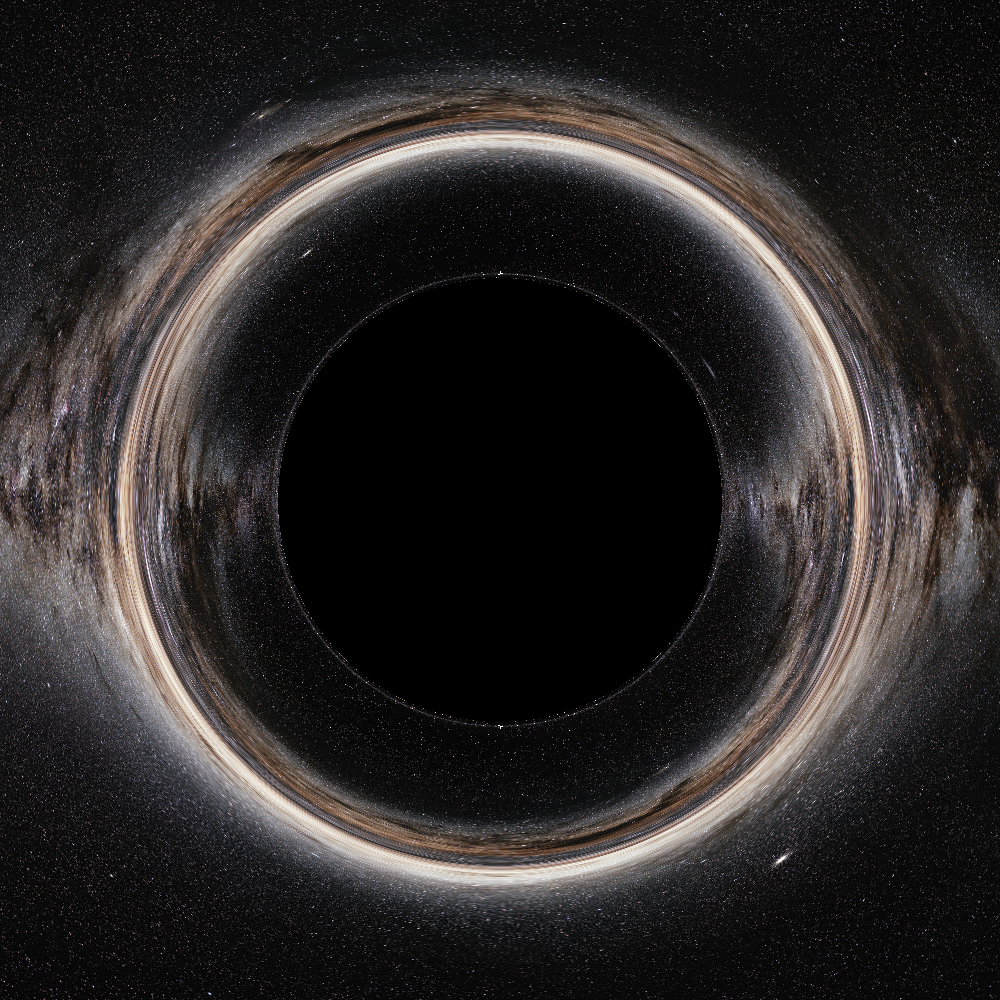}
		\caption{$q=0,a=0.0$}
	\end{subfigure}
	\hfill
	\begin{subfigure}[b]{0.32\textwidth}
		\centering
		\includegraphics[width=\textwidth]{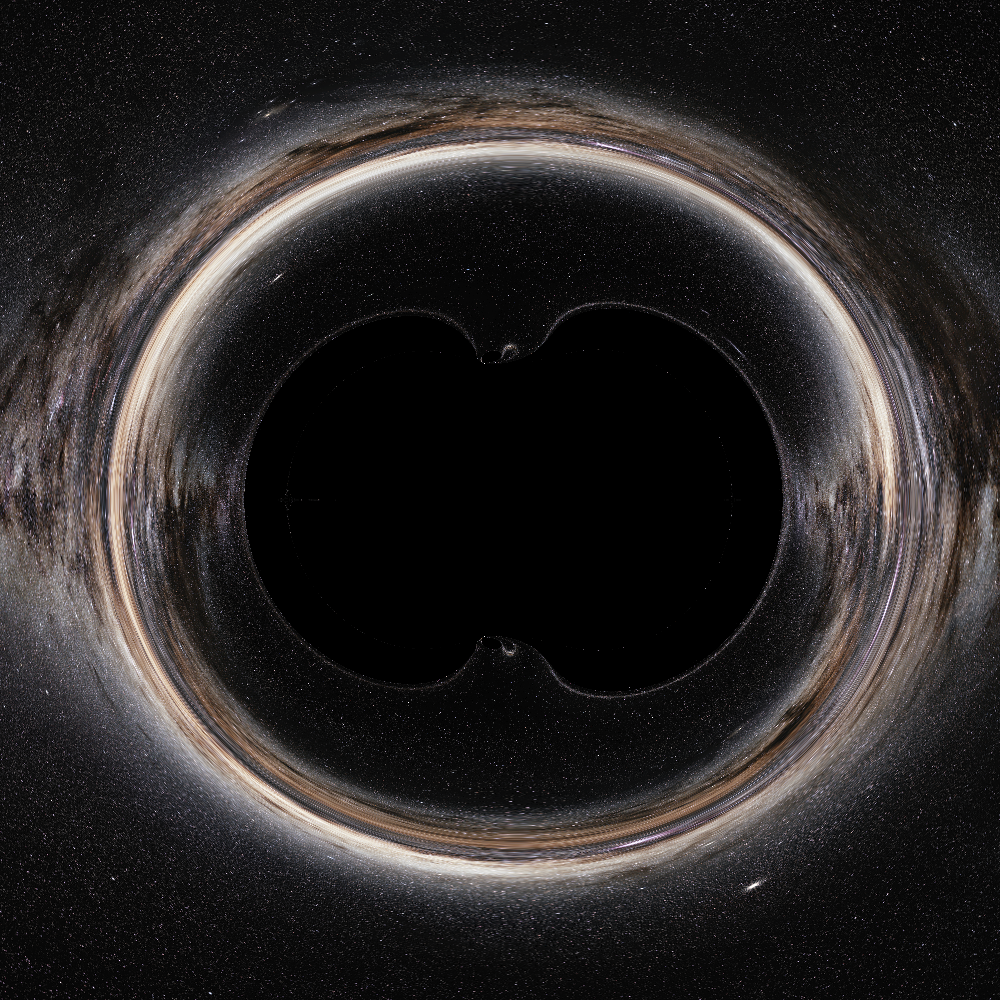}
		\caption{$q=2,a=0.1$}
	\end{subfigure}
	\begin{subfigure}[b]{0.32\textwidth}
		\centering
		\includegraphics[width=\textwidth]{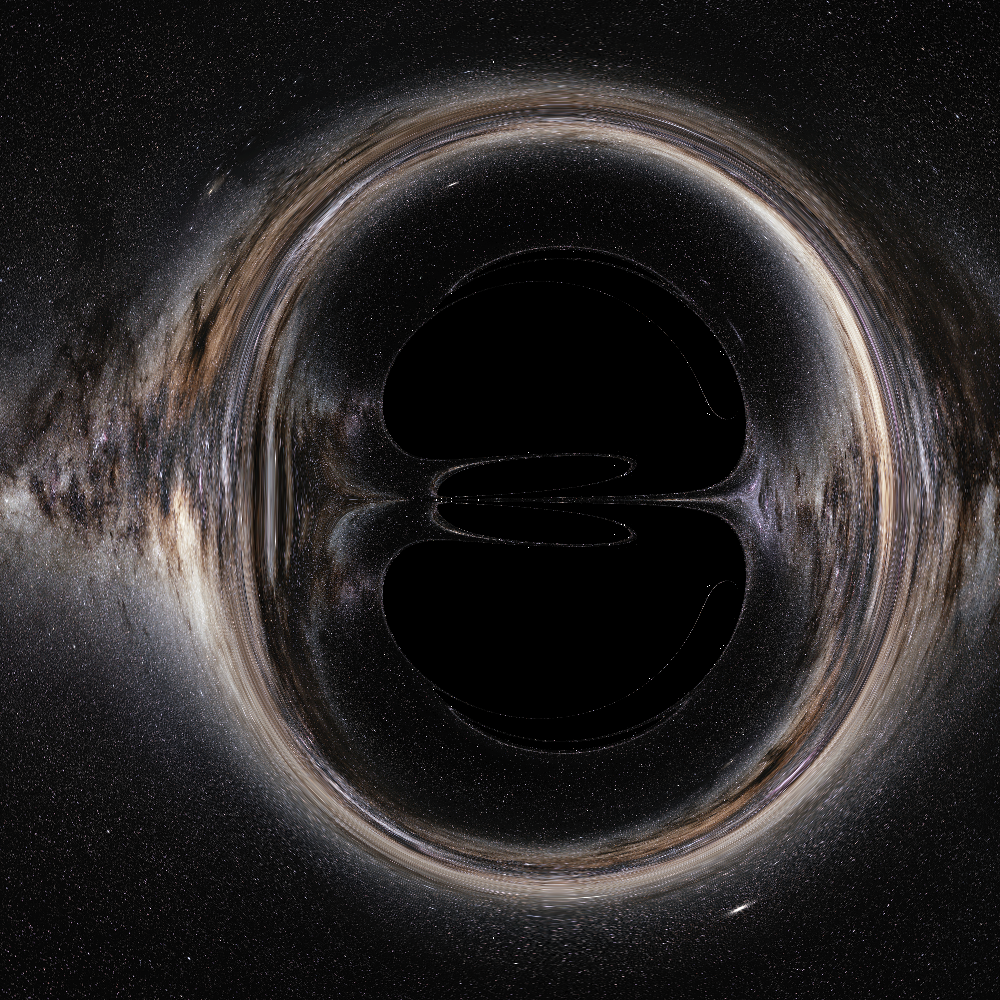}
		\caption{$q=-2,a=0.98$}
	\end{subfigure}
	\hfill
	\begin{subfigure}[b]{0.32\textwidth}
		\centering
		\includegraphics[width=\textwidth]{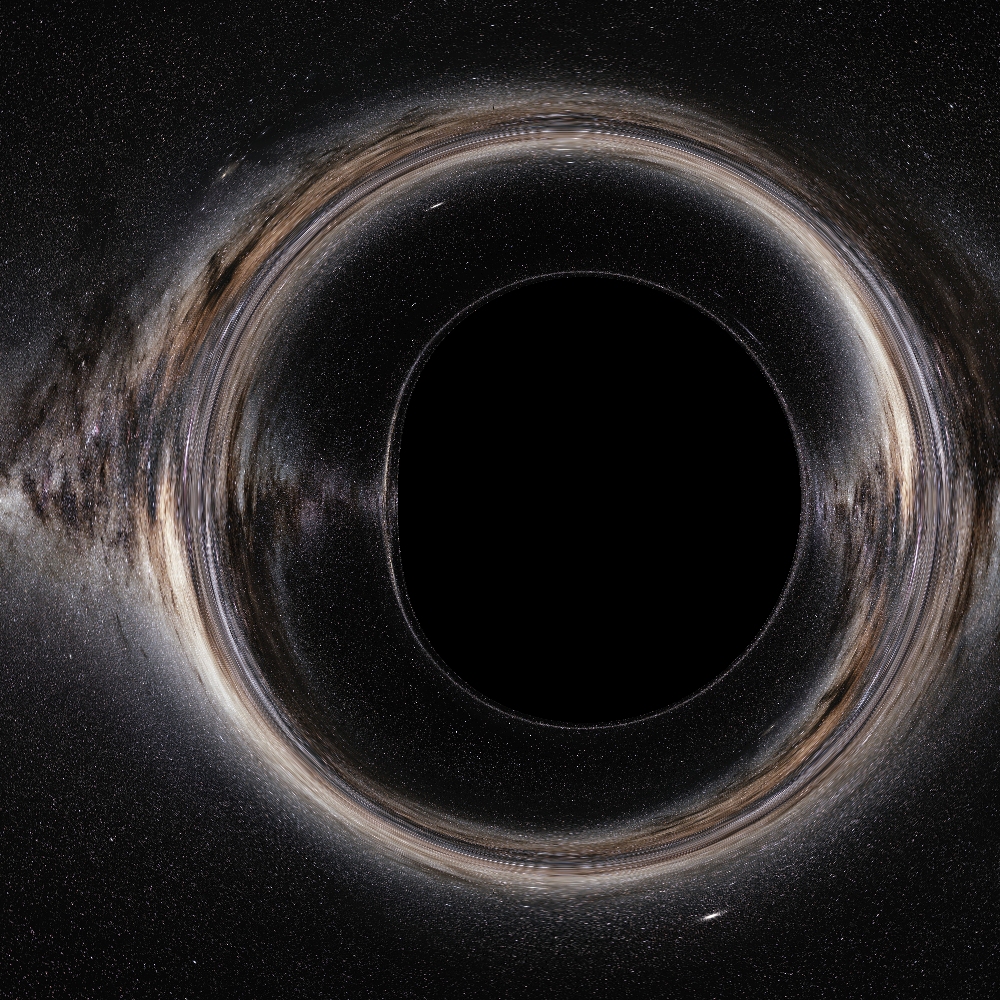}
		\caption{$q=0,a=0.98$}
	\end{subfigure}
	\hfill
	\begin{subfigure}[b]{0.32\textwidth}
		\centering
		\includegraphics[width=\textwidth]{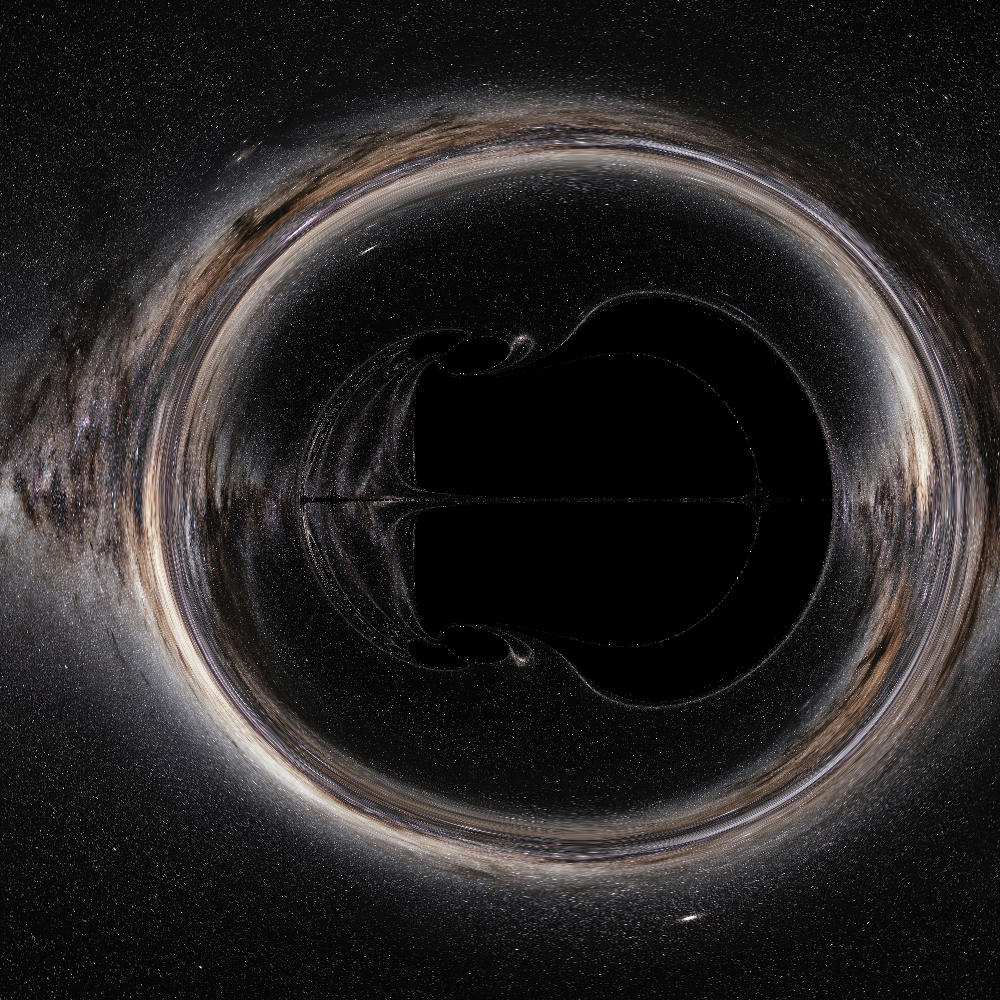}
		\caption{$q=2,a=0.98$}
	\end{subfigure}
	\caption{For these images the three lines on the background were replaced with a texture. Bilinear texture filtering was used to make the image look less pixelated.}
	\label{BHzooTexture}
\end{figure}

Instead of putting lines on the background, we can also use an image. The image I chose to use was a Milkyway panorama by the European Southern Observatory \cite{panorama}. This image and the visual results are seen in figure \ref{BHzooTexture}. To get the image less grainy, without needing to increase the resolution a simple anti-aliassing technique called bi-linear texture filtering was used.

\begin{figure}[h]
     \centering
     \begin{subfigure}[b]{0.32\textwidth}
         \centering
         \includegraphics[width=\textwidth]{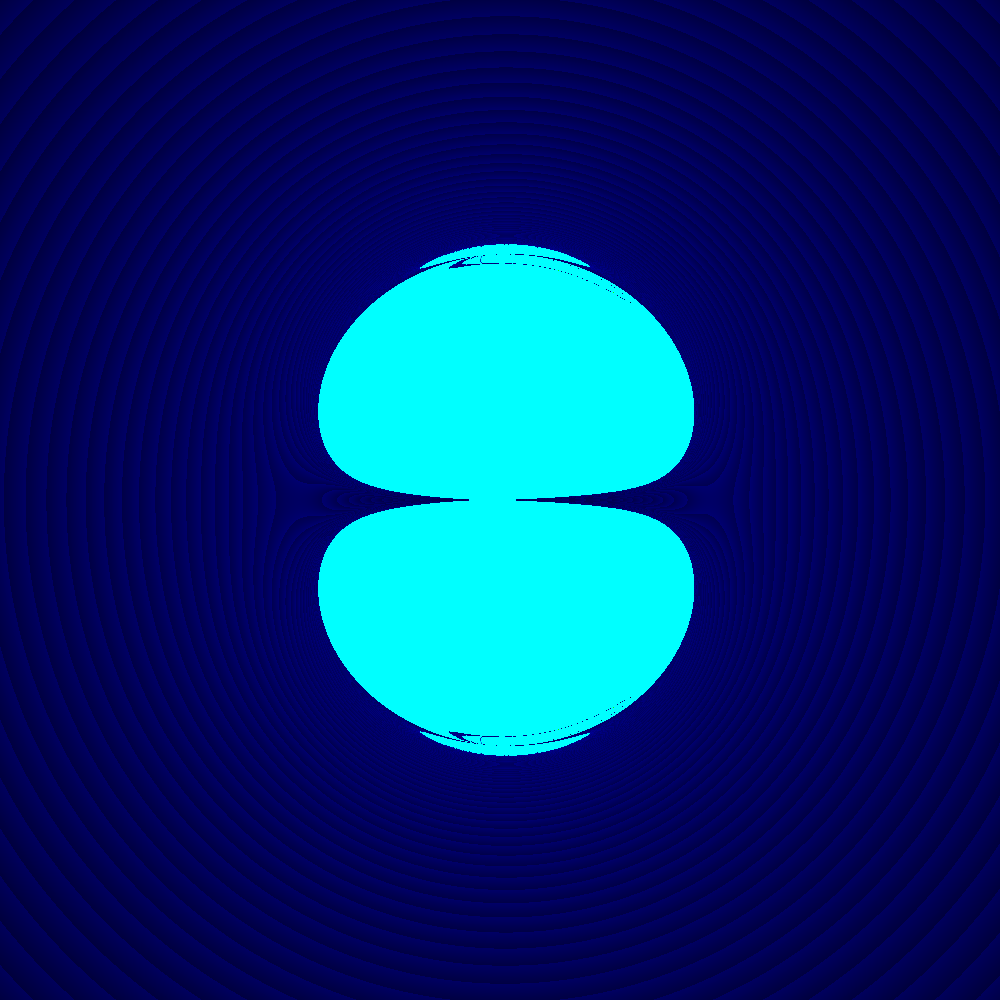}
         \caption{$q=-2,a=0.1$}
     \end{subfigure}
     \hfill
     \begin{subfigure}[b]{0.32\textwidth}
         \centering
         \includegraphics[width=\textwidth]{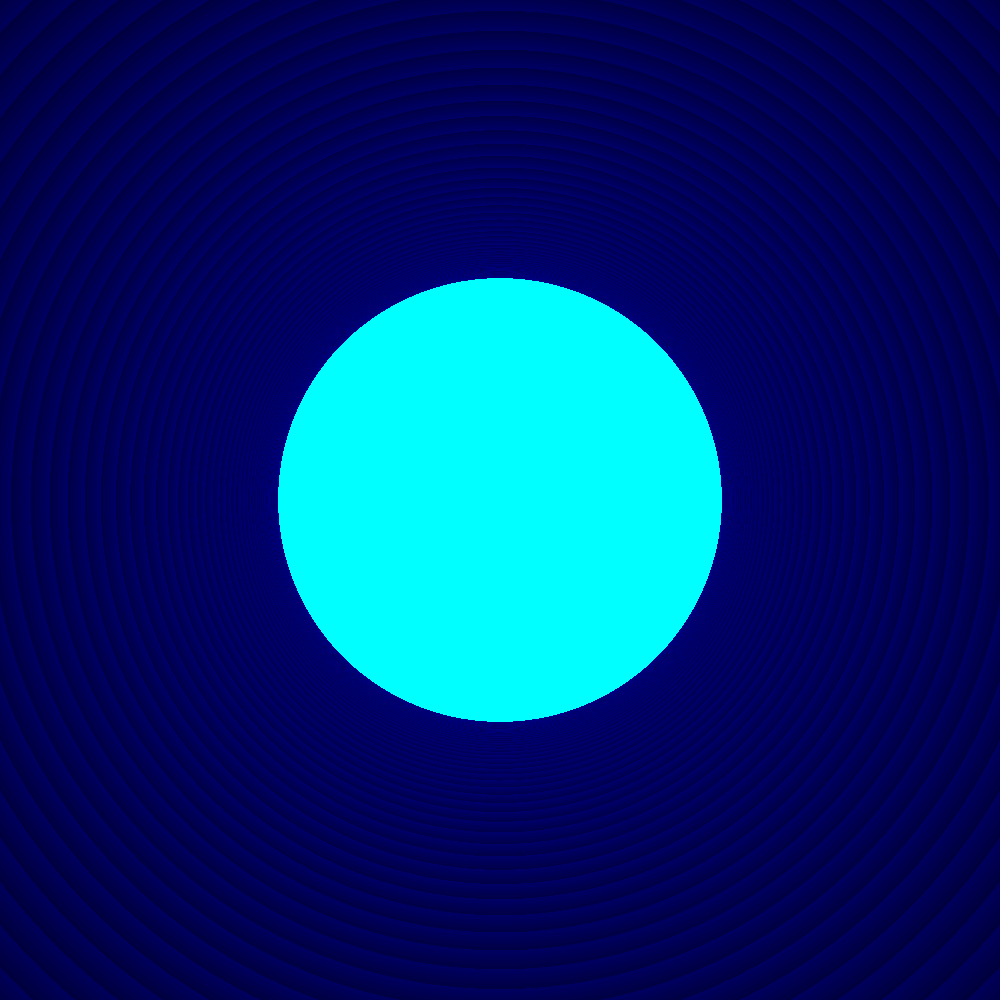}
         \caption{$q=0,a=0.0$}
     \end{subfigure}
     \hfill
     \begin{subfigure}[b]{0.32\textwidth}
         \centering
         \includegraphics[width=\textwidth]{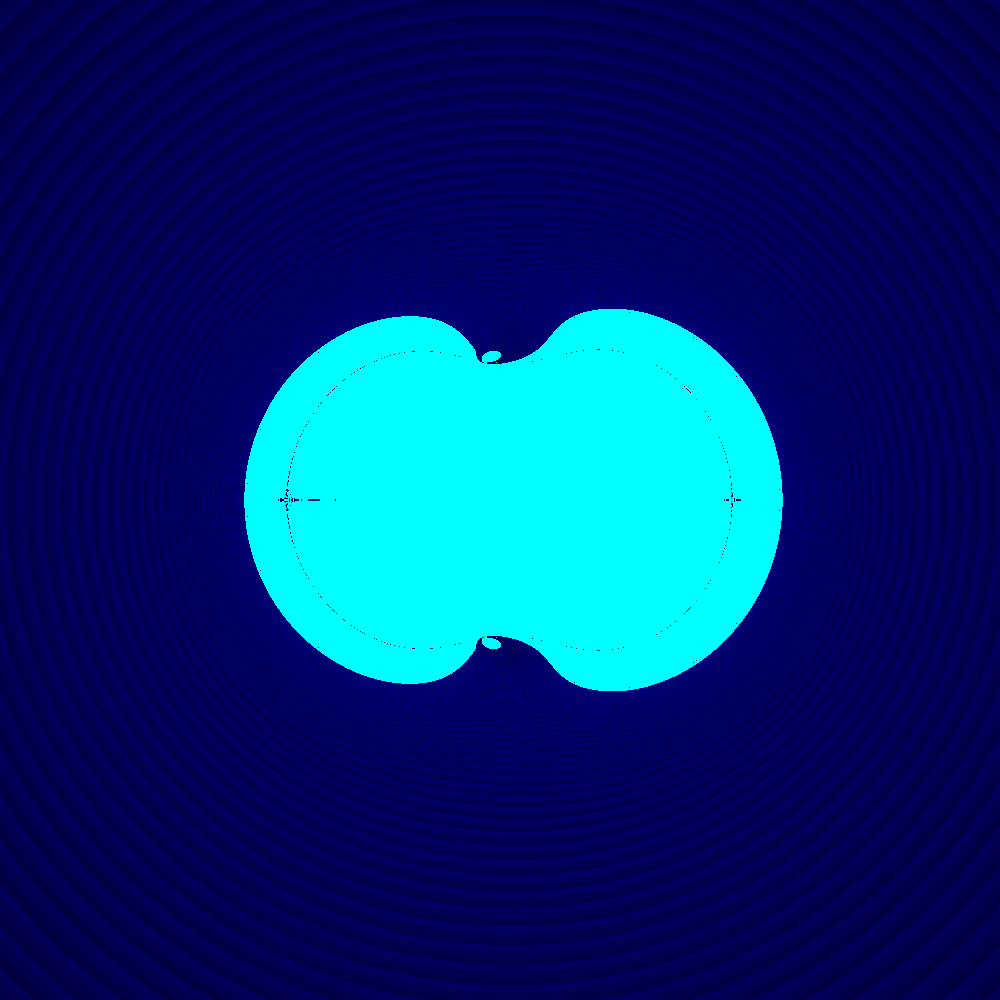}
         \caption{$q=2,a=0.1$}
     \end{subfigure}
      \begin{subfigure}[b]{0.32\textwidth}
         \centering
         \includegraphics[width=\textwidth]{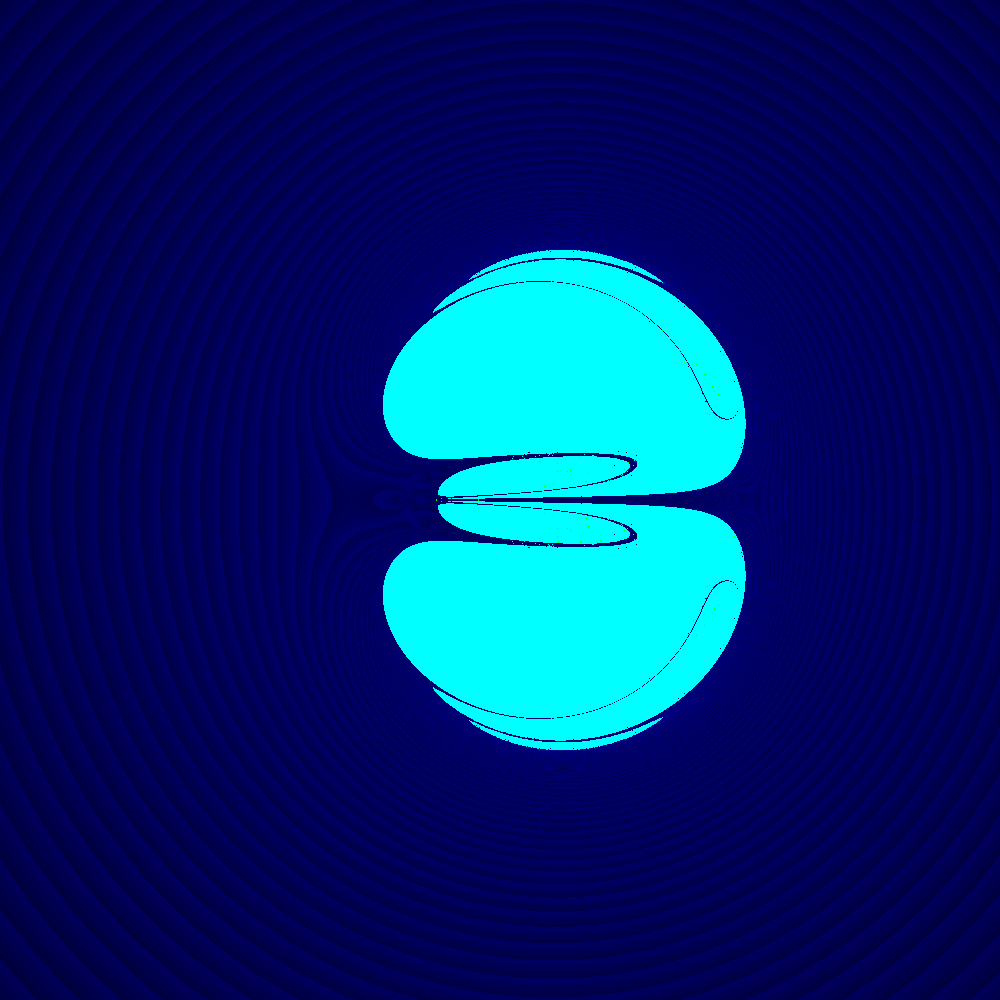}
         \caption{$q=-2,a=0.98$}
     \end{subfigure}
     \hfill
     \begin{subfigure}[b]{0.32\textwidth}
         \centering
         \includegraphics[width=\textwidth]{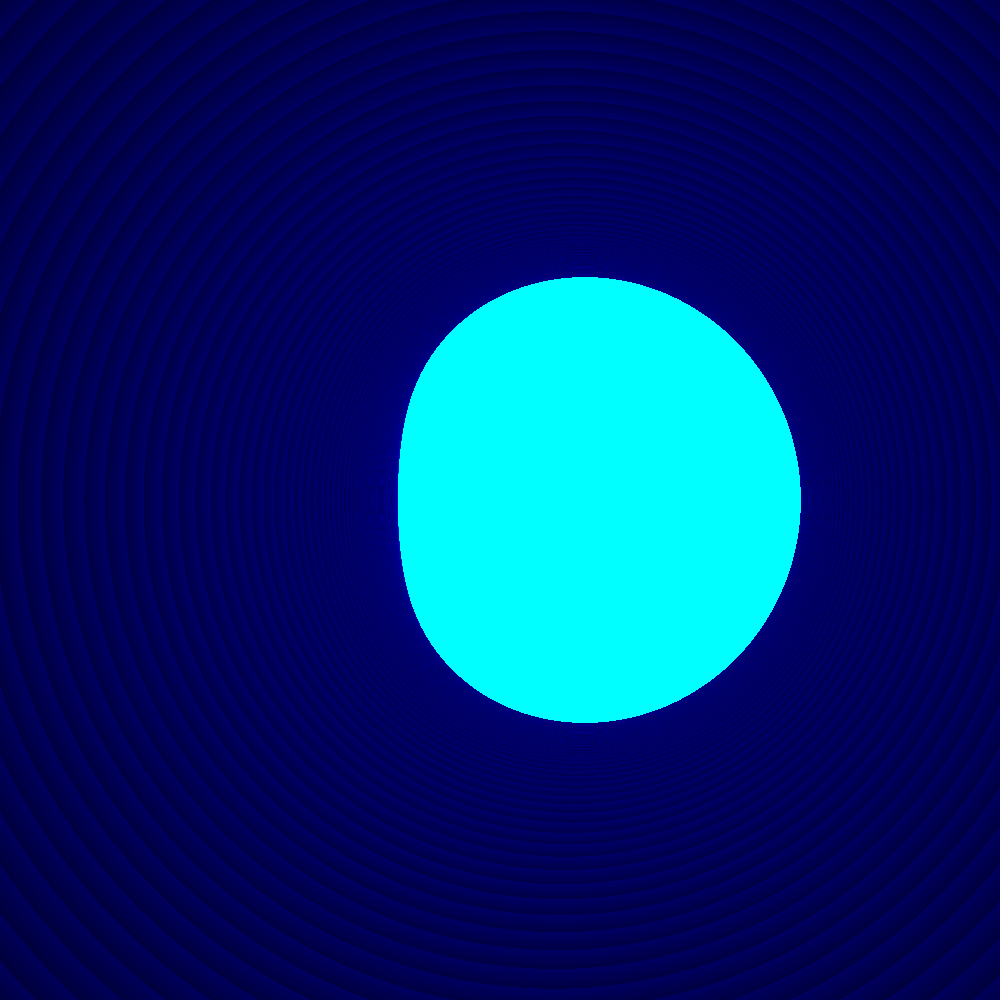}
         \caption{$q=0,a=0.98$}
     \end{subfigure}
     \hfill
     \begin{subfigure}[b]{0.32\textwidth}
         \centering
         \includegraphics[width=\textwidth]{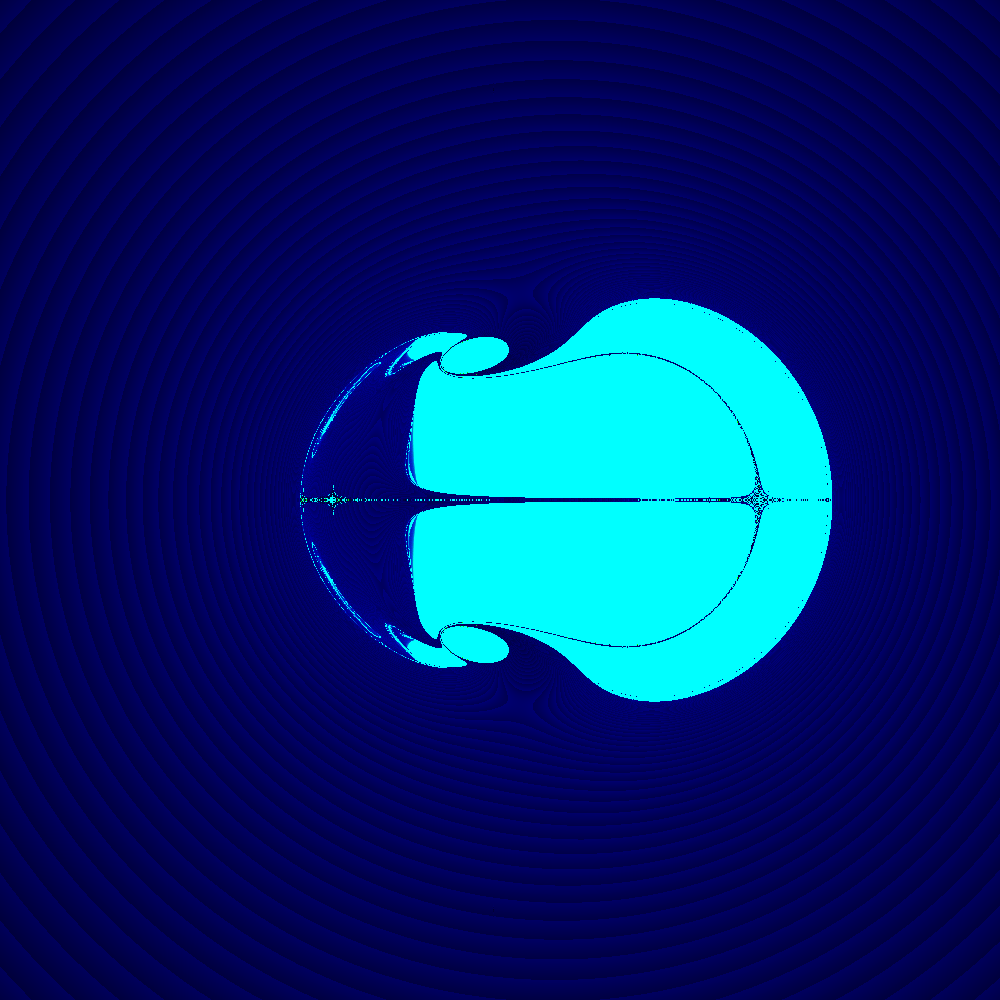}
         \caption{$q=2,a=0.98$}
     \end{subfigure}
     \caption{The calculation time for each pixel to reach the exit parameters of the simulation. Cyan is infinity (the simulation did not leave the bounding box before running out of time). Slight banding was added to the gradient to make it easier to see.}
     \label{BHzooTime}
\end{figure}

\newpage
Besides the spatial coordinate, we can also gain interesting information from the time coordinate, that is to say, how much (coordinate) time it takes for the light ray to leave the simulations boundary. This is plotted in figure \ref{BHzooTime}. In the plot we coloured every pixel based not on its spatial position as it exited the simulation, but its time position. Fast exit times are darker blue, slower exit times, or pixels that do not exit the simulation area, are coloured cyan. In the faster exit times some banding was added to the colour spectrum to more clearly see differences between pixels there. In this pictures the shadow seems to "glow" cyan. An interesting result we can see is that the shadow of the Manko-Novikov spacetimes becomes disconnected. 

\begin{figure}[h]
     \centering
     \begin{subfigure}[b]{0.48\textwidth}
         \centering
         \includegraphics[width=\textwidth]{figures/MNPlusSpaceRender.png}
         \caption{$q=2,a=0.98M$ spatial render.}
     \end{subfigure}
     \hfill
     \begin{subfigure}[b]{0.48\textwidth}
         \centering
         \includegraphics[width=\textwidth]{figures/MNPlusTimeRender.png}
         \caption{$q=2,a=0.98M$ exit time.}
     \end{subfigure}
     \hfill
     \begin{subfigure}[b]{0.48\textwidth}
         \centering
         \includegraphics[width=\textwidth]{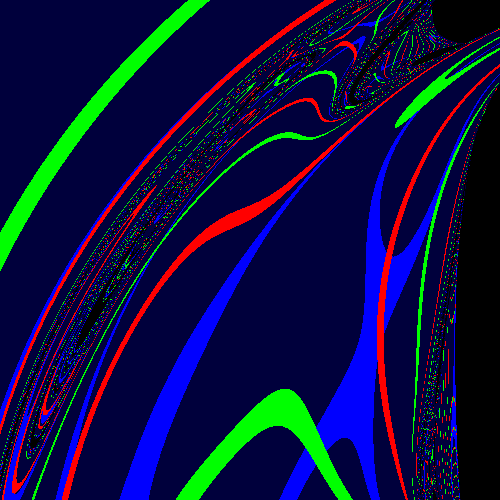}
         \caption{$q=2,a=0.98M$ spatial render, zoomed in.}
     \end{subfigure}
     \hfill
     \begin{subfigure}[b]{0.48\textwidth}
         \centering
         \includegraphics[width=\textwidth]{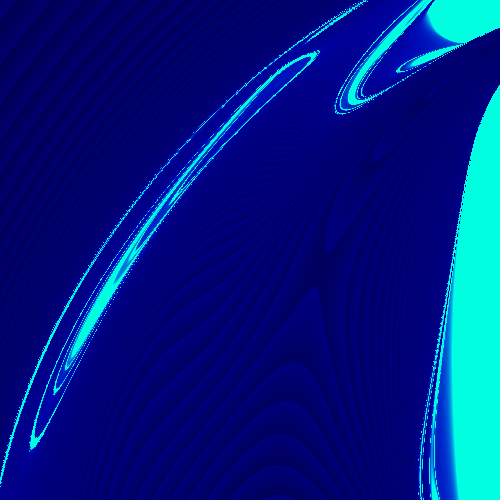}
         \caption{$q=2,a=0.98M$ exit time, zoomed in.}
     \end{subfigure}
     \caption{A render of the Manko Novikov, zoomed in on the area enclosed by the Halo above the horizon. In the bottom left picture we see that this region deflects rays very chaotically, as all colours are present meaning rays get deflected in every direction. In the bottom right we see that there is an extra part of the shadow, disconnected from the rest of the structure, which seems to be shaped like concentric rings, or a similar shape.}
     \label{MNzoomed}
\end{figure}
\begin{figure}[h]
     \centering
     \begin{subfigure}[b]{0.42\textwidth}
         \centering
         \includegraphics[width=\textwidth]{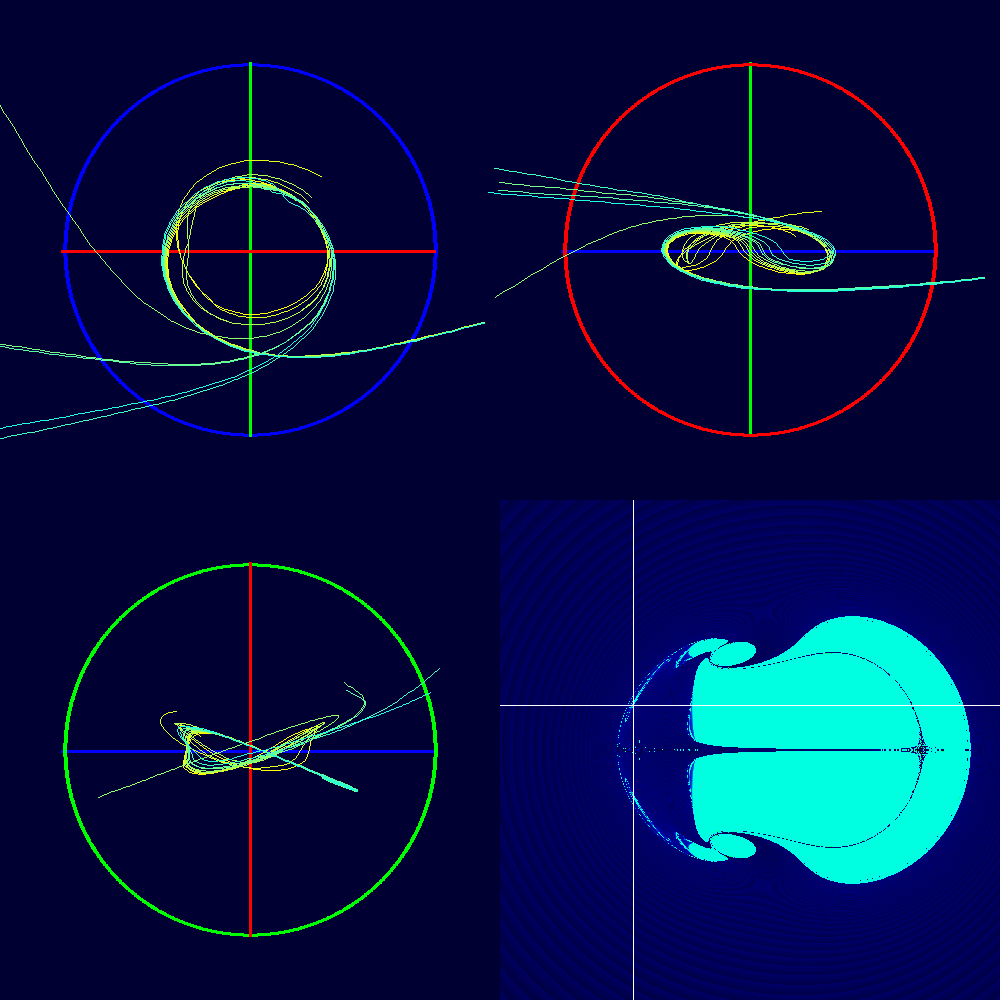}
     \end{subfigure}
     \hfill
     \begin{subfigure}[b]{0.42\textwidth}
         \centering
         \includegraphics[width=\textwidth]{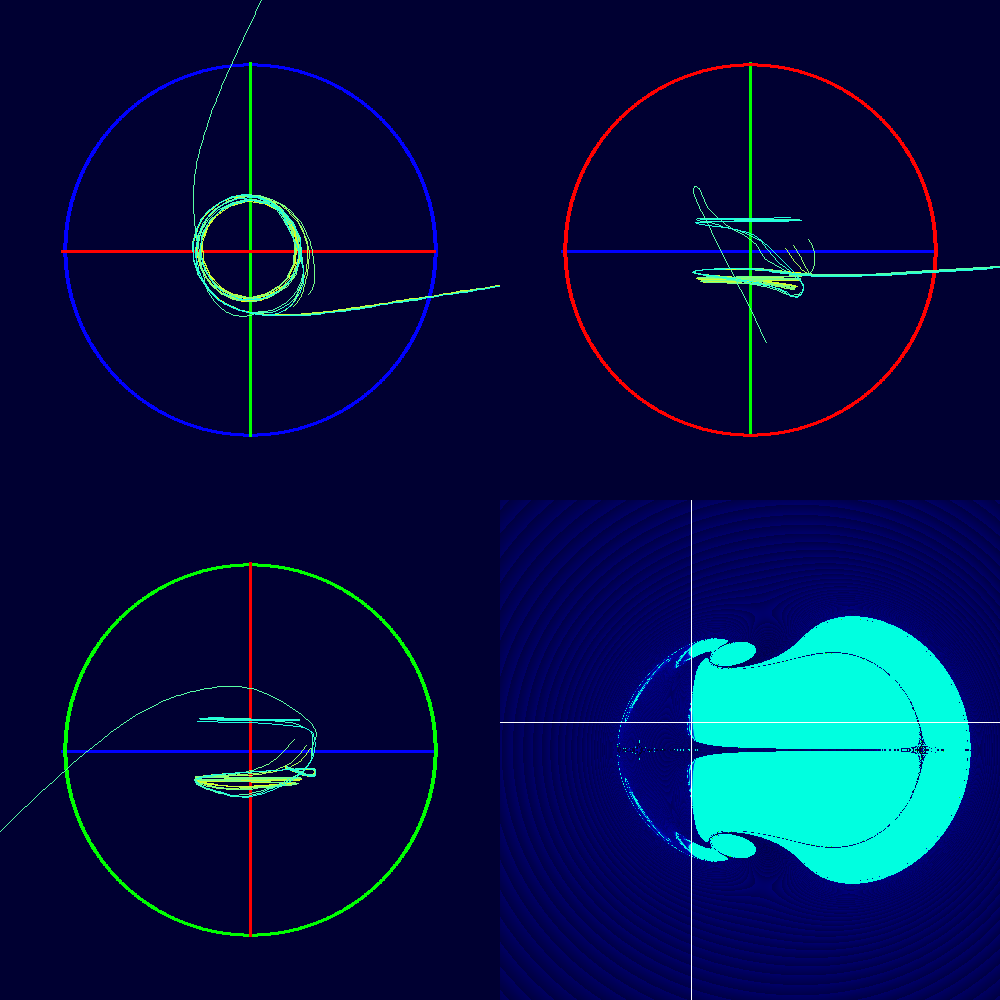}
     \end{subfigure}
     \hfill
     \begin{subfigure}[b]{0.42\textwidth}
         \centering
         \includegraphics[width=\textwidth]{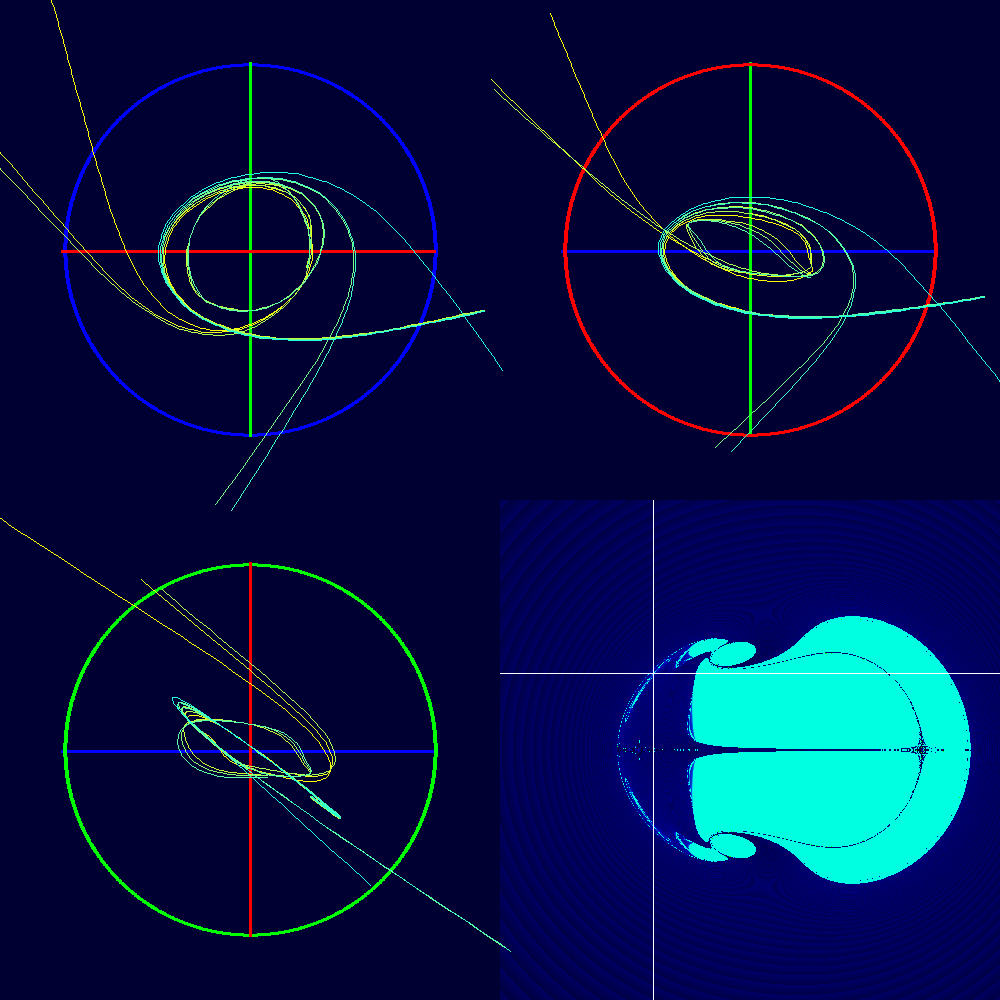}
     \end{subfigure}
     \hfill
     \begin{subfigure}[b]{0.42\textwidth}
         \centering
         \includegraphics[width=\textwidth]{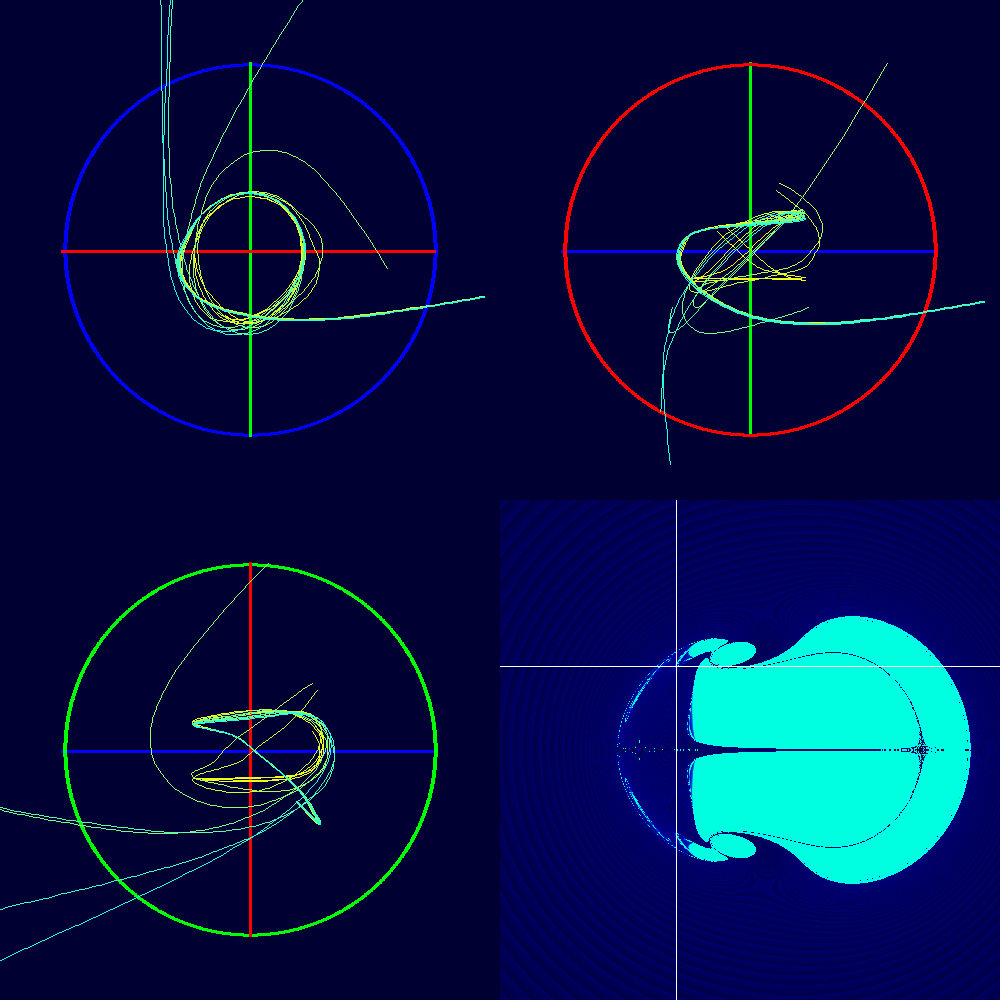}
     \end{subfigure}
     \hfill
     \begin{subfigure}[b]{0.42\textwidth}
         \centering
         \includegraphics[width=\textwidth]{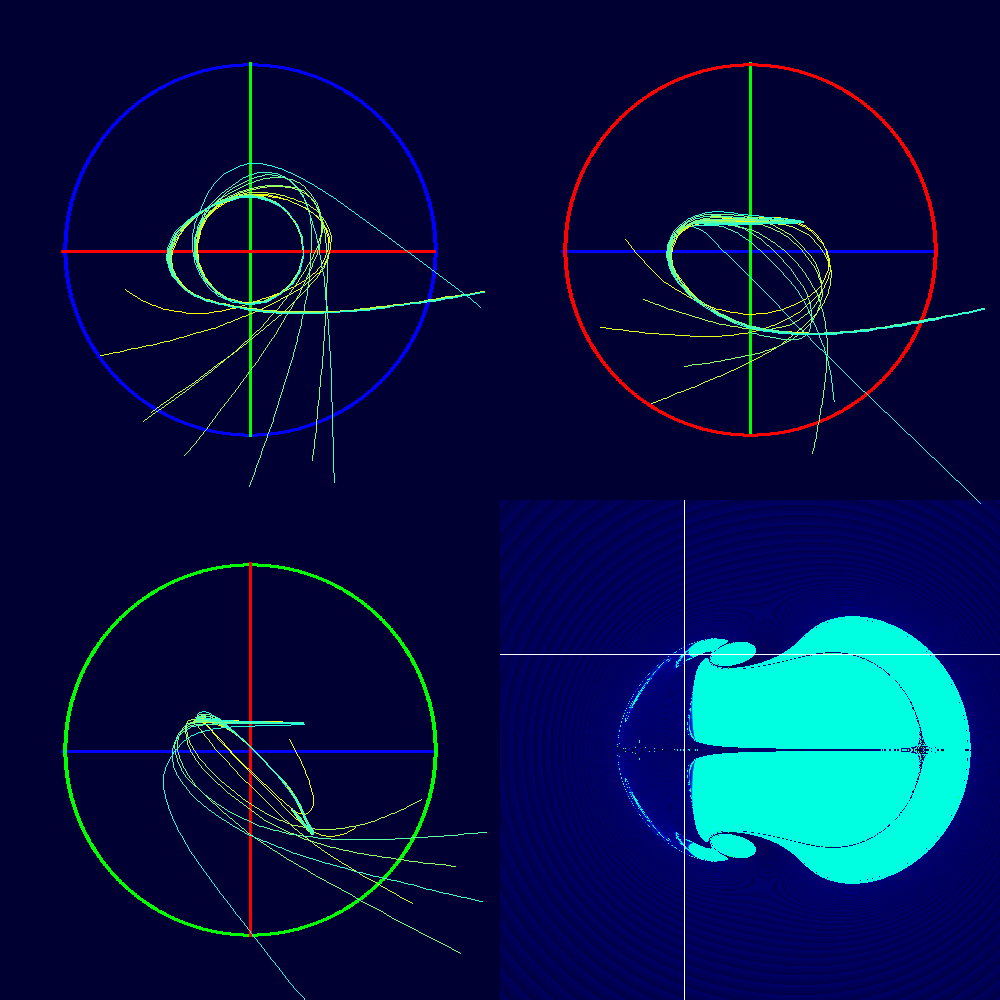}
     \end{subfigure}
     \hfill
     \begin{subfigure}[b]{0.42\textwidth}
         \centering
         \includegraphics[width=\textwidth]{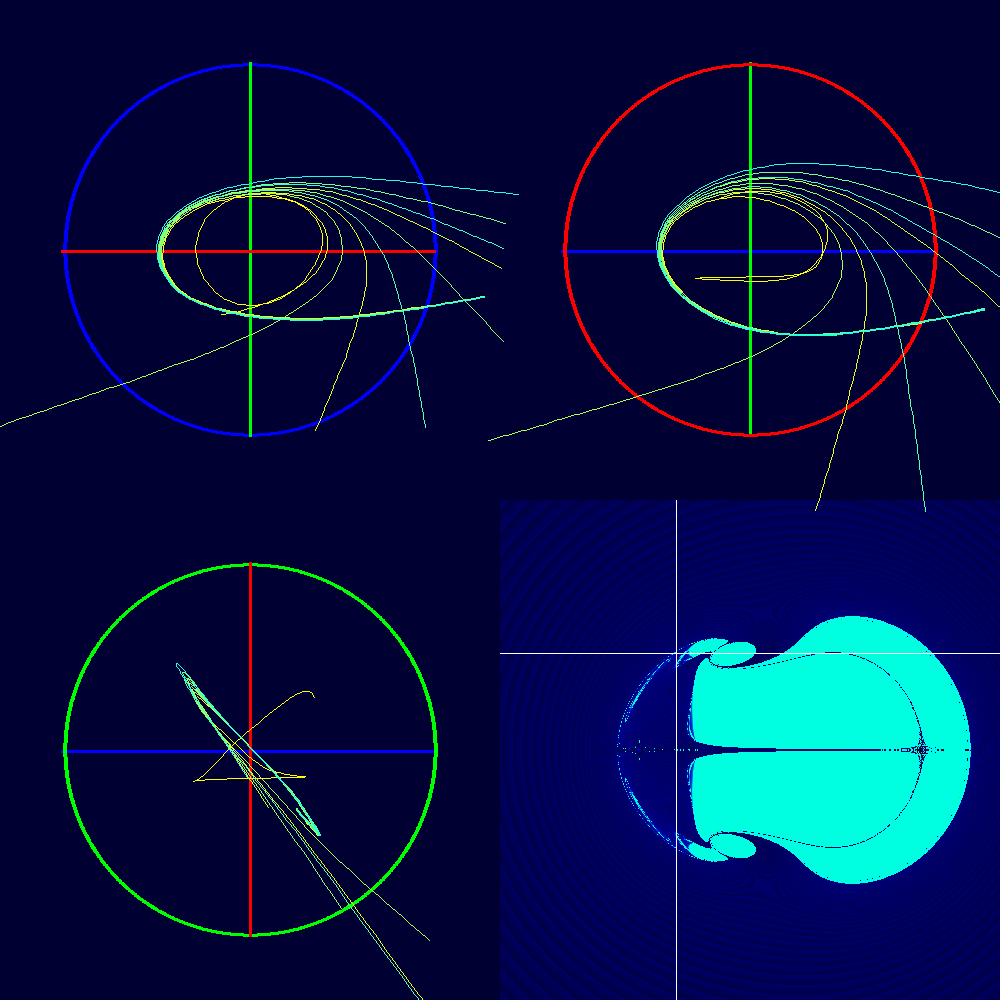}
     \end{subfigure}
     \caption{Samples of orbits from the generated shadows of the Manko-Novikov black hole with $a=0.98,q=2$. The red, green and blue lines on the background are drawn. They are drawn at a lower radius of $r=6M$ to allow for a more zoomed image. Each image has 9 orbits, the central orbit is given by the pixel indicated on the bottom right of the image, while the 8 around it are each offset along the image by a fraction of a pixel.}
     \label{MNorbits1}
\end{figure}

\begin{figure}[h]
	\centering
	\begin{subfigure}[b]{0.32\textwidth}
		\centering
		\includegraphics[width=\textwidth]{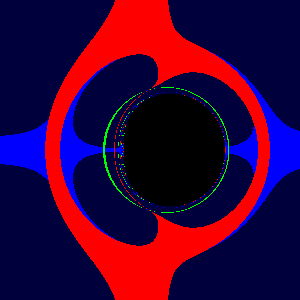}
	\end{subfigure}
	\hfill
	\begin{subfigure}[b]{0.32\textwidth}
		\centering
		\includegraphics[width=\textwidth]{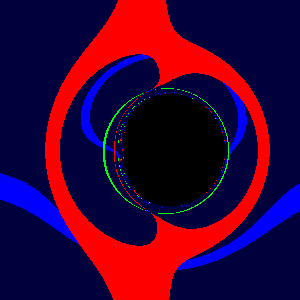}
	\end{subfigure}
	\hfill
	\begin{subfigure}[b]{0.32\textwidth}
		\centering
		\includegraphics[width=\textwidth]{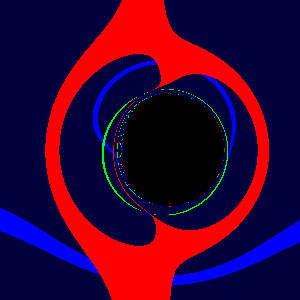}
	\end{subfigure}
	\hfill
	\begin{subfigure}[b]{0.32\textwidth}
		\centering
		\includegraphics[width=\textwidth]{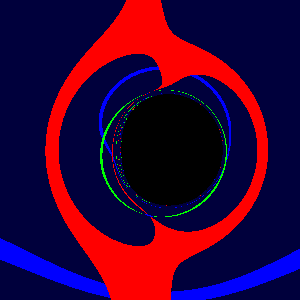}
	\end{subfigure}
	\hfill
	\begin{subfigure}[b]{0.32\textwidth}
		\centering
		\includegraphics[width=\textwidth]{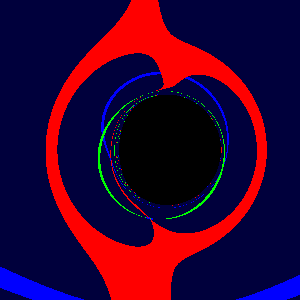}
	\end{subfigure}
	\hfill
	\begin{subfigure}[b]{0.32\textwidth}
		\centering
		\includegraphics[width=\textwidth]{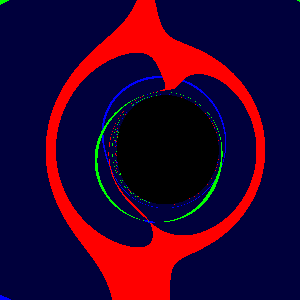}
	\end{subfigure}
	\hfill
	\begin{subfigure}[b]{0.32\textwidth}
		\centering
		\includegraphics[width=\textwidth]{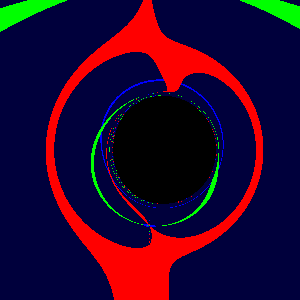}
	\end{subfigure}
	\hfill
	\begin{subfigure}[b]{0.32\textwidth}
		\centering
		\includegraphics[width=\textwidth]{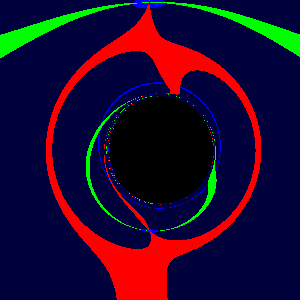}
	\end{subfigure}
	\hfill
	\begin{subfigure}[b]{0.32\textwidth}
		\centering
		\includegraphics[width=\textwidth]{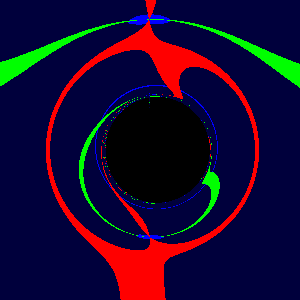}
	\end{subfigure}
	\hfill
	\begin{subfigure}[b]{0.32\textwidth}
		\centering
		\includegraphics[width=\textwidth]{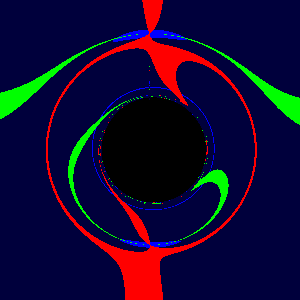}
	\end{subfigure}
	\caption{The Shadow of the Kerr black hole (M = 0.5, a = 0.98M), seen from a different $\theta$, starting from $0$ to $0.9*\pi/2$ in steps of $0.1*\pi/2$. The exact vertical angle is not added as the simulation breaks at the axis of the Boyer-Lidnquist coordinates.}
	\label{MNrotT}
\end{figure}

\begin{figure}[h]
	\centering
	\begin{subfigure}[b]{0.32\textwidth}
		\centering
		\includegraphics[width=\textwidth]{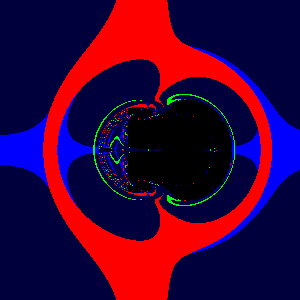}
	\end{subfigure}
	\hfill
	\begin{subfigure}[b]{0.32\textwidth}
		\centering
		\includegraphics[width=\textwidth]{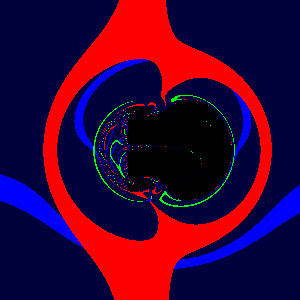}
	\end{subfigure}
	\hfill
	\begin{subfigure}[b]{0.32\textwidth}
		\centering
		\includegraphics[width=\textwidth]{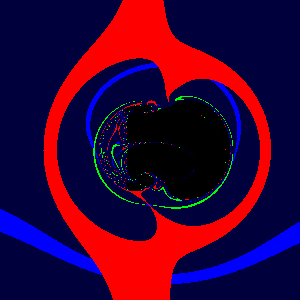}
	\end{subfigure}
	\hfill
	\begin{subfigure}[b]{0.32\textwidth}
		\centering
		\includegraphics[width=\textwidth]{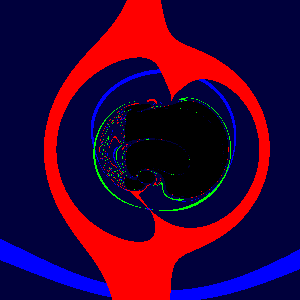}
	\end{subfigure}
	\hfill
	\begin{subfigure}[b]{0.32\textwidth}
		\centering
		\includegraphics[width=\textwidth]{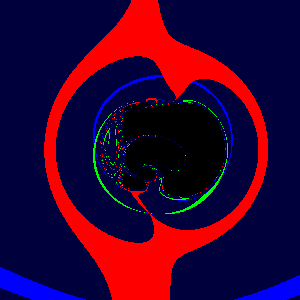}
	\end{subfigure}
	\hfill
	\begin{subfigure}[b]{0.32\textwidth}
		\centering
		\includegraphics[width=\textwidth]{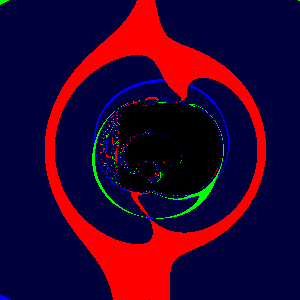}
	\end{subfigure}
	\hfill
	\begin{subfigure}[b]{0.32\textwidth}
		\centering
		\includegraphics[width=\textwidth]{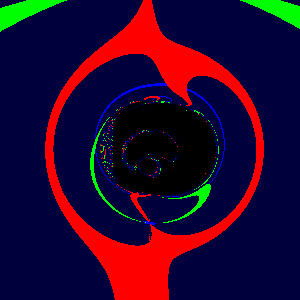}
	\end{subfigure}
	\hfill
	\begin{subfigure}[b]{0.32\textwidth}
		\centering
		\includegraphics[width=\textwidth]{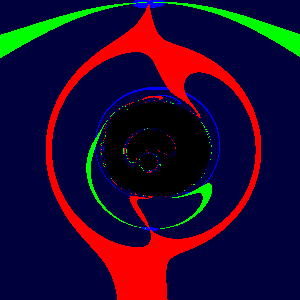}
	\end{subfigure}
	\hfill
	\begin{subfigure}[b]{0.32\textwidth}
		\centering
		\includegraphics[width=\textwidth]{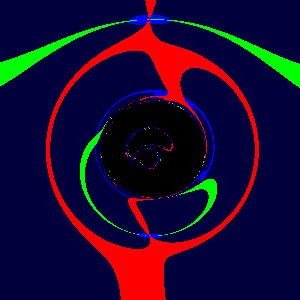}
	\end{subfigure}
	\hfill
	\begin{subfigure}[b]{0.32\textwidth}
		\centering
		\includegraphics[width=\textwidth]{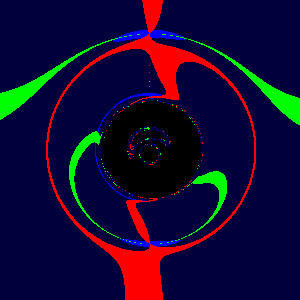}
	\end{subfigure}
	\caption{The Shadow of the MN Black hole (M=0.5,a=0.98M,q=2.0), seen from a different $\theta$.}
	\label{MNRot}
\end{figure}

\begin{figure}[h]
	\centering
	\begin{subfigure}[b]{0.32\textwidth}
		\centering
		\includegraphics[width=\textwidth]{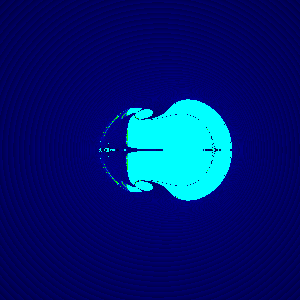}
	\end{subfigure}
	\hfill
	\begin{subfigure}[b]{0.32\textwidth}
		\centering
		\includegraphics[width=\textwidth]{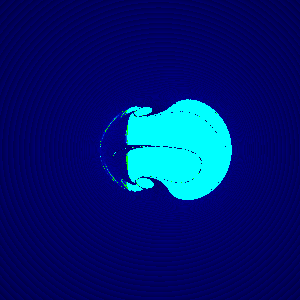}
	\end{subfigure}
	\hfill
	\begin{subfigure}[b]{0.32\textwidth}
		\centering
		\includegraphics[width=\textwidth]{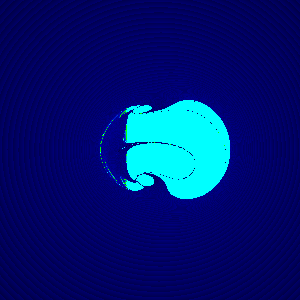}
	\end{subfigure}
	\hfill
	\begin{subfigure}[b]{0.32\textwidth}
		\centering
		\includegraphics[width=\textwidth]{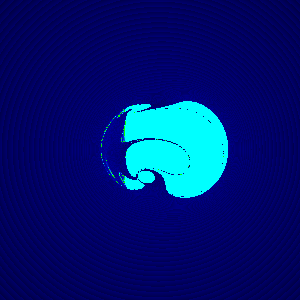}
	\end{subfigure}
	\hfill
	\begin{subfigure}[b]{0.32\textwidth}
		\centering
		\includegraphics[width=\textwidth]{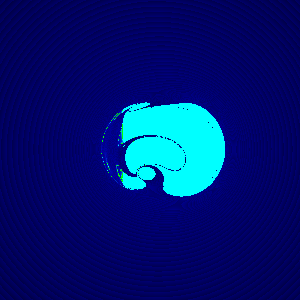}
	\end{subfigure}
	\hfill
	\begin{subfigure}[b]{0.32\textwidth}
		\centering
		\includegraphics[width=\textwidth]{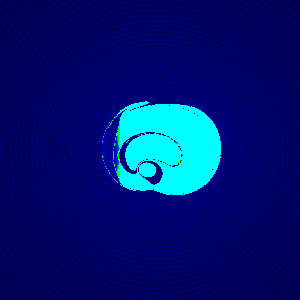}
	\end{subfigure}
	\hfill
	\begin{subfigure}[b]{0.32\textwidth}
		\centering
		\includegraphics[width=\textwidth]{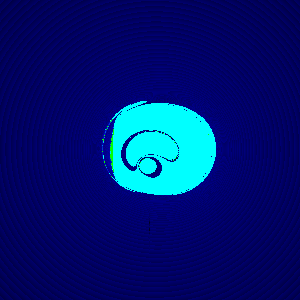}
	\end{subfigure}
	\hfill
	\begin{subfigure}[b]{0.32\textwidth}
		\centering
		\includegraphics[width=\textwidth]{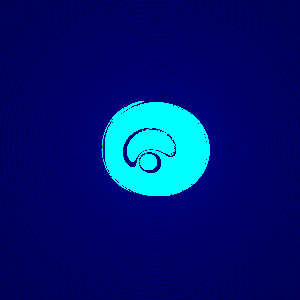}
	\end{subfigure}
	\hfill
	\begin{subfigure}[b]{0.32\textwidth}
		\centering
		\includegraphics[width=\textwidth]{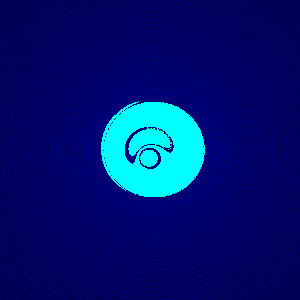}
	\end{subfigure}
	\hfill
	\begin{subfigure}[b]{0.32\textwidth}
		\centering
		\includegraphics[width=\textwidth]{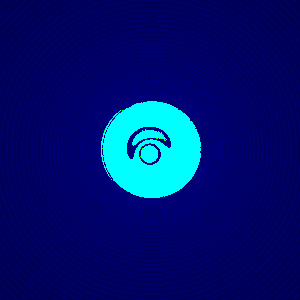}
	\end{subfigure}
	\caption{The Shadow of the MN Black hole (M=0.5,a=0.98M,q=2.0), seen from a different $\theta$.}
	\label{MNrotT}
\end{figure}

\begin{figure}[h]
	\centering
	\begin{subfigure}[b]{0.32\textwidth}
		\centering
		\includegraphics[width=\textwidth]{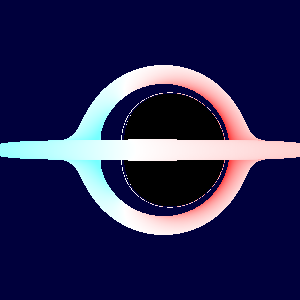}
	\end{subfigure}
	\hfill
	\begin{subfigure}[b]{0.32\textwidth}
		\centering
		\includegraphics[width=\textwidth]{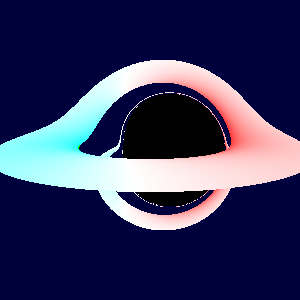}
	\end{subfigure}
	\hfill
	\begin{subfigure}[b]{0.32\textwidth}
		\centering
		\includegraphics[width=\textwidth]{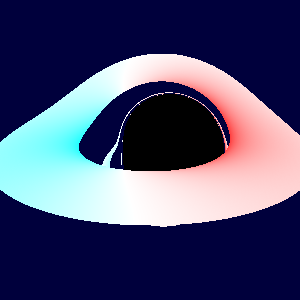}
	\end{subfigure}
	\hfill
	\begin{subfigure}[b]{0.32\textwidth}
		\centering
		\includegraphics[width=\textwidth]{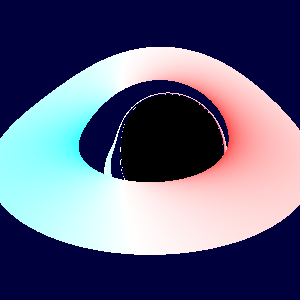}
	\end{subfigure}
	\hfill
	\begin{subfigure}[b]{0.32\textwidth}
		\centering
		\includegraphics[width=\textwidth]{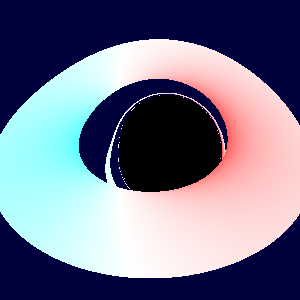}
	\end{subfigure}
	\hfill
	\begin{subfigure}[b]{0.32\textwidth}
		\centering
		\includegraphics[width=\textwidth]{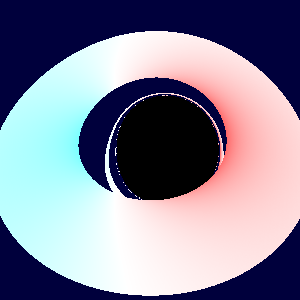}
	\end{subfigure}
	\hfill
	\begin{subfigure}[b]{0.32\textwidth}
		\centering
		\includegraphics[width=\textwidth]{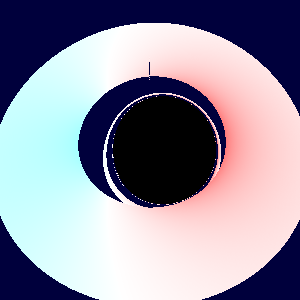}
	\end{subfigure}
	\hfill
	\begin{subfigure}[b]{0.32\textwidth}
		\centering
		\includegraphics[width=\textwidth]{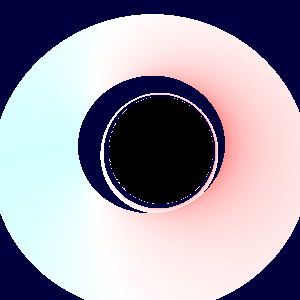}
	\end{subfigure}
	\hfill
	\begin{subfigure}[b]{0.32\textwidth}
		\centering
		\includegraphics[width=\textwidth]{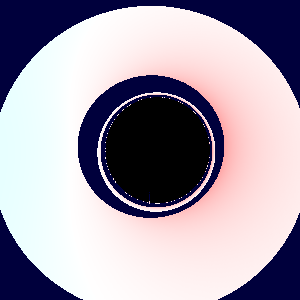}
	\end{subfigure}
	\hfill
	\begin{subfigure}[b]{0.32\textwidth}
		\centering
		\includegraphics[width=\textwidth]{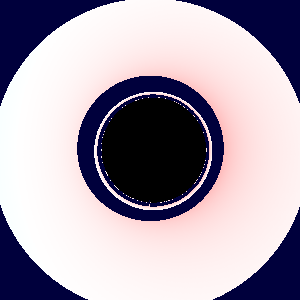}
	\end{subfigure}
	\caption{A Kerr Black hole with a ring (M = 0.5, a = 0.98M). Redshift is calculated for the rays intersecting the ring, and colours of the ring are based on this redshift. The ring is a region in space around the equator (of slightly above non-zero thickness). If a ray intersects this bounding box it will be put through the redshift calculation described. White means the redshift factor is around $1$, red means a factor higher than $1$ and blue means a factor lower than $1$.}
	\label{KerrDisk}
\end{figure}

\begin{figure}[h]
	\centering
	\begin{subfigure}[b]{0.32\textwidth}
		\centering
		\includegraphics[width=\textwidth]{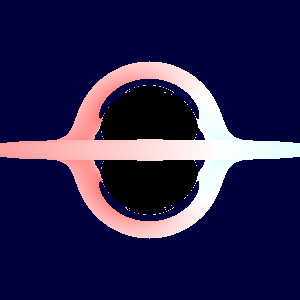}
	\end{subfigure}
	\hfill
	\begin{subfigure}[b]{0.32\textwidth}
		\centering
		\includegraphics[width=\textwidth]{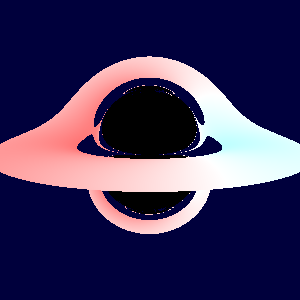}
	\end{subfigure}
	\hfill
	\begin{subfigure}[b]{0.32\textwidth}
		\centering
		\includegraphics[width=\textwidth]{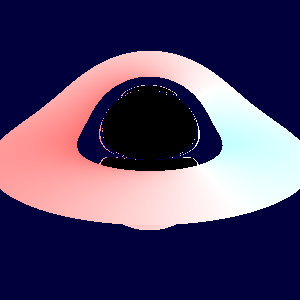}
	\end{subfigure}
	\hfill
	\begin{subfigure}[b]{0.32\textwidth}
		\centering
		\includegraphics[width=\textwidth]{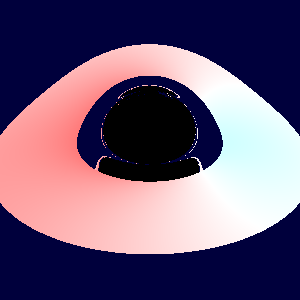}
	\end{subfigure}
	\hfill
	\begin{subfigure}[b]{0.32\textwidth}
		\centering
		\includegraphics[width=\textwidth]{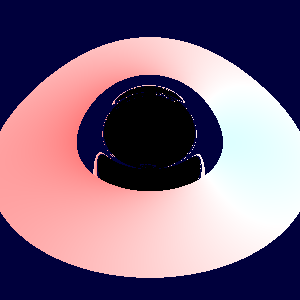}
	\end{subfigure}
	\hfill
	\begin{subfigure}[b]{0.32\textwidth}
		\centering
		\includegraphics[width=\textwidth]{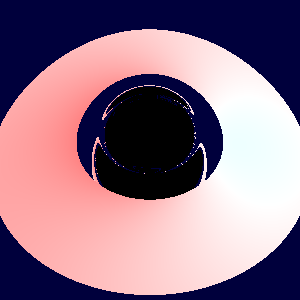}
	\end{subfigure}
	\hfill
	\begin{subfigure}[b]{0.32\textwidth}
		\centering
		\includegraphics[width=\textwidth]{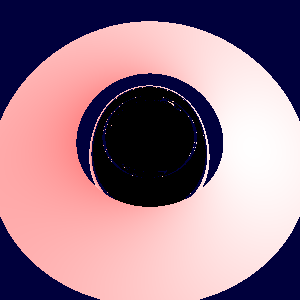}
	\end{subfigure}
	\hfill
	\begin{subfigure}[b]{0.32\textwidth}
		\centering
		\includegraphics[width=\textwidth]{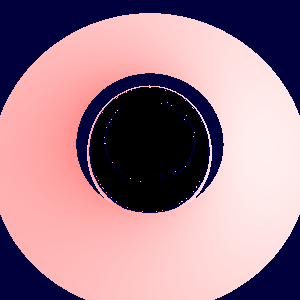}
	\end{subfigure}
	\hfill
	\begin{subfigure}[b]{0.32\textwidth}
		\centering
		\includegraphics[width=\textwidth]{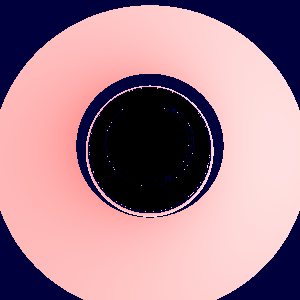}
	\end{subfigure}
	\hfill
	\begin{subfigure}[b]{0.32\textwidth}
		\centering
		\includegraphics[width=\textwidth]{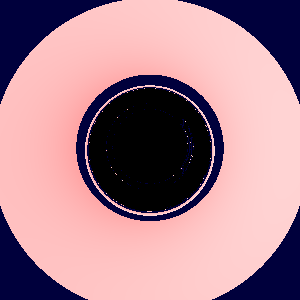}
	\end{subfigure}
	\caption{A Manko-Novikov Black hole with a ring (M = 0.5, a = 0.1M, q = -2). Redshift is calculated for the rays intersecting the ring, and colours of the ring are based on this redshift. The ring is a region in space around the equator (of slightly above non-zero thickness). If a ray intersects this bounding box it will be put through the redshift calculation described. White means the redshift factor is around $1$, red means a factor higher than $1$ and blue means a factor lower than $1$.}
	\label{MNMinDisk}
\end{figure}

\begin{figure}[h]
	\centering
	\begin{subfigure}[b]{0.32\textwidth}
		\centering
		\includegraphics[width=\textwidth]{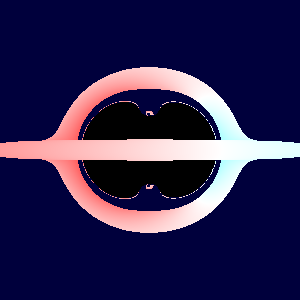}
	\end{subfigure}
	\hfill
	\begin{subfigure}[b]{0.32\textwidth}
		\centering
		\includegraphics[width=\textwidth]{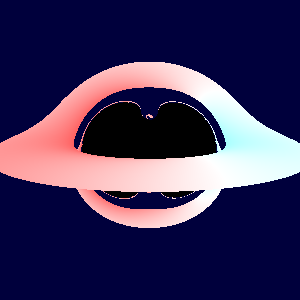}
	\end{subfigure}
	\hfill
	\begin{subfigure}[b]{0.32\textwidth}
		\centering
		\includegraphics[width=\textwidth]{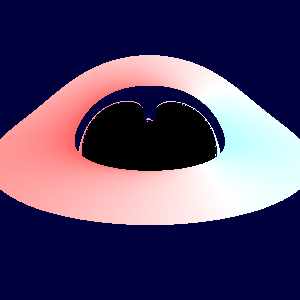}
	\end{subfigure}
	\hfill
	\begin{subfigure}[b]{0.32\textwidth}
		\centering
		\includegraphics[width=\textwidth]{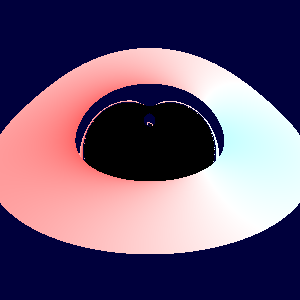}
	\end{subfigure}
	\hfill
	\begin{subfigure}[b]{0.32\textwidth}
		\centering
		\includegraphics[width=\textwidth]{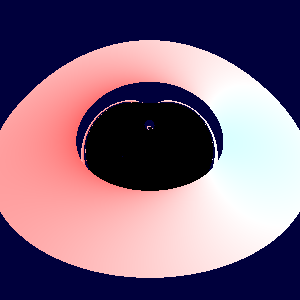}
	\end{subfigure}
	\hfill
	\begin{subfigure}[b]{0.32\textwidth}
		\centering
		\includegraphics[width=\textwidth]{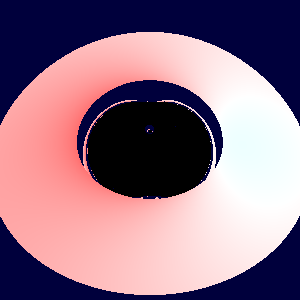}
	\end{subfigure}
	\hfill
	\begin{subfigure}[b]{0.32\textwidth}
		\centering
		\includegraphics[width=\textwidth]{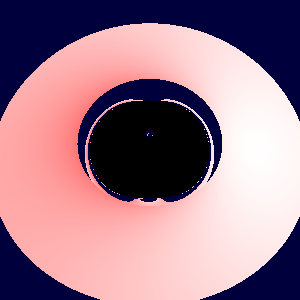}
	\end{subfigure}
	\hfill
	\begin{subfigure}[b]{0.32\textwidth}
		\centering
		\includegraphics[width=\textwidth]{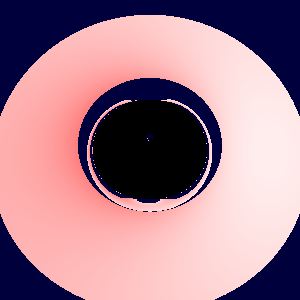}
	\end{subfigure}
	\hfill
	\begin{subfigure}[b]{0.32\textwidth}
		\centering
		\includegraphics[width=\textwidth]{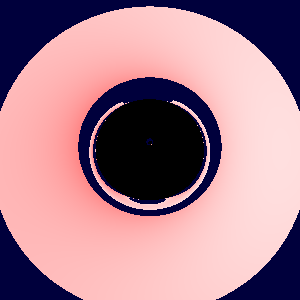}
	\end{subfigure}
	\hfill
	\begin{subfigure}[b]{0.32\textwidth}
		\centering
		\includegraphics[width=\textwidth]{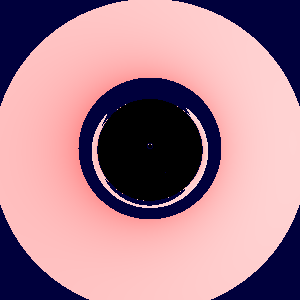}
	\end{subfigure}
	\caption{A Manko-Novikov Black hole with a ring (M = 0.5, a = 0.1M, q = -2). Redshift is calculated for the rays intersecting the ring, and colours of the ring are based on this redshift. The ring is a region in space around the equator (of slightly above non-zero thickness). If a ray intersects this bounding box it will be put through the redshift calculation described. White means the redshift factor is around $1$, red means a factor higher than $1$ and blue means a factor lower than $1$.}
	\label{MNPlusDisk}
\end{figure}

%% file: 04_code.tex
% Add a new section: The enumeration of this section will continue in the mainfile, even though it's alone here. So if this is the 5th section including earlier subfiles, this will print: '5  First section' in the main file.
\newpage % Use newpage to start on a new page
\section{Code and Requirements}
\subsection{Libaries and dependencies}
The python code makes use of several libraries. The most important and problematic of these is the CuPy library. This is a modified version of NumPy which allows calculations to be done on GPU kernels. This library is based on the CuDa language and thus will only run on CuDa compatible GPUs, and requires the correct CuDa drivers, and C compiler to be installed. The other important library used is Pillow, which is used for image processing. However this library has no such outside dependencies, and as such is much easier to get running. Furthermore NumPy is used for some of the setup calculations such as creating arrays, and being able to save the raw data arrays.
\subsection{Code Setup}
The code is split up into 2 files. One generates the rays, propagates them, and exports them as NumPy arrays, this is the script that generates the data. The other script part draws the visual images from these arrays, this is for data analysis. These are separated to not have to generate new data every time the visual analysis is changed. 

The data generating file consists of 4 sections. The first section defines the starting variables, these include the resolution, black hole parameters, focal length and $\theta$ of observation. The second section contains the main functions. These are the functions that generate the starting points, and propagate them. These are some python functions to generate an array of $3$ vectors which give the direction of the lightbeams in $3$-dimensional space, and two GPU kernels. One of these kernels turns a set of $3$-vectors to initial positions for light, the other propagates them. To use a different metric, the metricAtPoint ($g_{\mu\nu}$) and invMetricAtPoint ($g^{\mu\nu}$) in the first and second kernel need to be editted to the desired metric. 

Using different coordinates from Boyer-Lindquist will also require the coordinate change in the first kernel to be correctly changed. Derivatives of metrics are calculated numerically, and thus do not need to be calculated by hand. Objects in the space, such as a disk around the black hole are solid objects, and cannot be infinitely thin, as their collision is based on the ray propagating to a point inside volume of the object.

The drawing file contains a single GPU kernel, which turns output positions into colours. Specifically in the case of adding elements that require redshift calculations, this kernel includes a metric function, and black hole parameters. These are not read automatically from the datafiles, and need to be correctly filled in by hand.

There are different versions of these files for black holes with and without objects, as without objects much less calculation needs to take place in the drawing script specifically. All this code can be found on github \cite{myCode}.

%% file: 05_conclusion.tex
\newpage % Use newpage to start on a new page
\section{Conclusions}
The code discussed in this document was used for generating and propagating geodesics on a GPU. Using this kind of hardware meant we were able to attain much greater resolution than comparable CPU simulations in similar time. We demonstrated the code on the application of drawing shadows and lensing of black holes, as well as calculating the redshift of objects moving around them such as an accretion disk. Here we were able to attain greater resolution and accuracy than previous studies of the shadow of Manko-Novikov spacetimes, allowing us to study the strange chaotic regions of these shadows more closely.

The main goal of the code is to be widely applicable, and easy to use. For this reason most of it was written in python, and only the GPU-simulation itself was written in CUDA. Besides replacing the function defining the metric nothing has to be changed in these files to render other alternative black holes, such as Kerr spacetimes in alternate gravity, or other solutions. To allow others to use this code in their research it was made publicly available on github.

\section{Acknowledgements}
I would like to thank my masters thesis supervisors Tanja Hinderer and Álvaro del Pino Gómez. This code is an extension of the code I wrote as part of my thesis at Utrecht University \cite{myThesis}.

%% file: A_geometry.tex
% Add a new section: The enumeration of this section will continue in the mainfile, even though it's alone here. So if this is the 5th section including earlier subfiles, this will print: '5  First section' in the main file.
\newpage % Use newpage to start on a new page
\section{Mathematical Background}\label{appendixMath}
The actual thesis of which this code was a part was written from both a theoretical physics and a mathematics background. I have left the most important parts of this math. This appendix includes the proof that our Hamiltonian correctly describes the behavior of geodesics in our situation in, along with some of the notation it uses.
\subsection{Differential Geometry}
To be able to define our geometrical structures, we will be using the framework of differential geometry. We assume knowledge of basic manifold structures such as tangent- and cotangent spaces. We will also briefly use charts, and their pullback/pushforward of elements of (co)tangent spaces, these charts are often implied in physical coordinates. For an extensive treatment of the basics of differential manifolds one can check \cite{lee}. It is important to note that the Einstein index-notation often used in physics does translate directly to the language of differential geometry. Specifically we will use this notation as follows:

\begin{definition}
	Given a manifold $M$ with a vectorfield $X\in TM$ and a chart $\psi:\mathbb{R}\supset U\rightarrow M$ we define:
	\begin{align}
		X^\mu = \psi^*X \in T\mathbb{R}^n
	\end{align}
	Similarly for a $1$-form $\alpha\in T^*M$ we define
	\begin{align}
		\alpha_\mu = \psi^*\alpha \in T^*\mathbb{R}^n
	\end{align}
	Important to note is that a lower index means a $1$-form and a higher index means a vector. Multiple indices of different types give a tensor, for example:
	\begin{align}
		Y_\mu^\nu = \psi^*Y \in (T^*M\otimes TM)^*
	\end{align}
	Note that the chart is noted implicitly, as in this notation we often only consider a single chart whose image covers the part of the manifold we are interested in.
	
	For derivatives this works analogously, we write
	\begin{align}
		\partial_\mu f &= \psi^*df\\
		X^\mu \partial_\mu f &= \psi^*\mathcal{L}_X f
	\end{align}
\end{definition}
It is important to note that often charts are used that do not cover the entire manifold, for example polar coordinates can be used to describe the plane minus the origin. So all use of this index notation is local.
\subsection{Symplectic Geometry}

For our treatment of symplectic manifolds we will use definitions as written in the lecture notes by Fabian Ziltener \cite{ziltener}. The symplectic structure is encoded in the symplectic form, which we define as follows:
\begin{definition}Given a manifold $M$, we call a $2$-form $\omega$ on $M$ {\it symplectic} if it is anti-symmetric, closed and non degenerate.
	The anti-symmetric property means that at any $x\in M$, for $v,w\in T_xM$ we have $\omega_x(v,w) = -\omega_x(w,v)$.
	Closedness means that $d\omega = 0$. 
	Lastly non-degenerateness means that at a point $p\in M$ we have no $v\in T_pM$ such that $\iota_v\omega_p = \omega(v,\cdot) = 0$. This leads us to define the map
	\begin{align}
		\omega^\flat:TM&\rightarrow T^*M\\
		v&\mapsto \iota_v\omega
	\end{align}
	The non-degenerateness property now is equivalent to this map being of maximal rank and thus inverteble. We call the inverse $\omega^\sharp$.
\end{definition}

The structure of symplectic manifolds is its entire own field of mathematics, however we are specifically interested in Hamiltonian dynamics. The endeavour to rigorously define the mechanics of a Hamiltonian system on a general phase-space manifold was one of the roots of symplectic geometry. 
\begin{definition}
	Given a sympelctic manifold $(M,\omega)$ a Hamiltonian is given by a function $H:M\rightarrow\mathbb{R}$. Its dynamics are given by a vectorfield $X_H\in TM$ defined implicitly by 
	\begin{align}
		dH = \iota_{X_H}\omega = \omega^\flat X_H
	\end{align}
	Or, using the fact that $\omega$ is non-degenerate, we can invert $\omega^\flat$, so we can define it explicitly as
	\begin{align}
		X_H = \omega^\sharp(dH)
	\end{align}    
	The fact that this $X_H$ always exists and is unique now follows immediately from existence of $\omega^\sharp$.
\end{definition}

In physics, this symplectic manifold $M$ describes our phase space. Most commonly this phase space will take the shape of a cotangent bundle $T^*Q$ for a given manifold $Q$. 

To define the canonical symplectic form on this bundle we first define the Liouville form $\lambda$ on $T^*Q$ at a given point $x$. We take $q:T^*Q\rightarrow Q$ the standard projection, and define $\lambda = p\circ dq$, where $p\in T_{q(x)}^*Q$ is the element such that $x = (p,q)$. By slight abuse of notation we will usually write $x = (p,q)$ as local coordinates in which $T^*Q$ is equivalent to $\mathbb{R}^n\times \mathbb{R}^n$ where the first component corresponds to the cotangent fibre. 

We now define $\omega = -d\lambda$. It is immediately clear that this form is indeed anti-symmetric and closed by the definition of the external derivative $d$. For non-degenerateness one can rewrite the form in local coordinates $p$ and $q$, as we see in those coordinates $\omega = dq\wedge dp$, which is non-degenerate.

In general relativity we take $Q$ to be our spacetime manifold, which will later be equipped with a pseudo-Riemannian metric. Then $T^*Q$ becomes the phase space. In local coordinates we call $q$ the position and $p$ the momentum.

\subsection{Pseudo-Riemannian Geometry}
Much like symplectic geometry, Riemannian geometry is its own entire branch of differential geometry based on a structure induced by a $2$-tensor. This $2$-tensor is called the Riemannian metric, this metric locally defines distance on the manifold, and it connects naturally to the definition of connections and parallel transport. In our later physical applications we use pseudo-Riemannian metrics, which drop some of the conditions of Riemannian metrics, but maintain many of their useful properties.

\subsection{The metric}
Given a manifold $Q$, a Riemannian metric $g$ is a $2$-tensor that defines an inner product on $T_pQ$ for each point $p$. We remember that an inner product $g\langle \cdot,\cdot\rangle$ on a vectorspace is defined as a symmetric bilinear function that is positive definite. This means for our $2$-form $g$ that it is symmetric and positive definite at each point $x\in M$. From the fact that it is positive definite, we can deduce that the form is also non-degenerate, and thus like with $\omega$ we have flat map 
\begin{align}
	g^\flat : TQ&\rightarrow T^*Q\\
	g^\flat(X) &= g\langle X,\cdot\rangle
\end{align}

and its inverse $g^\sharp$. These two maps are especially important in index notation, as they are what we use to lower and raise indices of vectors and covectors. We call (again for a chart $\chi$):
\begin{align}
	X_\mu &= X^\nu g_{\mu\nu} = \chi^*g^\flat(X)\\
	\alpha^\mu &= \alpha_\nu g^{\mu\nu} = \chi^*g^\sharp(\alpha)
\end{align}

Riemannian metrics are a great tool, they can also allow one to define distance on manifolds, which a priori is not a property manifolds possess, and they give a strong link between the tangent and cotangent spaces by serving as an inner product. For our purposes however, the Riemannian metric is a slightly too narrow definition. To be able to study general relativity we need to use Pseudo-Riemannian metrics. For this we will drop the positive definiteness demand of the inner product.
\begin{definition}
	A pseudo-Riemmanian metric on a manifold $M$ is given by a $2$-form $g$ which is symmetric and non-degenerate. Symmetry means that for $x\in M$ and $v,w\in T_xM$ we have that $g_x(v,w) = g_x(w,v)$. Non-degenerateness means that for $x \in M$ there is no $v\in T_xM$ such that $\iota_v(g) = 0\in T^*_xM$.
\end{definition}

\subsection{Geodesics}
The part of Riemannian and Pseudo-Riemannian geometry that we will focus on is the calculation of geodesics. On a Riemannian manifold such geodesics are described as the locally shortest paths. This means that they are the curved space equivalent of straight lines.

We want to find the equations of motion of a geodesic on a pseudo-Riemannian manifold, in a given set of coordinates. We will not go into the entire principle of least action here, but we will give a short description. For a particle moving a long a path given by $q:\mathbb{R}\rightarrow Q$, the action is given by:
\begin{align}
	S = \int \mathcal{L}(q(\tau),\dot{q}(\tau)) d\tau
\end{align}
Where $\mathcal{L}$ is the Lagrangian, this is in general a function on $TQ$.

It is useful for us to know the conclusion, which is that the action is locally minimal for $q$ if
\begin{align}
	\frac{\partial\mathcal{L}}{\partial q} - \frac{d}{d\tau}\frac{\partial\mathcal{L}}{\partial \dot{q}} = 0
\end{align}
This is called the Euler Lagrange equation, and it describes a vectorfield in $TM$.

The Lagrangian description of a system is related to the Hamiltonian one by the Legendre transform. We will first define this transform on a vectorspace, after which we will extend it to the tangent space $TQ$ of a manifold $Q$. 

\begin{definition}
	Given a vectorspace $V$ and a function $L:V\rightarrow \mathbb{R}$ we define a derivative $DL:V\rightarrow V^*$ via the expected way
	\begin{align}
		DL(v)(w) = \frac{d}{d\tau}\Bigg|_{\tau=0} L(v+\tau w)
	\end{align}
	We now ask that this $DL$ is injective, this is equivalent with asking that the Hessian of $L$ is non-degenerate. It is then invertible (as $V^*$ and $V$ share dimension, both being finite). We define:
	\begin{align}
		L^* = L\circ DL^{-1}:V^*\rightarrow \mathbb{R}
	\end{align}
	As the Legendre transform of $L$.
\end{definition}
Next we want to translate this from vectorspaces to tangent spaces. We take a manifold $Q$, with a Lagrangian $\mathcal{L}:TQ\rightarrow \mathbb{R}$. Next we define a map $d\mathcal{L}_q:T_qQ\rightarrow T_q^*Q$ in much the same way as above, however here do it pointwise:
\begin{align}
	D\mathcal{L}_q(v)(w) = \frac{d}{d\tau}\Bigg|_{t=0} \mathcal{L}(q,v+\tau w)
\end{align}
We then take the pointwise inverse of $D\mathcal{L}$ and define $\mathcal{L}^* = \mathcal{L}\circ D\mathcal{L}^{-1}$. Next we want to show that the Hamiltonian dynamics of this Legendre transform indeed coincide with those of the Lagrangian. To see this we will show that the dynamics of Euler-Lagrange equations coincide with those of the Hamiltonian vector field, when restricted to $Q$. We will however only do this for the specific case of pseudo-Riemannian geodesics.

% Like the Lagrangian description of a system there is also a Hamiltonian description. Mathematically this description is given by how we defined Hamiltonian vectorfields on symplectic manifolds in the chapter before. It is however not immediately clear what kind of physics such a system would describe, so we will link their dynamics to that of a Lagrangian system. Given such a system, in physics we define:
% \begin{align}
	%     p &= \frac{\partial\mathcal{L}}{\partial \dot{q}}\\
	%     H &= p \dot{q} - \mathcal{L}
	% \end{align}
% the equations of motion are now given by
% \begin{align}
	%     \dot{p} &= -\frac{\partial H}{\partial q}\\
	%     \dot{q} &= \frac{\partial H}{\partial p}    
	% \end{align}
% Since we are not going into the calculus of variations here this will be the extent of the general description of these systems. We will now limit ourselves to only those systems that we call free particles in general relativity.

\subsection{The Legendre Transform for Pseudo-Riemannian geodesics}
We take $(Q,g)$ a Riemannian manifold, and $(T^*Q,\omega)$ a symplectic manifold with $\omega$ the canonical symplectic form. The Lagrangian description of a free particle gives us a map
\begin{align}
	\mathcal{L}:TQ&\rightarrow \mathbb{R}\\
	(q,v)&\mapsto \tfrac{1}{2}g_q(v,v)
\end{align}
We will now show that we can use a Legendre transform and turn this Lagrangian into a Hamiltonian system. First we define our Legendre transform. We start by calculating $D\mathcal{L}$. We see that
\begin{align}
	D{\mathcal{L}}_q(v)(w)&= g_q(v,\cdot)(w)\\
	&= p(w)
\end{align}
We call $p$ the momentum. We now see, using invertibility of $g$
\begin{align}
	H(q,p) &= \mathcal{L}L\circ D{\mathcal{L}}^{-1}_q(p)\\
	&= \mathcal{L}(q,g^\sharp(p))\\
	&= \tfrac{1}{2} g_q(g^\sharp(p),g^\sharp(p))
\end{align}

% Suppose we look in canonical coordinates $(p_i,q_i)$ around a point where the canonical symplectic form on $T^*Q$ is given by $dq\wedge dp$. We see that
% $$
% X_H = \sum_i \partial_{p_i}H\partial_{q_i} - \partial_{q_i}H\partial_{p_i}
% $$
% We want to show that the above follows from the Euler-Lagrange equation
% $$
% \partial_q \mathcal{L} - \frac{d}{dt}\partial_{\dot{q}}\mathcal{L} = 0
% $$

We will now show that the Hamiltonian defined above actually encodes the dynamics of the geodesics on a pseudo-Riemannian manifold. We note that the dynamics of vectorfields is a local property, and we can thus work in charts. It only remains for us to show that the Hamiltonian and Lagrangian describe the same dynamics on $\mathbb{R}^4$ with an arbitrary Pseudo-Riemannian metric $g$.

To start, we see that the Liouville form on $T^*\mathbb{R}^4 = (\mathbb{R}^4)^*\oplus\mathbb{R}^4$ is given by
\begin{align}
	\lambda: \big((\mathbb{R}^4)^*\oplus\mathbb{R}^4\big)\oplus \big((\mathbb{R}^4)^*\oplus\mathbb{R}^4\big) &\rightarrow \mathbb{R}\\
	(\alpha,v,p,q)&\mapsto p(v)
\end{align}
Here $(\mathbb{R}^4)^*$ is the fibre term. This means we can define the symplectic form $\omega = -d\lambda$ as:
\begin{align}
	\omega: \big((\mathbb{R}^4)^*\oplus\mathbb{R}^4\big)\oplus\big((\mathbb{R}^4)^*\oplus\mathbb{R}^4\big)\oplus \big((\mathbb{R}^4)^*\oplus\mathbb{R}^4\big) &\rightarrow \mathbb{R}\\
	(\alpha,v,\beta,w,p,q)&\mapsto \alpha(w)-\beta(v)
\end{align}

We now look again at the Lagrangian of the geodesic motion on $T\mathbb{R}^4$, we also translate to index notation as we are working in coordinates on $\mathbb{R}^4$:
\begin{align}
	\mathcal{L}(q,v) = \tfrac{1}{2} g(v,v) = \tfrac{1}{2}g_{\mu\nu} v^\mu v^\nu
\end{align}
Using the Euler-Lagrange equations we get the geodesic equation, for the full calculation see \cite{carroll}:
\begin{align}
	\frac{d^2}{d\tau^2}q^\mu + \Gamma^\mu_{\rho\sigma}\frac{d}{d\tau}q^\rho\frac{d}{d\tau}q^\sigma = 0
	\label{geodesicEq}
\end{align}
Where
\begin{align}
	\Gamma^\mu_{\rho\sigma} = \tfrac{1}{2}g^{\mu\nu}(\partial_\rho g_{\mu\sigma} + \partial_\sigma g_{\mu\rho} - \partial_\mu g_{\rho\sigma})
\end{align}

We compare this to the dynamics of the Hamiltonian on $T^*\mathbb{R}^4$, given by
\begin{align}
	H(p,q) = \frac{g(g^\sharp(p),g^\sharp(p))}{2} = \frac{g^*(p,p)}{2}
\end{align}
% We will first show this is the Legendre transform of the Lagrangian above. We see that
% \begin{align*}
	% DL_q(\dot{q})(\dot{q}') &= \frac{d}{dt}\Bigg|_{t=0,q} L(\dot{q} + t\dot{q}')\\
	% g(\dot{q},\dot{q}')\\
	% \end{align*}
% so
% $$
% DL^{-1}_q(p) = g_q^\sharp(p)
% $$
% and from that it follows
% $$
% H(p,q) = \frac{g(g^\sharp(p),g^\sharp(p))}{2} = \frac{g^*(p,p)}{2}
% $$

We calculate the Hamiltonian vectorfield $X_H$. We remember the definition
\begin{align}
	\omega_{p,q}(X_H,\cdot) = dH_{p,q}
\end{align}
Now let $(\alpha,v)\in T(T_{p,q}^*\mathbb{R}^4)$. We calculate
\begin{align}
	dH_{p,q}(\alpha,v) &= \mathcal{L}_{(\alpha,v)}H|_{(p,q)}\\
	&= \mathcal{L}_{(\alpha,0)}H|_{(p,q)} + \mathcal{L}_{(0,v)}H|_{(p,q)}\\
	&= \frac{d}{d\tau}\bigg|_{\tau=0}H(p+\tau\alpha,q) + \frac{d}{d\tau}\bigg|_{\tau=0}H(p,q+\tau v)\\
	&= \frac{d}{d\tau}\bigg|_{\tau=0} \frac{g_q^*(p + \tau\alpha,p + \tau\alpha)}{2} + \frac{d}{d\tau}\bigg|_{\tau=0} \frac{g^*_{q+\tau v}(p,p)}{2}\\
	&= g^*(p,\alpha) + \frac{(\mathcal{L}_v g^*)(p,p)}{2}
\end{align}
So we can say that
\begin{align}
	dH_{p,q} = g^*(p,\pi_p) + \frac{(\mathcal{L}_q g^*)(p,p)}{2}\pi_q
\end{align}
where $\pi_p$ is the projection to the fibre and its tangent space, so to $ T(\mathbb{R}^4)^*\subset T(T^*\mathbb{R}^4)$, meanwhile $\pi_q$ is the projection to the manifold part of the chart, so to $ T\mathbb{R}^4\subset T(T^*\mathbb{R}^4)$.
\begin{align}
	X_{H,(p,q)} = \omega_{p,q}^\sharp\Bigg(g^*(p,\pi_p) + \frac{(\mathcal{L}_{\pi_q} g^*)(p,p)}{2}\Bigg)
\end{align}
Here we use $\tau$ as the variable of the parametrisation:
\begin{align}
	\partial_\tau (q,p) = X_{H,(q,p)}
\end{align}
and so
\begin{align}\label{hamGeodesic}
	\frac{d}{d\tau} q &= g^\sharp(p)\\
	\frac{d}{d\tau} p &= -\frac{(\mathcal{L}_{q} g^*)(p,p)}{2}
\end{align}
% Here $(\mathcal{L}_{q} g^*)$ is what in GR is generally called the Christoffel symbols. We use the definition of the Lie derivative
% \begin{align*}
	% (\mathcal{L}_{V} g^*)_q(p,p) &= \frac{d}{ds}\bigg|_{s=0}(\Phi^s_{V,*}g^*)_q(p,p)\\
	% &= \frac{d}{ds}\bigg|_{s=0}(g^*)_{\Phi^t_{V}(q)}(\Phi^{s,*}_{V}p,\Phi^{s,*}_{V}p)
	% \end{align*}
We now want to translate this to index notation, to get the geodesic equation \ref{geodesicEq}.

Rewriting the above into index notation we are given:
\begin{align}
	\frac{d}{d\tau} q^\mu &= g^{\mu\nu}p_\nu\\
	\frac{d}{d\tau} p_\mu &= -\frac{(\partial_\mu g^{\rho\sigma})p_\sigma p_\rho}{2}
\end{align}
We will now show that these align with the usual definition for the geodesic equation:
\begin{align}
	\frac{d^2}{d\tau^2}q^\mu &= \frac{d}{d\tau}(g^\sharp p)^\mu\\
	&= \frac{d}{d\tau}(g^{\mu\sigma}p_\sigma)\\
	&= \frac{d}{d\tau}(g^{\mu\sigma})p_\sigma + g^{\mu\sigma}\frac{d}{dt}p_\sigma\\
	&= \Big(\frac{d}{d\tau}q^\rho\Big)(\partial_\rho g^{\mu\sigma})p_\sigma - g^{\mu\sigma}\partial_\sigma H\\
	&= (\tfrac{d}{d\tau}q^\rho)(\partial_\rho g^{\mu\sigma})g_{\nu\sigma}\tfrac{d}{d\tau}q^\nu  - \tfrac{1}{2}g^{\mu\sigma}\partial_\sigma g^{\rho\nu}p_\rho p_\nu\\
	&= \tfrac{1}{2}g^{\mu\nu}\big(
	-\partial_\nu g_{\rho\sigma} + \partial_\sigma g_{\rho\nu} + \partial_\rho g_{\sigma\nu}
	\big) \tfrac{d}{d\tau}q^\sigma \tfrac{d}{d\tau}g^\rho
\end{align}
In the last line the the first term was made symmetric in $\rho\sigma$, as the $p_\rho p_\sigma$ is also symmetric, and indeces were lifted/lowered to conform with the usual definition of $\Gamma$ without torsion. 

We have now shown that the Hamiltonian and Lagrangian description give the same dynamics in coordinates, and since this is a local property we can state the same for general pseudo-Riemannian manifolds.